\documentclass[preprint2,times,tighten]{aastex6}

\usepackage{color}
\usepackage{enumitem}
\usepackage{hyperref}
\usepackage{verbatim}
\usepackage{amsmath,amstext,mathrsfs}
\usepackage[all]{hypcap} 
\usepackage{afterpage}
\usepackage{color}
\usepackage{soul}

\begin{document}
\title{Tracing magnetic fields with spectroscopic channel maps}

\author{A. Lazarian\altaffilmark{1}, Ka Ho Yuen\altaffilmark{1}}
\email{lazarian@astro.wisc.edu }
\altaffiltext{}{Department of Astronomy, University of Wisconsin-Madison, USA}

\begin{abstract}
We identify velocity channel map intensities as a new way to trace magnetic fields in turbulent media. This work makes use of both of the modern theory of MHD turbulence predicting that the magnetic eddies are aligned with the local direction of magnetic field, and also the theory of the spectral line Position-Position-Velocity (PPV) statistics describing how the velocity and density fluctuations are mapped into the PPV space.  In particular, we use the fact that, the fluctuations of intensity of thin channel maps are mostly affected by the turbulent velocity, while the thick maps are dominated by density variations.  We study how contributions of the fundamental MHD modes affect the Velocity Channel Gradients (VChGs) and demonstrate that the VChGs arising from Alfven and slow modes are aligned perpendicular to the local direction of magnetic field, while the VChGs produced by the fast mode are aligned parallel to the magnetic field. The dominance of the Alfven and slow modes in the interstellar  media will therefore allow a reliable magnetic field tracing using the VChGs.  We explore ways of identifying self-gravitating regions that do not require polarimetric information. In addition, we also introduce a new measure termed "Reduced Velocity Centroids" (RVCGs) and compare its abilities with the VChGs.  We employed both measures to  the GALFA 21cm data and successfully compared thus magnetic field directions with the {\it PLANCK} polarization. The applications of the suggested techniques include both tracing magnetic field in diffuse interstellar media and star forming regions, as well as removing the galactic foreground in the framework of cosmological polarization studies. 
\end{abstract}

\section{Introduction}

Interstellar media (ISM) of spiral galaxies is both magnetized and turbulent (see \citealt{Armstrong1995ElectronMedium, Chepurnov2010ExtendingData}) with turbulent magnetic fields playing a critical role for many key processes, including the process of star formation (see \citealt{MO07,MK04}), the propagation and the acceleration of cosmic rays (see \citealt{J66,YL08}) and the regulating heat and the mass transfer between different ISM phases (see \citealt{D09} for the list of the different ISM phases). In addition, galactic magnetic fields are responsible for the polarized radiation that presents a serious obstacle for the studies of the polarization of cosmological origin.
Therefore it is essential to have a reliable way to study the properties of magnetic fields in the ISM.  

Magnetic fields make turbulence anisotropic, with turbulent eddies elongated along the magnetic field (see \citealt{2013SSRv..178..163B}, for a review). As a result, observed velocity correlations are expected to be elongated along the underlying magnetic field with this property, which was demonstrated with synthetic observations in previous studies (\citealt{2002ASPC..276..182L,2005ApJ...631..320E}, henceforth EL05). Later the study of anisotropies was performed using the Principal Component Analysis (PCA) and applied to observations \cite{2008ApJ...680..420H}.\footnote{The idea behind the studies, which is employing the PCA is the same and, in fact, our study in \cite{2016ApJ...818..118C} shows that there is no practical advantage of using the PCA compared to the velocity centroids. On the contrary, the anisotropies of centroids, unlike the PCA eigen-images, are analytically related to the properties of the underlying turbulence, i.e. to the properties of the Alfven, slow and fast modes that constitute the MHD cascade (see \citealt{2017MNRAS.464.3617K}). This opens prospects of separating the contribution of the compressible (slow and fast) and the in-compressible (Alfven) modes using velocity centroids.}     

There is another way to employ properties of MHD turbulence in order to study magnetic fields. The aforementioned turbulent eddies are aligned with magnetic field which entails that the velocity gradients should have larger values for gradients calculated in the direction perpendicular to the magnetic field. This property of magnetic turbulence was employed in \citeauthor{GL17} (\citeyear{GL17}, henceforth GL17), in which the approach of using velocity centroids gradients (VCGs) to trace magnetic field orientations was proposed. The technique was further extended and elaborated in \citeauthor{YL17} (\citeyear{YL17}, henceforth YL17a) and the new way of magnetic field tracing was successfully compared with observations on polarization from {\it PLANCK}. The subsequent studies in \citeauthor{YL17b} (\citeyear{YL17b}, henceforth YL17b) revealed the synergy of simultaneous use of VCGs and intensity gradients (IGs). \footnote{By itself, the IGs were shown to be inferior to the VCGs in the ability of tracing magnetic field (GL17, YL17a,b), but synergistic in terms of studying shocks and regions dominated by self-gravity.} These papers introduced a new way of studying magnetic fields with spectroscopic data as well as studying other key processes taking place in the ISM. 

In a separate development, the same idea of tracing magnetic fields with gradients was employed with synchrotron intensity maps in \citeauthor{LYLC17} (\citeyear{LYLC17}, henceforth LYLC). Using synthetic data which LYLC showed that synchrotron intensity gradients (SIGs) can reliably trace the magnetic field in the ISM, and the study also confirmed the conclusion by comparing {\it PLANCK} synchrotron polarization maps with the SIGs maps. The synergies of SIGs, IGs and VCGs in self-gravitating media were all demonstrated in \cite{YL17b}, showing that the relative rotations between different types of gradient vectors are informative of the stage of collapse of a piece of self-gravitating cloud.

Velocity centroid is a way of representing ISM velocities using observations, and this way has its limitations. For example, centroids reflect the contributions arising from both the velocity and the density fluctuations along the line of sight (see EL05). However, the density fluctuations are not as well aligned with magnetic fields as velocity (see \citealt{CL03,2005ApJ...624L..93B,2007ApJ...658..423K}, or the comparison of VCGs to IGs in YL17b ). At the same time, the analytical study in \cite{2000ApJ...537..720L,2004ApJ...616..943L} revealed that channel maps are sensitive only to velocity fluctuations if the corresponding turbulent density fluctuations are dominated by large scale contributions. In addition, studies of different velocity channels in some cases may allow separated contributions which are coming from spatially different regions along the line of sight, and therefore enabling one to study the 3D structure of magnetic field as well as other effects that the velocity gradients are sensitive to. This suggests that in a number of cases the VChGs may have advantages compared to the VCGs. This motivates our present study to explore the ability of velocity channel gradients (VChGs) in tracing the orientations of the magnetic field.   

In some aspects the VChGs have similarities with the technique based on studies of filaments in the velocity HI channel maps in \cite{Clark15}, where these filaments were shown to be correlated with the magnetic field directions as revealed the {\it PLANCK} polarimetry. On the basis of \cite{2000ApJ...537..720L} one can conclude that the filaments observed by \cite{Clark15} in thin channel maps can be identified with caustics caused by velocity crowding. The relationship between the VChGs and the underlying velocity gradients is more straightforward than for the filaments, and therefore, we expect a better correlation between the VChGs and the magnetic field than with the filaments. Not to mention, the gradient technique shows its capability identifying shock and self-gravitating regions (YL17b), while no similar conclusion has ever been reported using filaments. Whether or not there can be any synergistic usage of the simultaneous use of the VChGs and tracing the channel filaments should be answered by further additional studies. 

In what follows, we discuss in \S \ref{sec:theory} the theoretical motivation of this work. \S \ref{sec:numerics} discusses the numerical methods on simulations and the ways of doing analysis. We explore the gradients in the channel maps from turbulent velocities in \S \ref{sec:gradients-in-fsa}.
We examine the performance of channel map gradients and correlation anisotropies in \S \ref{sec:anisotropy}. In \S \ref{sec:fluctuations} we compare the performance of density fluctuations and velocities from channel maps. We explore the reduced centroid gradient in \S \ref{sec:reducedcentroid}. We testify our method with observational data in \S \ref{sec:obs}. We discuss our result in \S \ref{sec:discussion}, and make our conclusion in \S \ref{sec:conclusion}.

\section{Theoretical motivation and expectations}
\label{sec:theory}

 \subsection{Anisotropy of MHD turbulence: illustration}
 
MHD turbulence theory is an old subject that has been boosted recently by the ability of performing high resolution 3D numerical simulations. Before that there was no way of testing theoretical constructions and many competing theories describing MHD turbulence were able to co-exist. \footnote{We are talking about realistic 3D MHD turbulence theories. MHD turbulence in 2D is very different from the one in 3D (see \citealt{2011ApJ...743...51E}).}  For instance, the original studies of Alfvenic turbulence by \cite{I64} and \cite{K65} were based on a hypothetical model of isotropic MHD turbulence, while the later studies (see \citealt{1981PhFl...24..825M, 1983PhRvL..51.1484M, 1983JPlPh..29..525S,1984ApJ...285..109H}) demonstrated the anisotropic nature of the MHD cascade.

\begin{figure}[t]
\centering
\includegraphics[width=0.48\textwidth]{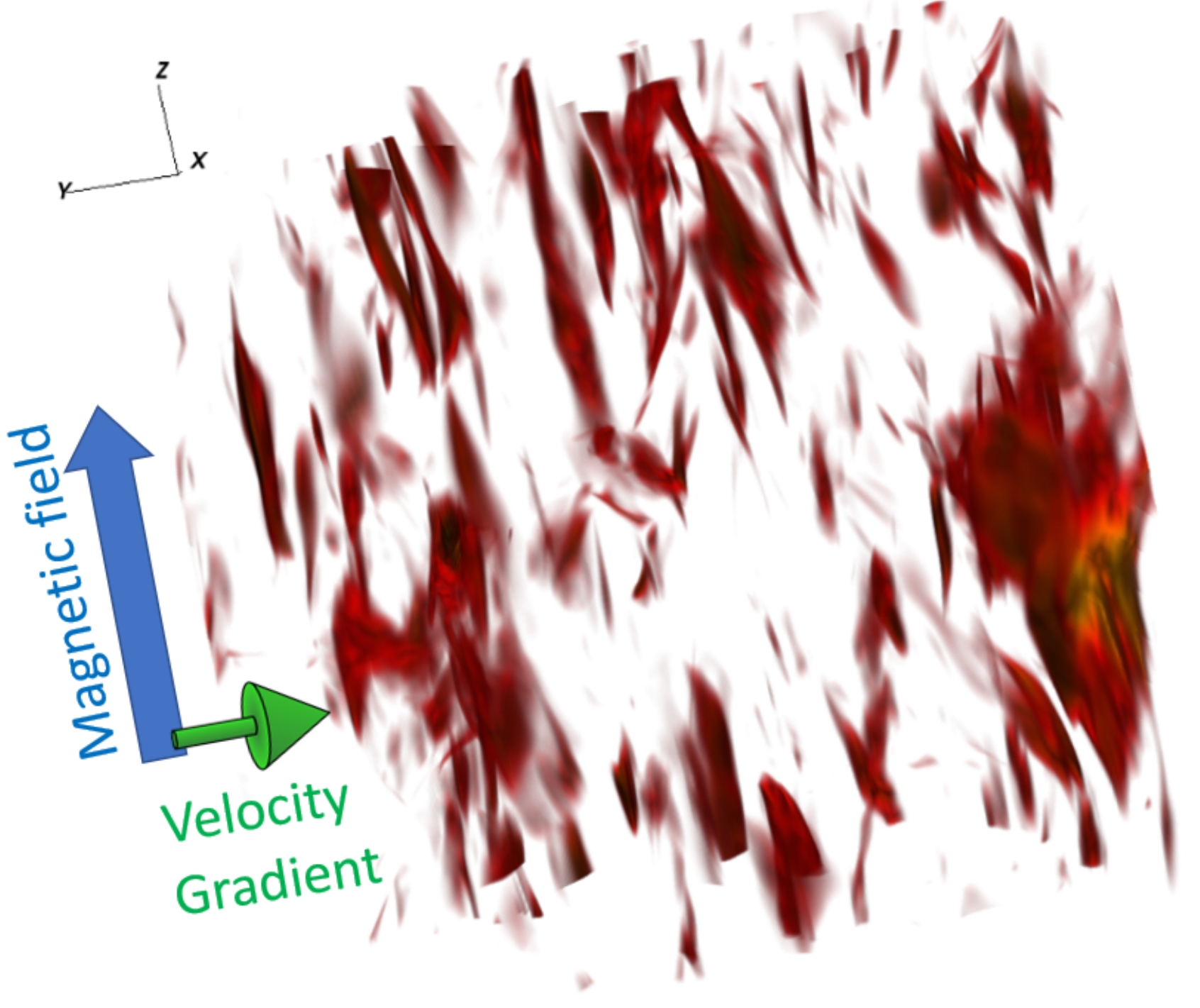}
\caption{\label{fig:illus0}A 3D snapshot of velocity iso-contours (red structures) from a sub-Alfvenic simulation showing they are all elongated with the magnetic field directions (Blue shows the mean directions, which is along the z-axis). One can imagine if gradients are calculated in this snapshot, they will be perpendicular to the field directions.  }
\end{figure}

Figure  \ref{fig:illus0} shows the visualization\footnote{Produced using Visit 2.8.1: https://wci.llnl.gov/simulation/computer-codes/visit} of our numerical simulations. In particular, the iso-contours of velocity in for the sub-Alfvenic turbulence are shown. It is obvious that the velocity gradients are directed perpendicular to the magnetic field and therefore by studying the direction of velocity gradients one can study the direction of magnetic field. In what follows, we explain that, in fact, one is expected not only to trace mean magnetic field in the sub-Alfvenic turbulence, but trace the magnetic field in its complexity both in sub-Alfvenic and super-Alfvenic turbulence.

\subsection{Anisotropy of MHD turbulence: local directions of magnetic field}

The modern theory of MHD turbulence originates from the prophetic work by (\citealt{GoldreichP.Sridhar1995GS95IITurbulence}, henceforth GS95). Given originally rather lukewarm acceptance by the MHD turbulence community, this theory nevertheless was supported by further theoretical and numerical studies (\citealt{Lazarian1999ReconnectionField}, henceforth LV99, \citealt{2000ApJ...539..273C,2001ApJ...554.1175M,Lithwick2001CompressiblePlasmas,Cho2001SimulationsMedium,Cho2002CompressiblePlasmasb,CL03,Kowal2010VelocityScalingsb}, see \citealt{2013SSRv..178..163B} for a a review) that extended the theory and provided their rigorous testing. Our present study is based on the modern understanding of the MHD turbulence cascade and the statistical properties of MHD turbulence that are confirmed numerically.

In the sub-Alfvenic regime, i.e. for the injection velocity $V_L$ being less than the Alfven velocity $V_A$, the Alfven modes initially evolve by increasing the perpendicular wavenumber  while keeping the parallel wavenumber the same (see LV99, \citealt{Gal2005}). The increase of the perpendicular wave number makes the Alfvenic wave-vectors more and more perpendicular to magnetic field. Therefore the gradients of velocity structures tend to be perpendicular to the magnetic-field direction. Eventually at a scale $l_{trans}\approx L M_A^2$, where $L$ is the turbulence injection scale and $M_A=V_L/V_A$ is the Alfven Mach number (see LV99, \citealt{Lazarian2006}), the parallel scale of eddies starts changing, signifying the start of the GS95 regime. In the GS95 cascade both parallel and perpendicular wave-numbers of Alfvenic perturbations increase but the eddies get more and more elongated. To quantify this, one should adopt the system of reference aligned with the local magnetic field.\footnote{ In the works that followed groundbreaking GS95 paper (namely, LV99, \citealt{2000ApJ...539..273C,2001ApJ...554.1175M,Cho2002CompressiblePlasmasb}), it was shown that to describe MHD turbulence one should use the {\it local} magnetic field rather than the mean magnetic field.}  In such a system of reference the fluid motions perpendicular to magnetic fields are not constrained by magnetic tension. This is the consequence of the fast turbulent reconnection that was shown to change the topology of the interacting magnetic flux tubes within the eddy turnover time (LV99). Therefore it is not surprising that the turbulent energy is channeled along this path of the least resistance. It is not surprising either that in the absence of magnetic resistance to the mixing motions, the perpendicular eddies evolve along the Kolmogorov cascade with the Kolmogorov scaling $v_l\sim l^{1/3}$.

However, the motions mixing magnetic field in the perpendicular direction also induce wave propagation along magnetic fields. The period of these waves $l_{\|}/V_A$ is equal of the period of the magnetized plasma mixing in the perpendicular direction $l_{\bot}/v_l$. The equality
of the two time scales is usually referred as the {\it critical balance} and the scales $l_{\|}$ and $l_{\bot}$ are associated with the parallel and perpendicular scales of the eddies. The relation between these scales for sub-Alfvenic turbulence is (LV99):
\begin{equation}
l_{\|}\approx L \left(\frac{l_{\bot}}{L}\right)^{2/3} M_A^{-4/3},
\label{lpar}
\end{equation}
which testifies that, for $l_{bot}$ is much smaller than the injection scale $L$, and therefore the eddies are strongly elongated. Note that these eddies are aligned with the local magnetic field. For trans-Alfvenic turbulence $M_A=1$ and Eq. (\ref{lpar}) provides the original GS95 scaling. Note that the alignment of the motions in respect to the local system of reference is an essential part of MHD turbulence as we understand it now. It is easy to see that due to this peculiar property of MHD turbulence, the velocity gradients are expected to be aligned with the {\it local} magnetic field. It is easy to show that the gradients of the smallest eddies produce the strongest signal. Indeed, the gradient of the eddy aligned mixing up local magnetic field lines is proportional to $v_l/l_{\bot}\sim l_{\bot}^{-2/3}$. The smallest eddies are the eddies at the resolution scale of the telescope. Thus we can claim that the velocity gradients trace magnetic field at the scale of the instrument resolution. 

It was shown numerically (see Cho \& Lazarian 2002, 2003) that compressible MHD turbulence can be presented as a composition of three cascades that marginally exchange energy between them\footnote{This is not true for the relativistic MHD turbulence where the coupling between fast and Alfvenic fundamental modes can be significant \citep{2016ApJ...831L..11T}.} . These are cascades of Alfvenic, slow and fast modes. We use the word "modes" rather than "waves" as in strong MHD turbulence the properties of motions may not be wave-like. As we discussed earlier, Alfvenic modes are essentially eddies and they non-linearly decay within one period (see more in a review Brandenburg \& Lazarian 2013). Within the GS95 picture and its generalization to the compressible media (GS95, \citealt{Lithwick2001CompressiblePlasmas, Cho2002CompressiblePlasmasb, CL03, Kowal2010VelocityScalingsb}) the slow modes are slaved by Alfven modes. Indeed, Alfven modes shear and cascade slow mode perturbations, which is the process confirmed by numerical simulations. Thus, in agreement with numerical studies, the anisotropic scaling given by Eq. (\ref{lpar}) is valid for slow mode eddies. As a result, both the gradients of Alfven modes and of slow modes are expected to be {\it perpendicular} to the {\it local} directions of magnetic field. This is the key idea behind tracing magnetic-field directions with velocity gradients. 

\subsection{Intensity fluctuations in thin channel maps: effect of velocity fluctuations}

While velocity gradients are directly not available from astrophysical observations of diffuse media, a number of measures can be constructed using observational data. In \cite{GL17} we explored Velocity Centroid Gradients (VCGs) of the first order and considered the higher order VCGs in \cite{YL17}. In this paper we prove another way to probe turbulent velocities, namely,  to use velocity channel maps  that can be constructed using spectroscopic observations of Doppler shifted lines. The statistics of these maps has been described in \citeauthor{2000ApJ...537..720L} (\citeyear{2000ApJ...537..720L}, henceforth LP00) for the optically thin data and in \cite{2004ApJ...616..943L} for the observations in the presence of absorption. In what follows we concentrate on the optically thin case and only mention some the possible effects of the optically thick data. Note that when we discuss thin and thick velocity slices, we mean not the effects of absorption but the thickness of channel maps. The minimal thickness of the latter is determined by the spectral resolution $\Delta v$ of the instrument, and it can be increased by integrating the spectroscopic data over larger $\Delta v$.  

An important prediction in LP00 is that velocity caustics create fluctuations of intensity of the channel maps and the relative importance of the velocity and density fluctuations changes with the thickness of the channel maps. In particular, LP00 identified a regime of "thin velocity slices" and found that in this regime the intensity fluctuations in the slice are dominated by the velocity fluctuations, provided that the three dimensional density spectrum is steep, i.e. most of the energy is concentrated at the large scales. In what follows we use the terms "velocity slices" and "channel maps" interchangeably.

The aforementioned statement about the steep density spectrum can be expressed in terms of the 3D power spectrum $P(k)$. In terms of $P(k)$ the Kolmogorov cascade corresponds to $k^{-11/3}$ and it is steep. The borderline spectrum is $k^{-3}$, with turbulence having spectrum $k^{-\alpha}$, $\alpha<3$ containing more energy at the small scale and therefore being shallow. For subsonic flows the density spectra in MHD turbulence are steep (see \citealt{2007ApJ...658..423K}).\footnote{The traditionally used definition of the Kolmogorov spectrum is obtained via spatial integration of $P(k)$ in the k-space, which is equivalent to multiplying the spectrum by $k^2$. Thus the usually referred value of the Kolmogorov spectrum is $k^{-5/3}$ and the border line spectrum between the shallow and the steep is $k^{-1}$.} Thus for such flows the intensity fluctuations in thin velocity slices of Position-Position-Velocity (PPV) data are influenced only by the turbulent velocity statistics. Naturally, the thermal broadening interferes with the minimal slice of the PPV for which the fluctuations can be studied. This means that for studying thin slices of subsonic flows one should use heavier species that those of the main hydrogen astrophysical flow. On the other hand, if the density spectrum is shallow, i.e. $\alpha<3$, the contribution of density and velocities to the statistics of intensity fluctuations of the velocity slices was evaluated in LP00. For thin slices, both velocity and density are important, while the contribution of velocity decreases as the slice thickness increases. For instance, when the integration is performed over the entire line, only density fluctuations determine the fluctuations of the resulting intensity distribution. In fact, no matter how steep the spectrum is, velocity contribution can be enhanced by reducing the channel width, which significantly simplifies the interpretation of the VChGs' directions in terms of magnetic field tracing.

In view of the above, it is important to perform the study of the gradients of intensities within PPV slices. In terms of separating compressible and incompressible components in KLP16, such a study provides an additional test that the anisotropies are actually caused by turbulence. By varying the thickness of the slice one can study the variation of gradients as the relative contribution of density and velocity changes. This may be important as the density and velocity, in general, have different statistics and the fluctuations of these fields are aligned with magnetic field to a different degree. For instance, the strongly supersonic flows demonstrate an isotropic spectra of density. In addition, for studies of galactic flows the regular galactic shear opens a way to study different turbulent regions separately and therefore different channel maps can be associated with different locations within the galactic volume.  In addition, velocity and density gradients can behave differently, e.g in self-gravitating regions and shocks (See YL17b). In such cases comparing alignments of VChGs from different channel widths opens a new way on locating self-gravitating media, regardless of the steepness of density spectrum.

In what follows the advancements of the understanding of the theory of PPV anisotropies in KLP16 guide us in studying gradients in the PPV velocity slices. Note that the criterion for the velocity slice of being thin or thick as it given in LP00 is as follows: the velocity slice is thin if the square root of turbulent velocity dispersion on the scales that the slice is being studied $\sqrt{\langle \delta v_R^2\rangle}$ is {\it greater} than the thickness of the slice $\Delta v$, i.e.
\begin{equation}
\sqrt{\langle \delta v_R^2\rangle}>\Delta v,
\label{criterion}
\end{equation}
where $R$ is the separation of the correlating points over plane of the sky, i.e. the PP separation. In the following, we distinguish thin and thick slices depending on whether the ratio $\Delta v/\sqrt{\langle \delta v_R^2\rangle}$ smaller or larger than unity. In observations the thickness $\Delta_v$ can be constrained either by the velocity resolution of the instrument or by the thermal line width $\sigma_{th}$that according to LP00 acts similarly. Therefore to make sure that the gradients are measured in the thin slice regime one should make sure that the scale over which the velocity gradients are calculated are sufficiently large.\footnote{LP00 study shows that velocity fluctuations are still important for the slice thicknesses larger than that given by the Eq. (\ref{criterion}). However, their relative contribution compared to density is gradually decreasing. Incidentally, this regime was termed in LP00 "thick slice regime". In the present paper we do not use this terminology and refer to "thick slice" only to the situations where the velocity information is integrated out.} For an individual turbulent volume $\delta v_R$ can be measured using the dispersion of the structure functions.

We note that if $\sqrt{\langle\delta v_R^2\rangle}<\sigma_{th}$ then it is advantageous to use velocity centroids that can represent velocity statistics for subsonic turbulence (see Esquivel \& Lazarian 2005, 
Kandel et al. 2017b). The traditional centroids, however, have disadvantages to channels, e.g. one cannot isolate a particular region of the galaxy according to the rotation curve. To problem in the paper we  discuss the use of a new construction that we term "reduced velocity centroids" that make use of the part of the line only. Gradients of the reduced velocity centroids and the VChGs can be used together, with the VChGs applied to scales $R$ for which the criterion given by Eq. (\ref{criterion}) is valid and with the reduced velocity centroid gradients for smaller scales. 

This all suggestive that the scale of the data blocks for calculating the gradients is another important parameter for the calculating of gradients. In YL17a we dealt with the velocity centroids we showed that the uncertainty of the calculation of velocity centroid gradients (VCGs) decreases with the increase of the data block size. This was augmented in YL17b by a finding that there was an optimal size for the VCGs which provided the maximal resolution without compromising the accuracy. Below we prove that this property is also a part of the VChGs technique.

\section{Numerical Simulations}

\label{sec:numerics}
\subsection{MHD turbulence simulations and mode decomposition}

The numerical data is obtained by 3D MHD simulations using a single fluid, operator-split, staggered grid MHD Eulerian code ZEUS-MP/3D \citep{2006ApJS..165..188H} to set up a three-dimensional, uniform, isothermal turbulent medium. Periodic boundary conditions are applied to emulate a part of interstellar cloud. Solenoidal turbulence injections are employed. Our simulations employ various Alfvenic Mach numbers $M_A=V_L/V_A$ and sonic Mach numbers $M_s=V_L/V_s$, where $V_L$ is the injection velocity, while $V_A$ and $V_s$ are the Alfven and sonic velocities respectively, which they are listed in Table \ref{tab:simulationparameters}. The domain $M_A<M_s$ corresponds to the simulations of plasma with magnetic pressure larger than the thermal pressure, i.e. plasma with low $\beta/2=V_s^2/V_A^2<1$, while the domain $M_A>M_s$ corresponds to the pressure dominated plasma with $\beta/2>1$. Further we refer to the simulations in the table by their model name. For instance, our figures will have the model name indicating which data cube was used to plot the figure. The simulations are named with respect to a variation of $M_s$ \& $M_A$ in ascending values of $\beta$. The ranges of $M_s, M_A, \beta$ are selected so that they cover different possible scenarios of astrophysical turbulence from very subsonic to supersonic cases. In this study we devoted much of our analysis to sub- and trans- Alfvenic cases only, and postpone the discussion on super-Alfvenic simulations to our next paper (Yuen \& Lazarian). We expect the velocity gradients to successfully trace magnetic field in superAlfvenic turbulence after the appropriate filtering of low frequency spatial modes. The practical difficulty of such a study is that it is that the inertial range of the MHD turbulence is rapidly shrinking with the increase of $M_A$. If for $M_A>1$ the injection scale is $L$ the transition to the MHD regime happens at the scale $l_{A}\approx LM_A^{-3}$ (see Lazarian 2006), which requires very large cubes to study velocity gradients arising from MHD turbulence. 
\begin{table}[h]
 \centering
 \label{tab:simulationparameters}
 \caption {Simulations used in our current work. The magnetic criticality $\Phi = 2 \pi G^{1/2} \rho L/B$ is set to be 2 for all simulation data. Resolution of them are all $480^3$}
 \begin{tabular}{c c c c}
Model & $M_s$ & $M_A$ & $\beta=2(\frac{M_A}{M_s})^2$\\ \hline \hline
Ms0.2Ma0.02 & 0.2 & 0.02 & 0.02 \\
Ms0.4Ma0.04 & 0.4 & 0.04 & 0.02 \\
Ms0.8Ma0.08 & 0.8 & 0.08 & 0.02 \\
Ms1.6Ma0.16 & 1.6 & 0.16 & 0.02 \\
Ms3.2Ma0.32 & 3.2 & 0.32 & 0.02 \\
Ms6.4Ma0.64 & 6.4 & 0.64 & 0.02 \\ \hline
Ms0.2Ma0.07 & 0.2 & 0.07 & 0.22 \\
Ms0.4Ma0.13 & 0.4 & 0.13 & 0.22\\
Ms0.8Ma0.26 & 0.8 & 0.26 & 0.22\\
Ms1.6Ma0.53 & 1.6 & 0.53 & 0.22\\\hline
Ms0.2Ma0.2 & 0.2 & 0.2 & 2 \\
Ms0.4Ma0.4 & 0.4 & 0.4 & 2 \\
Ms0.8Ma0.8 & 0.8 & 0.8 & 2 \\\hline
Ms0.13Ma0.4 & 0.13 & 0.4 & 18 \\
Ms0.20Ma0.66 & 0.20 & 0.66 & 18 \\
Ms0.26Ma0.8 & 0.26 & 0.8 & 18 \\\hline
Ms0.04Ma0.4 & 0.04 & 0.4 & 200 \\
Ms0.08Ma0.8 & 0.08 & 0.8 & 200 \\
Ms0.2Ma2.0 & 0.2 & 2.0 & 200\\\hline \hline
\label{tt1}
\end{tabular}
\end{table}

The numerical simulations that we employ for the study are listed in Table \ref{tt1}. The names of the simulations reflect both the sonic and Alfven Mach numbers. For instance, $Ms0.4Ma0.04$ corresponds to $M_s=0.4$
and $M_A=0.04$. 

To investigate the detail structure of gradients from different wavemodes, we employ the wave mode decomposition method in \cite{Cho2002CompressiblePlasmasb,CL03} to extract Alfven, slow and fast modes from velocity data. The corresponding equations determining the basis for the decomposition into modes are:
\begin{subequations} \label{eq:fsa-decompositions} 
\begin{align} 
\hat{\zeta}_f &\propto (1+\frac{\beta}{2}+\sqrt{D}) k_\perp \hat{\bf{k}}_\perp +(-1+\frac{\beta}{2}+\sqrt{D}) k_{||}\hat{\bf{k}}_{||} \\
\hat{\zeta}_s &\propto (1+\frac{\beta}{2}-\sqrt{D}) k_\perp \hat{\bf{k}}_\perp +
(-1+\frac{\beta}{2}-\sqrt{D}) k_{||}\hat{\bf{k}}_{||}\\
\hat{\zeta}_a &\propto -\hat{\bf{k}}_\perp\times \hat{\bf{k}}_{||}
\end{align}\end{subequations}
where $D=(1+\beta/2)^2-2\beta \cos^2\theta$\footnote{The proportional constant for $\cos\theta$ is correct only in CL02, i.e, $\cos^2\theta$.} , $\beta=\frac{\bar{P}_g}{\bar{P}_B}=\frac{2M_A^2}{M_s^2}$, and $\cos\theta= \hat{k}_{||} \cdot \hat{B}$. 
We would only use the LOS component of the decomposed velocities for velocity channels calculations.That is to say, the three velocity modes can then be acquired by
\begin{equation}
v_{(f,s,a),z}= [\mathscr{F}^{-1}(\mathscr{F}(\bf{v})\cdot\hat{\zeta}_{f,s,a})](\hat{\zeta}_{f,s,a} \cdot \hat{\zeta}_{LOS})
\end{equation}
where $\mathscr{F}$ is the Fourier transform operator. 

\begin{figure*}[t]
\centering
\includegraphics[width=0.96\textwidth]{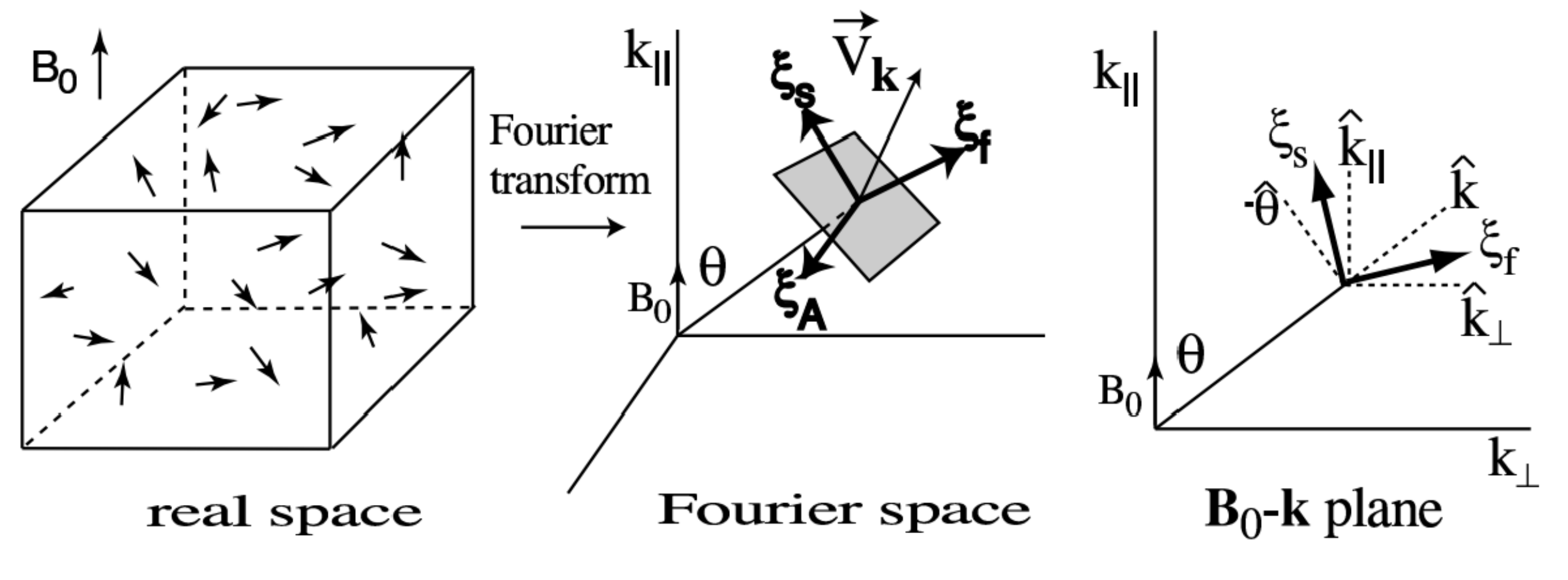}
\includegraphics[width=0.96\textwidth]{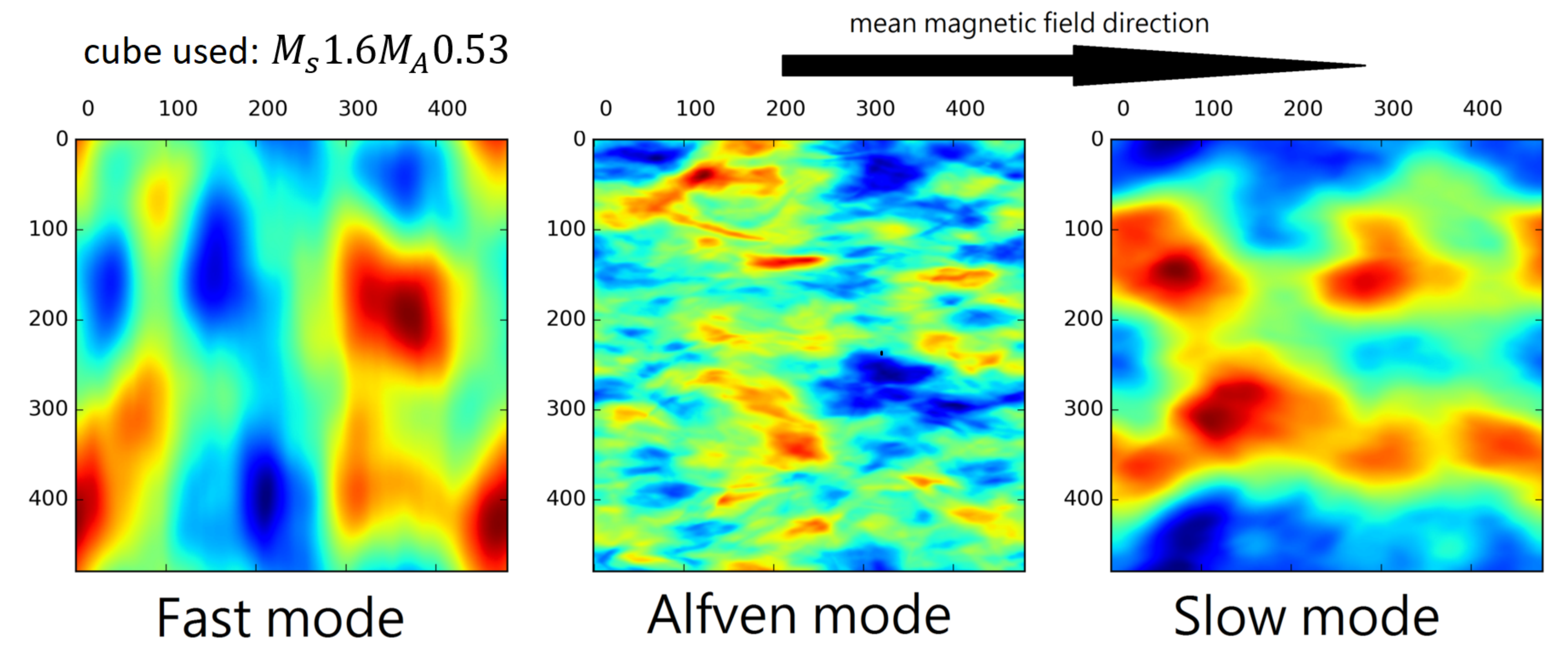}
\includegraphics[width=0.96\textwidth]{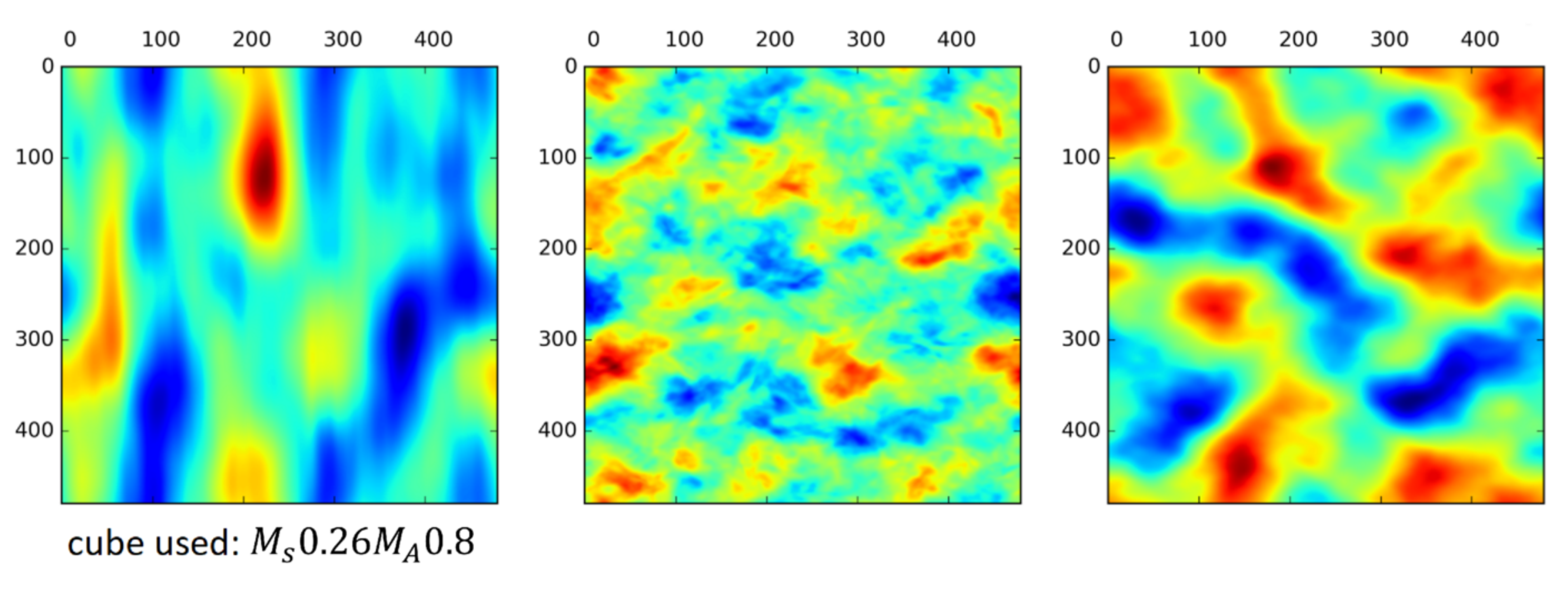}
\caption{\label{fig:fsa} (Top) Illustration of the  decomposition method. From \cite{CL03}. (Middle \& Bottom) The resultant projected maps from CL03-decomposition from two sample cubes with low (middle panel) and high  (low panel) $\beta$ }
\end{figure*}

The upper panel of Figure \ref{fig:fsa} illustrates the decomposition procedure that takes place in the Fourier space. The resulting 3 data cubes are dominated by the Alfven, slow and fast modes, respectively. In the middle and lower panels of Figure \ref{fig:fsa} we show the results of decomposed velocity cube which being projected along the x-axis. We illustrate the decomposition method using two cubes with low  and high $\beta$. The properties of the fast and slow modes differ in low and high $\beta$ plasmas, therefore we study these two cases separately in the following sections.

\subsection{Calculations of gradients in channel maps}

Gradients were calculated following the procedures described in YL17a, including the gradient calculation method and the {\it sub-block averaging}.\footnote{This procedure is very different from employed for gradient calculation by other authors, e.g. \cite{Soler2013}). It is very advantageous to apply this procedure for calculating other types of gradients.} In short, the gist of the {\it sub-block averaging} method is to define the most probable direction within a block by finding a local Gaussian-fitting peak for the distribution of gradients. The fitting error provides a quantitative estimate whether the block size is large enough for reliably determining the direction. 

The concept of thin channel maps was introduced in \cite{LP00}, where it was shown that the fluctuations within thin channel maps are mostly determined by velocity fluctuations. For determining whether the gradients are probing thin or thick channel map we use the following criterion: the channel is thin if the gradient is calculated over the scale of size $R$ for which the criterion given by Eq. (\ref{criterion}) is satisfied; otherwise, the channel is thick. To study velocity gradients over the scale larger than $R$ we construct the {\it velocity channel map}:
\begin{equation}
C(x,y) =\int dv \rho_{ppv} (x,y,v) e^{-\frac{|v-v_{0}|^2}{\delta v_R^2}}\\.
\end{equation}
As the thickness $v_R$ should not be less than the maximum of the thermal linewidth and spectrometer resolution, it means for images with high resolution the gradients at the smallest scales should be calculated using velocity centroids.

Readers should be reminded that channel map carries {\it intensity} information within a velocity channel. The reference velocity slice is convenient to define at the center of the spectral line. However, if the spectral line is broadened by the regular shear, as this is the case of Galactic atomic hydrogen, different velocity channels carry information about the 3D distribution of turbulence and therefore it is advantageous to study gradients in different bunches of channels in order 
magnetic fields and the interstellar medium (ISM) physical processes. 
 Similarly, in the presence of absorption, the wings of the lines are advantageous to used rather than the entire absorption lines. The latter are saturated and thus not informative at center of the line. 

To produce channel maps that contain only spatial frequencies for which the slice is thin we provide filtering of the high spatial frequencies for which the criterion given by Eq. (\ref{criterion}) is not satisfied. 
The filtered spacial frequencies can still be studied using the VCGs.  

\subsection{Uncertainties of magnetic field tracing}

\begin{figure}
\centering
\includegraphics[width=0.49\textwidth]{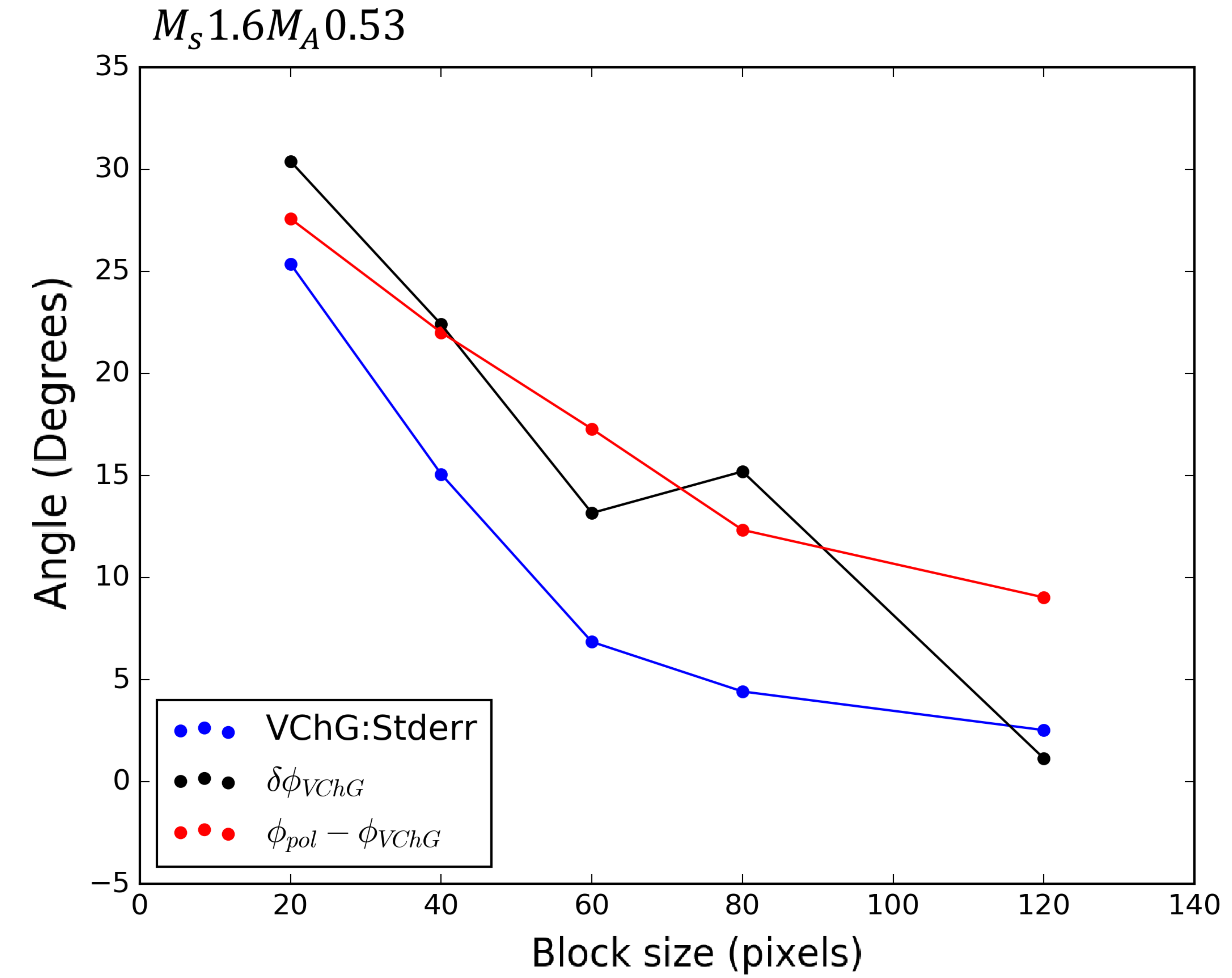}
\caption{\label{fig:vchgerr} The average error estimate from the Gaussian fitting function (Stderr), the angular differences between magnetic field directions from polarization and VChG orientation  ($\phi_{pol}-\phi_{VG}$) and the dispersion of VChGs ($\delta \phi_{VChG}$) for the velocity centroid map of cube $M_s1.6M_A0.53$ plotted against the change of block size. }
\end{figure}

Unlike the traditional technique of calculating gradients (see \cite{Soler2013}) our approach allows us provides us with the uncertainty of the determination of the magnetic field direction. To test how well this works in 
 Figure \ref{fig:vchgerr}  we show the error estimate for the VChG in the cube $M_s1.6M_A0.53$ that we calculate within our technique. As we fit the Gaussians into the distribution of the gradient directions we get an estimate of the fitting error, which is the lowest curve in Figure \ref{fig:vchgerr}. The half width of the Gaussians provides a significantly larger uncertainty shown by the black curve in Figure \ref{fig:vchgerr}. In numerical simulations we also can calculate the difference between the projected magnetic field and the measured gradient directions. This measurement is given by black curve in Figure \ref{fig:vchgerr}. These measures are changing are the function of the block size and our can observe our procedure of evaluating our error for magnetic field tracing is in reasonable correspondence with the actual measurements of the differences of the projected magnetic field direction and the direction given by the velocity gradients.  

\section{Alignment Measure and Density Effects for the VChGs}
\label{sec:gradients-in-fsa}
\begin{figure}
\centering
\includegraphics[width=0.48\textwidth]{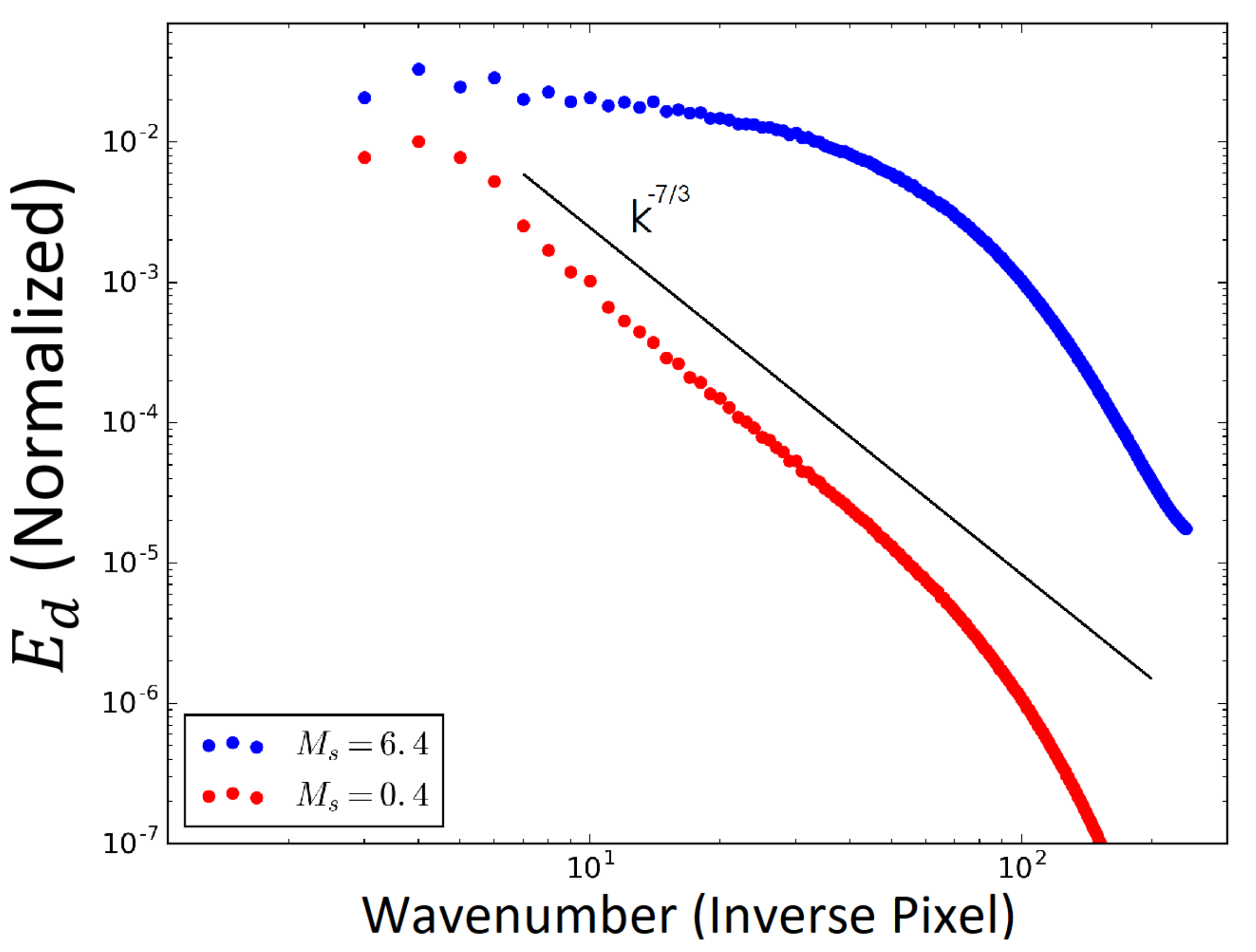}
\caption{\label{fig:illus-0} The 3D density spectrum from supersonic (blue,b15) and subsonic (red,b11) runs. While the subsonic spectrum has a steep slope, the supersonic one has a shallow one. The straight line shows the theoretically expected $-7/3$ density slope in subsonic strongly magnetized turbulence (see Kowal \& Lazarian 2010).}
\end{figure}

Before showing the gradient maps, we would like to illustrate the difference of supersonic and subsonic simulation in terms of density spectra. Figure \ref{fig:illus-0} shows the normalized density spectra from two simulation, one of it is supersonic and the other one is subsonic. The normalization is done by making the amplitudes of the spectra the same at the injection and it helps the reader to have easy comparison between the two spectra. The subsonic spectrum of density in strongly magnetized media scales as pressure and therefore follows not the Kolmogorov $k^{-5/3}$ law, but a steeper $k^{-7/3}$ law (see \cite{2007ApJ...658..423K}). For this definitions of spectra, the boderline between the shallow and the steep spectra corresponds to $k^{-1}$. It is very clear that the high $M_s$ spectra is shallow compared to the steep density spectra in subsonic systems. Our results on density studies agree well with those in \cite{2005ApJ...624L..93B} and \cite{2007ApJ...658..423K}.

\begin{figure}
\centering
\includegraphics[width=0.48\textwidth]{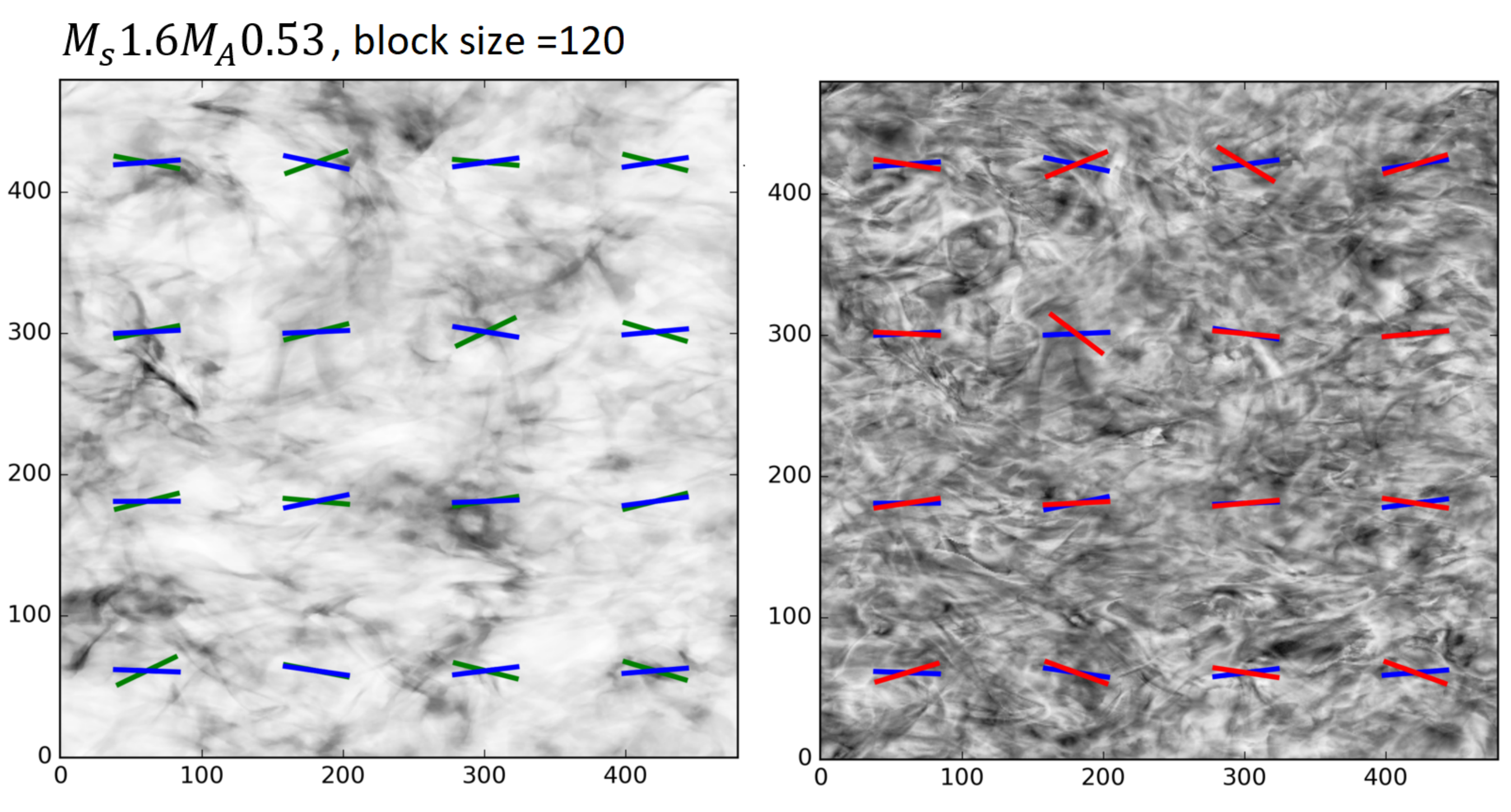}
\caption{\label{fig:illus-1} Intensity gradients (left panel, green color) and VChGs (right,red)  from a thin slice map with respect to the projected magnetic field (blue color). The maps prior to the CL03 decomposition into modes are used.}
\end{figure}

As the velocities, unlike densities, are directly related to MHD turbulence, we first consider only intensity fluctuations that arise only from turbulent velocity. For this purpose we create data cubes using the velocity field obtained from our 3D numerical simulations but substitute the actual densities by a constant density, which is just the {\it number density channel map} constructed above.  Figure \ref{fig:illus-1} illustrates the relative orientation velocity channel gradients (VChGs)  and the projected magnetic field on a {\it thin} slice setting. The gradients are all rotated for 90 degrees, which will be annotated as {\it rotated} gradients. The rotated gradients according to the theoretical considerations above correspond to the magnetic field directions.In the following sections, gradients are assumed to be rotated, unless emphasized specifically.  One can see from Figure \ref{fig:illus-1} that both number density and velocity channel map behave very good in terms of gradient alignment. That gives us confidence on using the velocity channel maps in our later analysis. 

The fact that the VChGs for thin maps are strongly influenced by the velocity field is suggesting that the tracing of magnetic fields using thin channel maps may be more accurate due to velocity being a better tracer of MHD turbulence. Below we consider the gradients arising from Alfven, slow and fast modes separately. The theoretical considerations giving the guidance for this study are provided in KLP16, where the anisotropies of the correlation of the intensities arising from velocities were studied. 

The  {\bf Alignment measure (AM)} is used by us to express quantitatively how well a certain type of gradient vectors trace magnetic field. AM is defined by the expression\footnote{This 2D measure is analogous to the 3D measure that is used in the theory of grain alignment (see Lazarian 2007 for a review).}:
\begin{equation}
AM=\langle2 cos^2\theta-1\rangle,
\end{equation}
and it ranges from -1 to 1. $\theta$ in the equation stands for the angle between the VChG and the projected magnetic field. The physical meaning of AM is as the following: if $AM \sim 1$, then the average angular difference in the directions of two vectors across the map is negligibly small; if $AM \sim 0$, then there is no alignment between two vectors. If the two directions tend to be perpendicular to each other, the $AM \rightarrow -1$. 

\section{Applying VChGs to Basic MHD modes}

It is important to understand how what is the effect of different modes on the VChGs. The composition of MHD turbulence in terms of the basic modes is changing with $M_A$ and plasma $\beta$ (see Cho \& Lazarian 2003). 

\subsection{Alfven modes}

Most important for the MHD turbulence are Alfven modes (see CL03). We will first demonstrate some properties of VChGs with these modes. 

The dynamics of Alfvenic modes in the strong MHD turbulence is very different from the dynamics of freely propagating Alfven waves. The modes cascade on the scale of the order of one period with the wavevector of the Alfvenic perturbations in strong turbulence being nearly perpendicular to the local direction of the magnetic field. As a result, the anisotropy and the iso-contours of intensity correlation are both elongated {\it parallel} to magnetic field. The latter is essential for tracing of magnetic fields, as the gradients that these modes are inducing are also perpendicular to the local direction of magnetic field. 

\begin{figure}
\centering
\includegraphics[width=0.48\textwidth]{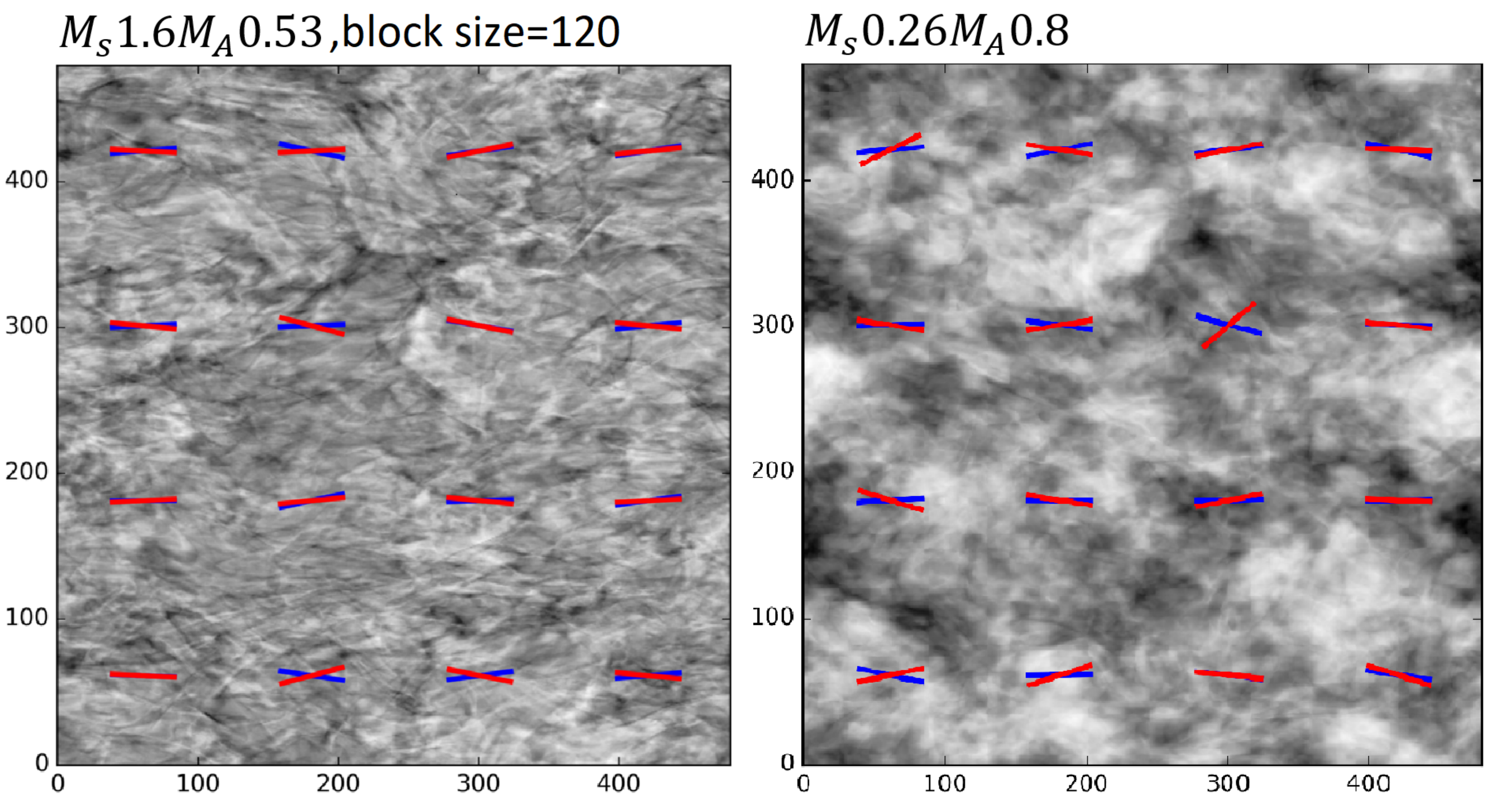}
\caption{\label{fig:alf-1} The velocity channel map for low (left) and high (right) $\beta$ from Alfven mode in a thin slice map. The rotated 90 degrees gradients (red for VChGs, green for IGs) are over-plotted over the projected magnetic field (blue).}
\end{figure}

\begin{figure}
\centering
\includegraphics[width=0.48\textwidth]{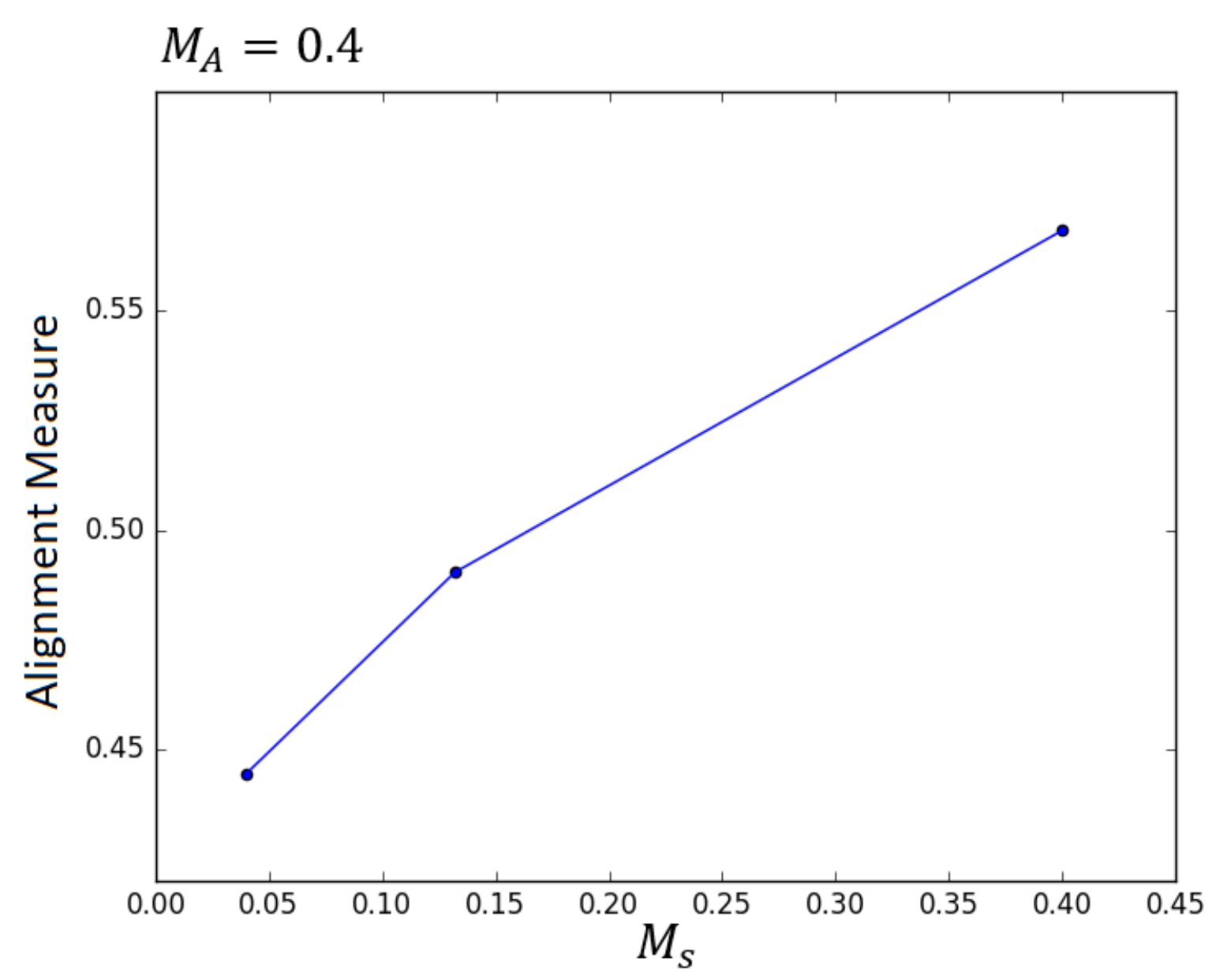}
\caption{\label{fig:alf-2} The change of $AM$ of Alfven mode VChGs with respect to $M_s$ }
\end{figure}

For Alfven modes the theoretical expectation is very natural: the anisotropy of the intensity correlations in a thin slice increases with the decrease of the Alfven Mach number, i.e. the weaker the velocity perturbations, the more turbulent motions are dominated by magnetic field which causes the eddies to be more anisotropic. Figure \ref{fig:alf-1} illustrates the channel maps for low and high $\beta$ and their gradients within a thin slice that are induced by Alfvenic modes.  According to \cite{CL03} the change of Alfven modes with $M_s$ is marginal. The corresponding change in terms gradients is shown in the left panel of Figure \ref{fig:alf-2}. 

\begin{figure*}
\centering
\includegraphics[width=0.48\textwidth]{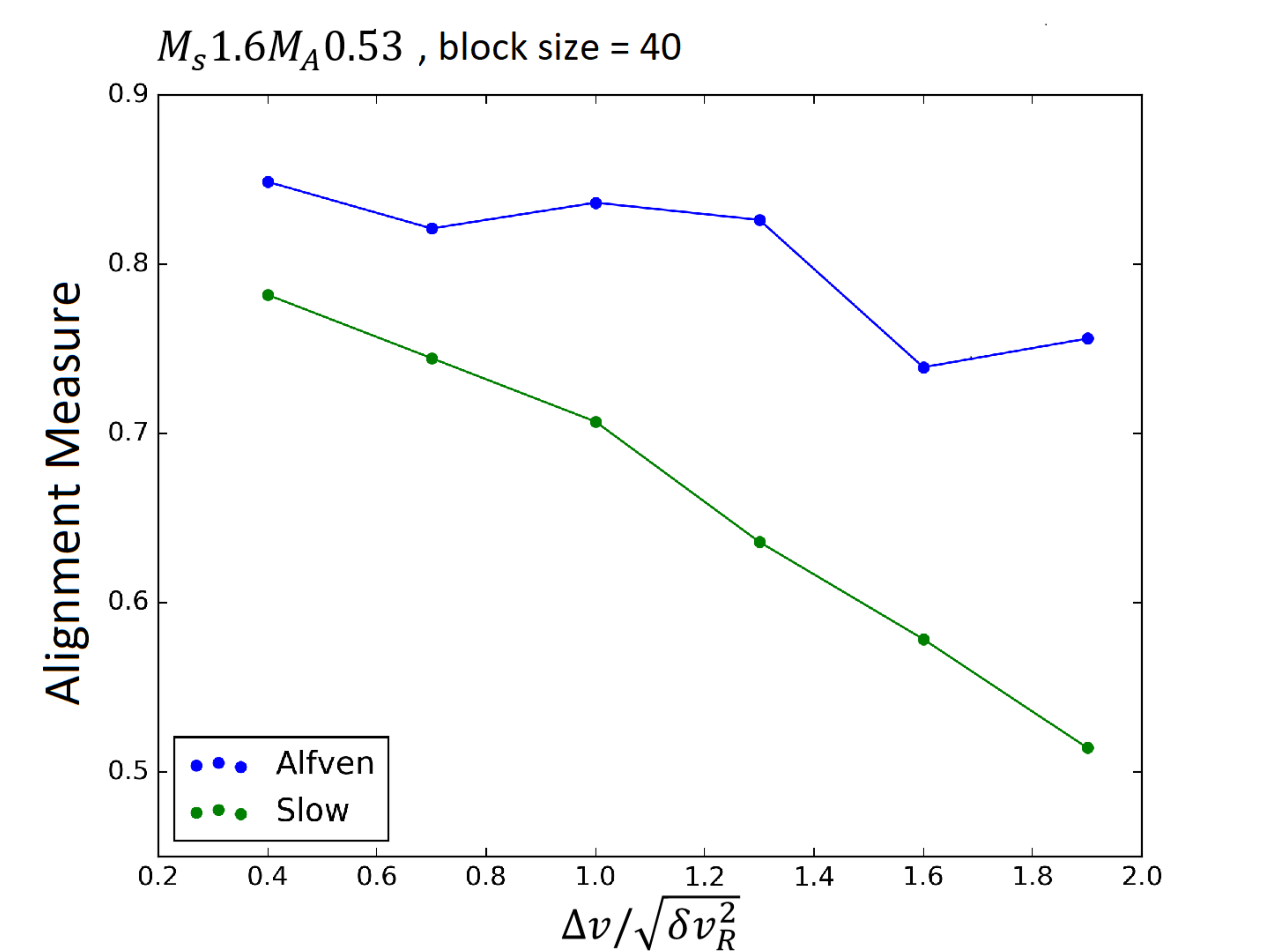}
\includegraphics[width=0.48\textwidth]{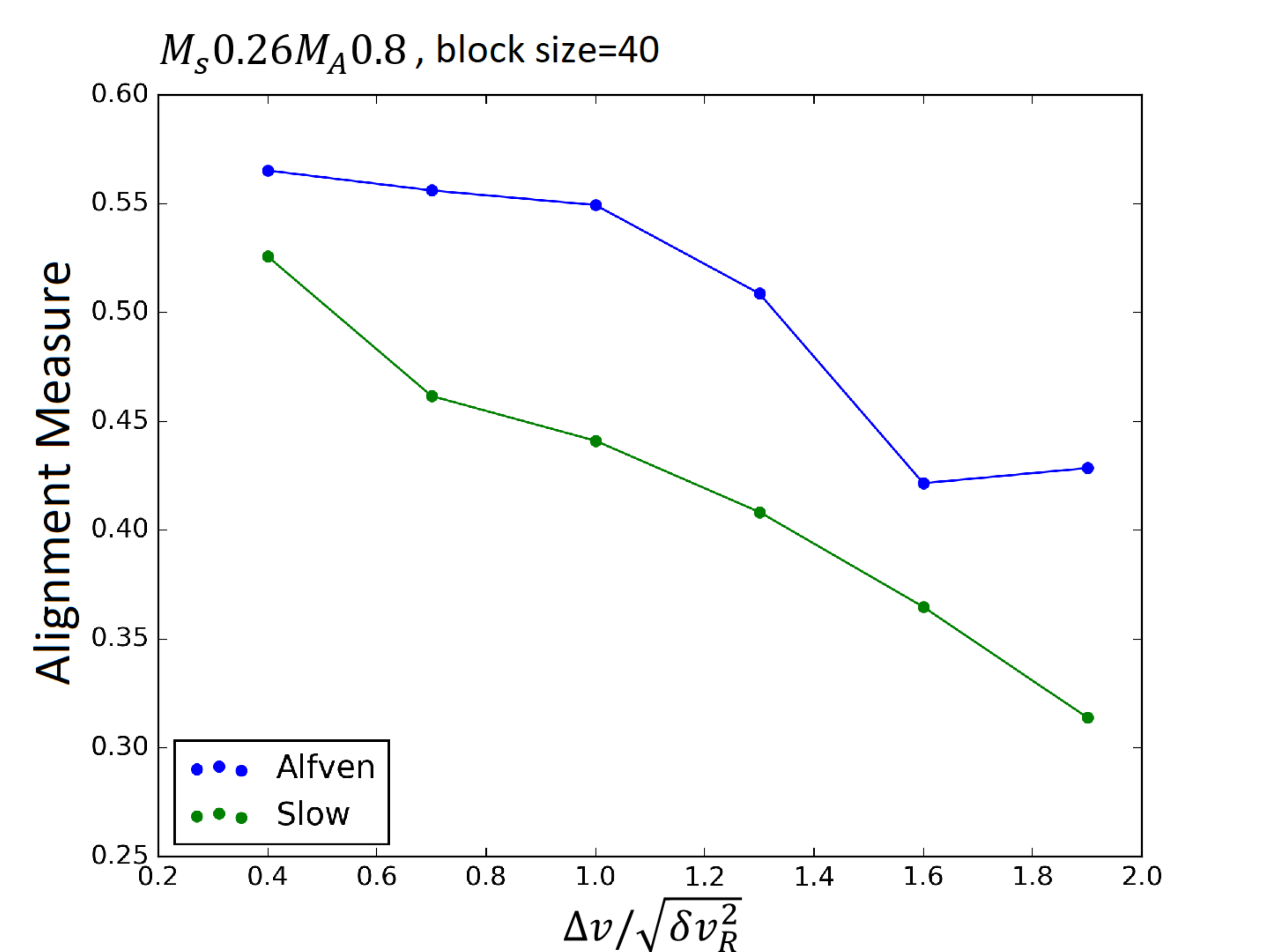}
\caption{\label{fig:alf-3} The change of $AM$ of slow (green) and Alfven (blue) modes for the VChGs  as the channel width  increases. (Left panel): The results for low-$\beta$ (left) simulations. (Right panel): The results for high-$\beta$ (right) systems simulations.}
\end{figure*}

As we increase the thickness of channel maps, the $AM$ of Alfven modes is decreasing, which is shown in the blue curves in Figure \ref{fig:alf-3} for both low and high $\beta$ cases. This is because the contribution from velocities decreases as the channel the channel width increases. However, we observe that the change is gradual (see also theory in LP00) rather than a sharp jump, which indicates even thin slices satisfying Eq. (\ref{criterion}) may not be available for some data, selection of slices as thin as possible can already help enhancing the contribution from velocity information in observation data. 

\begin{figure}
\centering
\includegraphics[width=0.48\textwidth]{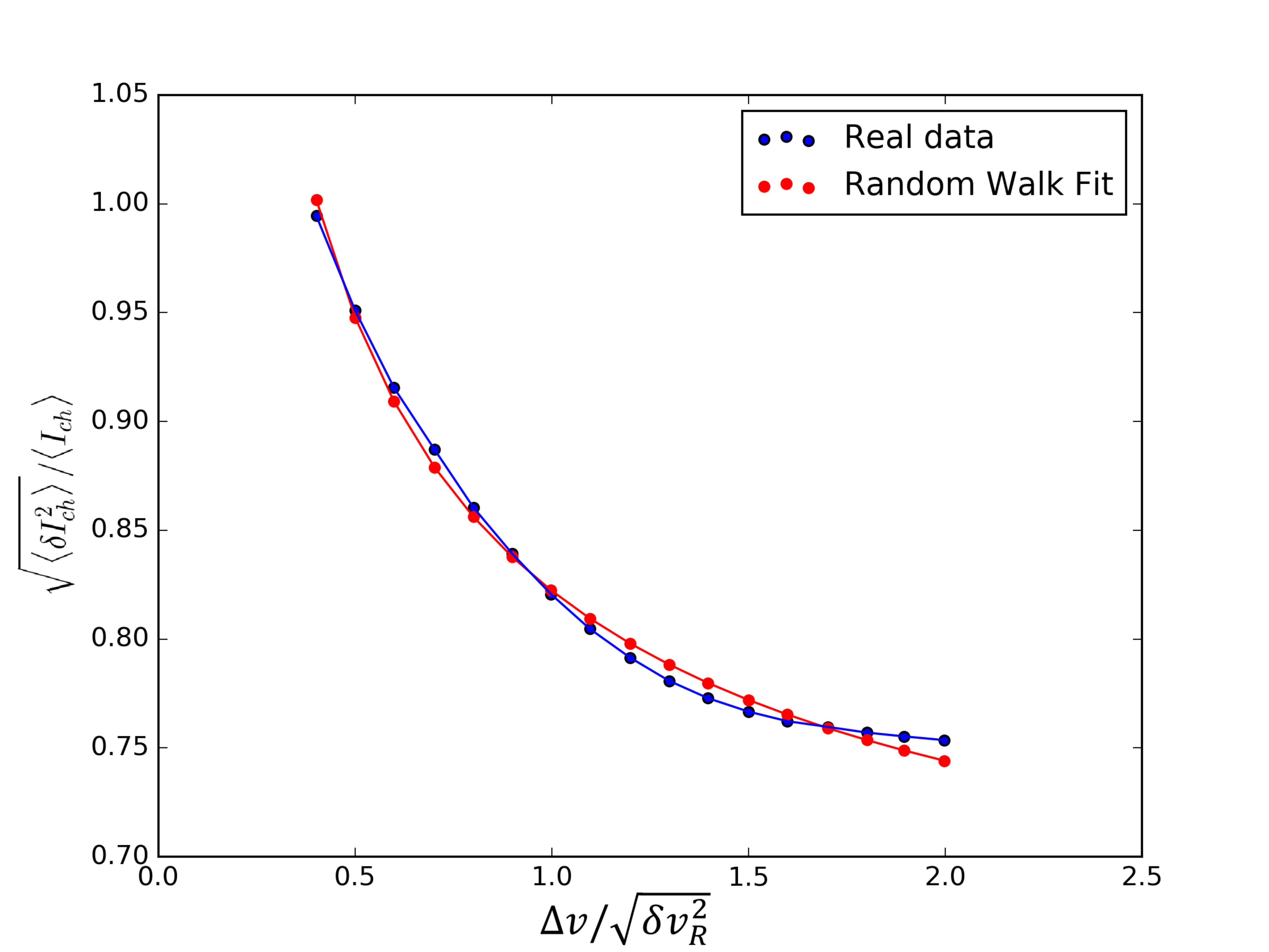}
\caption{\label{fig:alf-4} Variation of $\sqrt{\langle \delta I_{ch}^2\rangle}/\langle I_{ch}\rangle$ with respect to channel map thickness, for both density weighted (red) and constant density (blue) cases. The random walk fits are added accordingly. } 
\end{figure}

We can double check the above statement by testing the gradient variation dependence on the channel width. In this experiment, we use the density-weighted PPV cubes to illustrate the relative dominance of velocity over density fluctuations when one changes the channel width, as illustrated in Figure \ref{fig:alf-4}. When we increase the channel width, the VChGs variations, given by $\sqrt{\langle \delta I_{ch}^2\rangle}/ \langle I_{ch}\rangle$, are decreasing. Due to this decrease of gradient deviations, the role of velocity fluctuations in the channel map fluctuations is decreasing and the thick channels get dominated by density fluctuations. A more detailed discussion on the contribution of density effect will be in \S \ref{sec:fluctuations}.

One should remember, however, that whether the channels are thin or thick depends on the size of the eddies that are being studied (LP00). In our case, the block size is an additional factor limiting the size of the eddies that we can study, which usually have a large velocity dispersion than a small-block sampled channel maps. The verification of criterion of the thin slice (see Eq \ref{criterion}) has to be done every time in the case of small blocks. 

\begin{figure}
\centering
\includegraphics[width=0.49\textwidth]{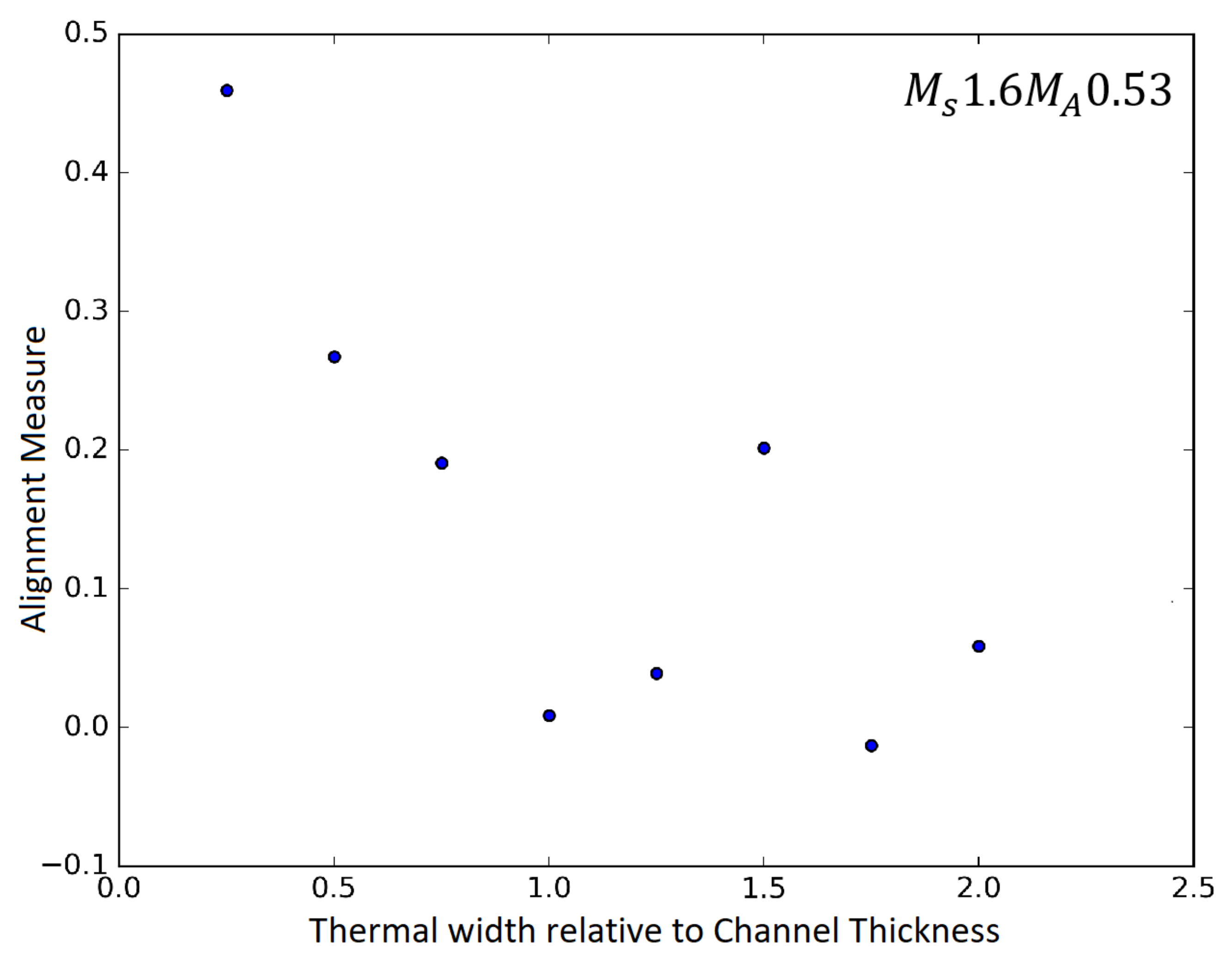}
\caption{\label{fig:thermal_channel} The thermal width relative to channel thickness versus the alignment measure value analyzed from data obtained in the cube Ms1.6Ma0.528. }
\end{figure}

According to LP00 the thermal line width acts similarly to the channel thickness. Here we used the data of the cube Ms1.6Ma0.528 by adding additional thermal dispersion of the velocities before constructing of the PPV cube. Figure ~\ref{fig:thermal_channel} shows the relationship between the AM versus the ratio thermal width over a channel width of $\delta v=0.2 \sqrt{\delta v_R^2}$ and the alignment measure. The decreasing trend resulted by an increase of thermal width is similar to the effect of increasing channel width illustrated in Figure \ref{fig:alf-3}. Moreover, when the ratio between thermal width to channel width is larger than 1, the thermal velocities dominates over the contribution from turbulent motions. Different from the effect from increasing channel width, the increase of thermal line width retain the velocity information yet washes away the fluctuations. This explains why the AM is oscillating around zero once the noise width is larger than the channel width. In spite of this, such an effect would not prevent the application of our technique in most of the region of the sky, which is mostly supersonic. This effect also sets a lower bound for observers to select the channel width when applying the method. 

\subsection{Slow modes}

Slow waves present perturbations that propagate along magnetic field lines. In the limit of incompressible media, slow waves are pure magnetic compressions that propagate along magnetic field lines. Formally the incompressible case corresponds to $\beta=\infty$ and in this limit the slow modes are frequently called preudo-Alfven modes. On the contrary, for $\beta\ll 1$ the slow waves are density perturbations propagating along magnetic field lines. 

In the presence of Alfvenic turbulence, slow modes do not evolve on their own, but are sheared by Alfven modes. As a result, the features of Alfvenic turbulence, e.g. spectrum and anisotropies, are imprinted on the slow modes (see GS95, \citealt{Lithwick2001CompressiblePlasmas,Cho2002CompressiblePlasmasb,CL03}).  This also means the perpendicular velocity gradients characteristic of Alfven modes are also inherited by slow modes.

\begin{figure}
\centering
\includegraphics[width=0.48\textwidth]{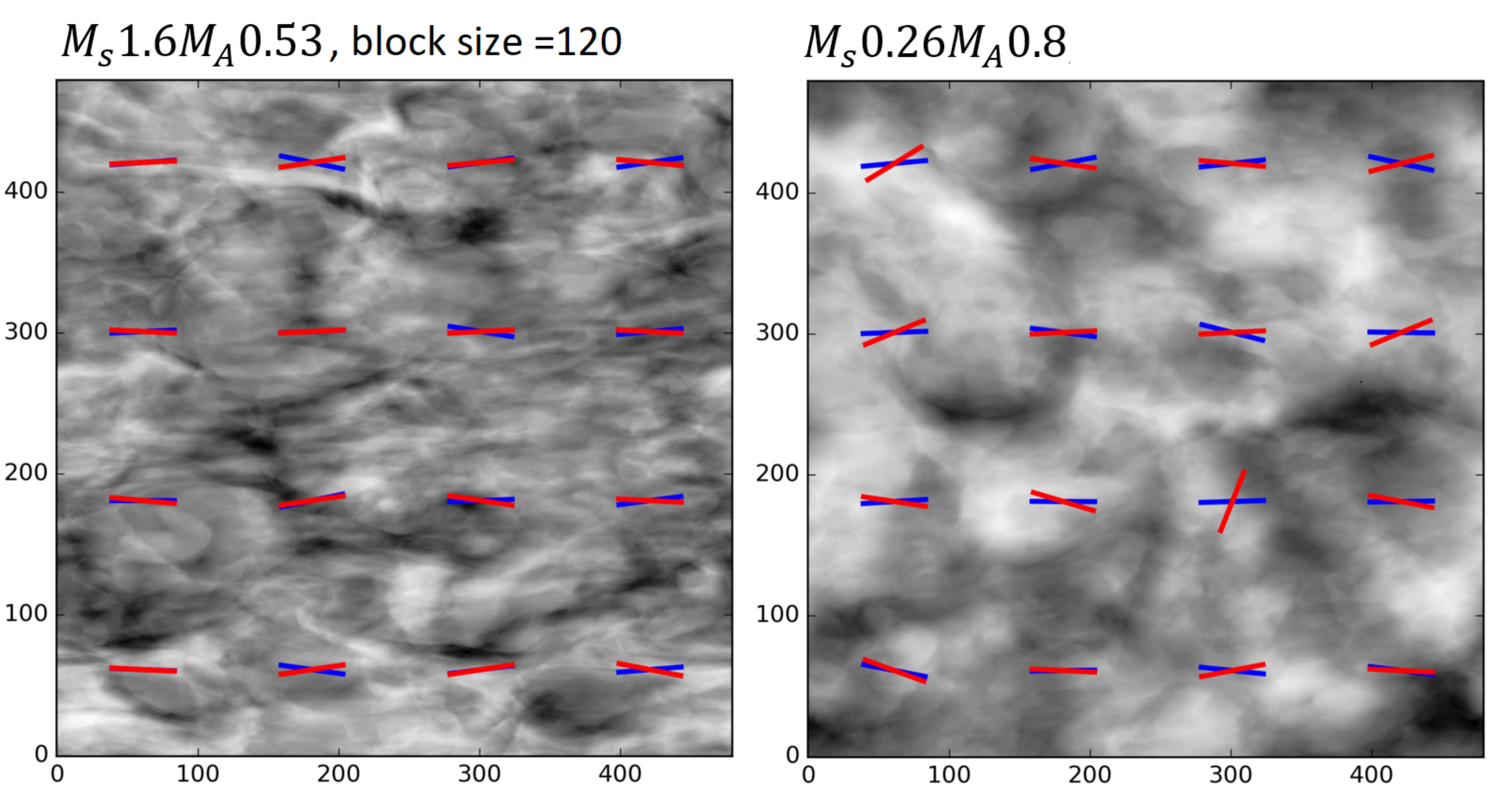}
\caption{\label{fig:slow-1} Slow modes:The {\it thin} slice velocity channel maps for low $\beta$ (left) and high $\beta$ (right) simulations.}
\end{figure}

Figure \ref{fig:slow-1} illustrates the channel maps for a thin slice for the case of low $\beta$ (left panel) and high $\beta$ (right panel). One can see that low $\beta$ VChGs performs better, which is also shown in Figure \ref{fig:alf-3}. In fact, when $\beta$ is small, the gradients of the corresponding channel maps are similar those of the Alfven mode channel maps. On the other hand, the slow mode channel maps from high $\beta$ systems are not as highly structured as the maps with low $\beta$. Moreover, the AM of slow modes from both cases decrease faster than that of Alfven modes as channel width increases. In reality both modes contribute to gradients of the channel maps. The result of Figure \ref{fig:alf-3} suggests that as the velocity channel thickness increases the AM decreases. The infinite channel thickness corresponds to using total intensities. It is obvious, that in terms of magnetic field tracing the intensities (thick channels) are inferior to the thin channels. 

\subsection{Fast modes}
\label{subsec:fastmodes}
Similar to the properties of the slow modes, the properties of fast modes are different in low and high $\beta$ plasmas.  For high $\beta$ plasmas the fast wave are similar to sound waves that propagate with sound speed irrespectively of the magnetic field direction. Similar to acoustic turbulence, the corresponding turbulence is expected to be isotropic. In low $\beta$ the fast modes correspond to magnetic field compressions that propagate with the Alfven velocity. In terms of correlation function anisotropy of fast modes, the elongation of the iso-contours of correlation is {\it perpendicular} to magnetic field. Therefore for fast mode dominated environments, one should expect the gradients to be parallel instead of perpendicular to local magnetic field.We note that the alignment of velocity gradients from the actual MHD turbulence contain contributions from the three modes with fast modes decreasing the alignment induced by the Alfven and slow modes. 

\begin{figure}
\centering
\includegraphics[width=0.48\textwidth]{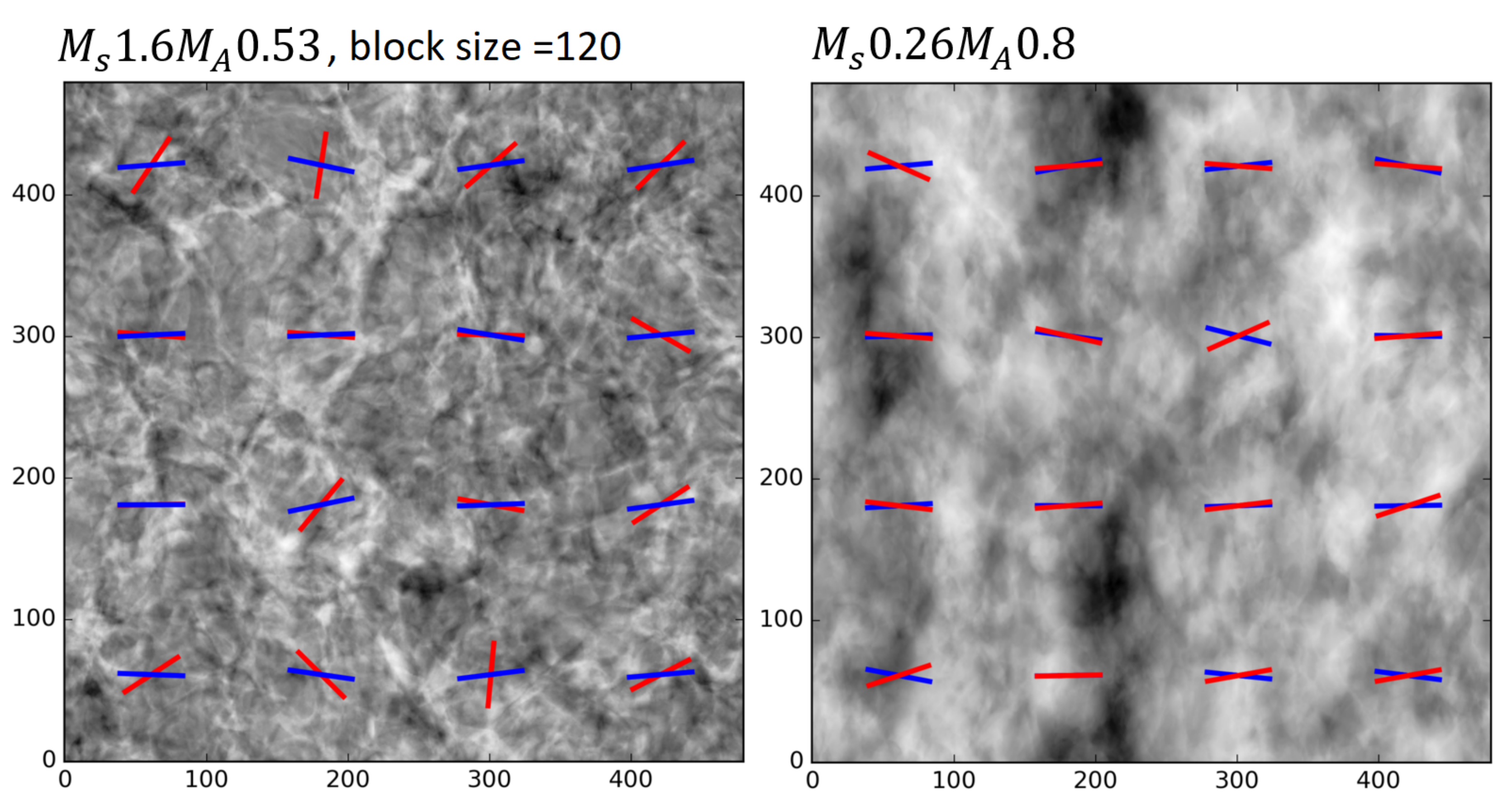}
\caption{\label{fig:fast-1}The {\it thin} slice  velocity channel maps with {\it unrotated} gradients for low $\beta$ (left) and right $\beta$ (right) for fast modes. The red vectors are the {\it unrotated} VChGs, and the blue vectors are projected magnetic field direction. }
\end{figure}
\begin{figure*}
\centering
\includegraphics[width=0.48\textwidth]{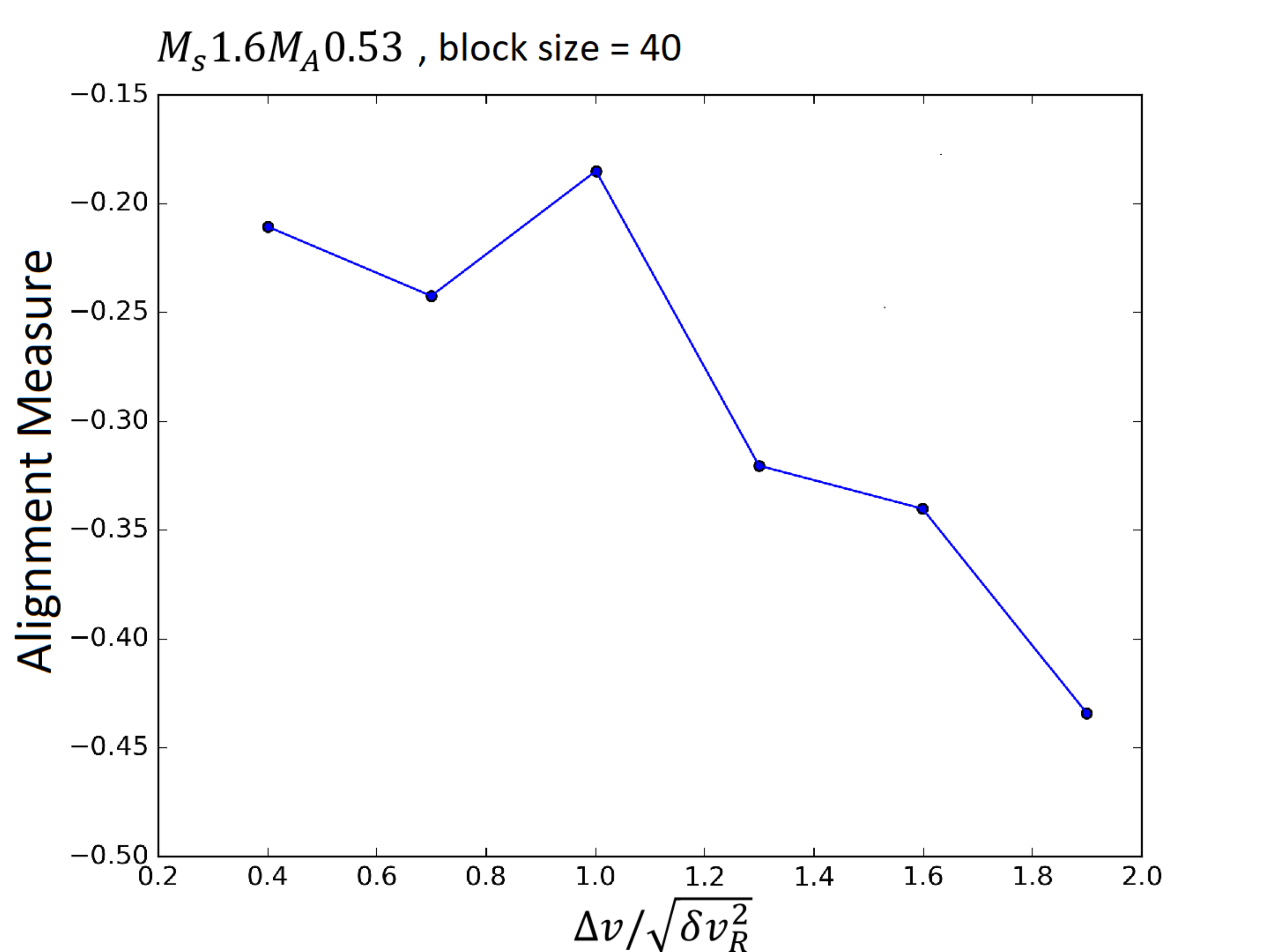}
\includegraphics[width=0.48\textwidth]{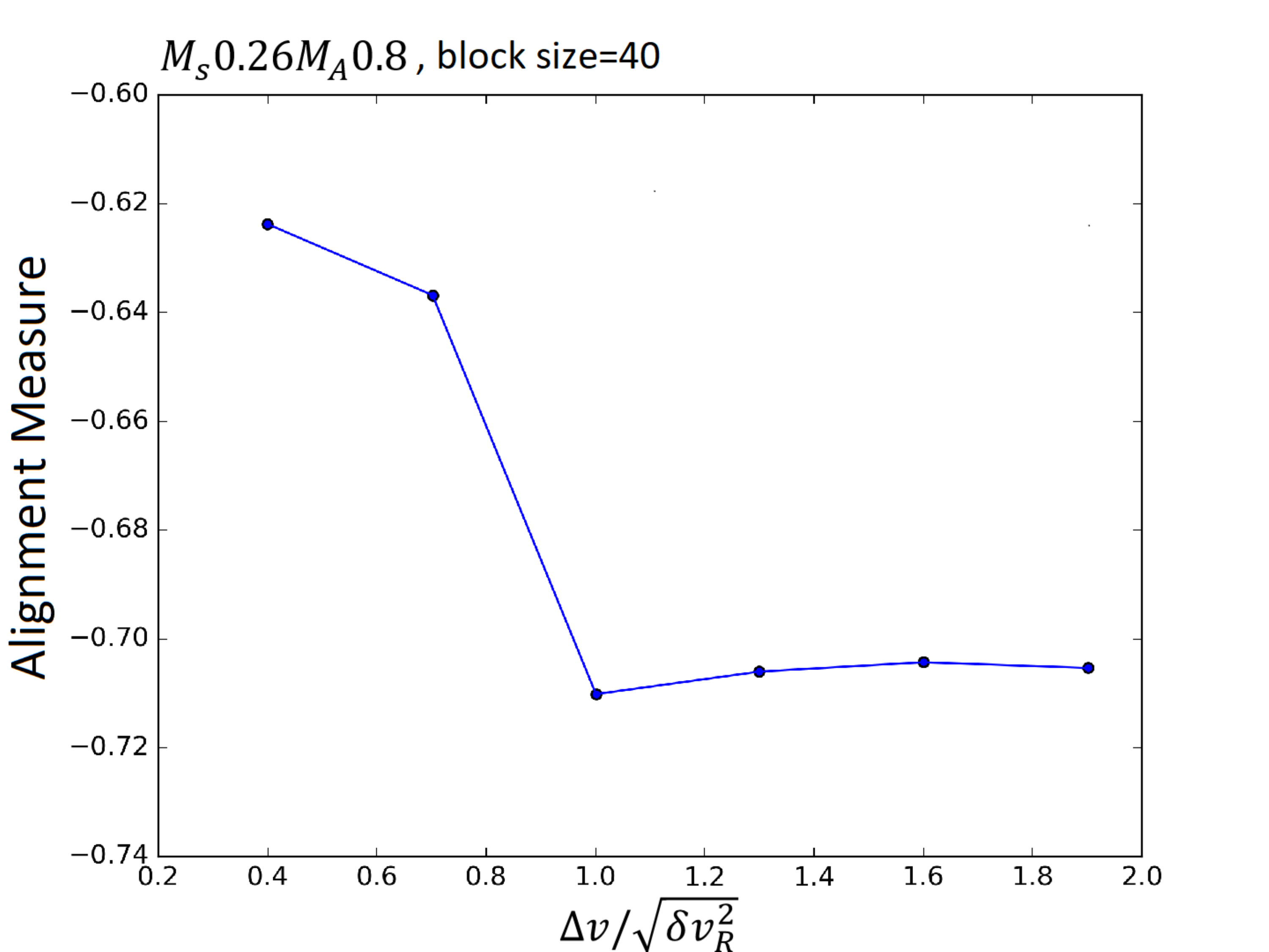}
\caption{\label{fig:fast-2} The change of $AM$ of fast modes {\it rotated} VChGs with respect to a change of channel width for low (left) and high $\beta$ (right), respectively. }
\end{figure*}

Figure \ref{fig:fast-1} shows, respectively, the channel maps with low and high $\beta$ plasmas. The low $\beta$ map carries more structures than that of high $\beta$ map, but they are obviously carrying different anisotropy direction when examining the contours in the maps. In the language of fast mode gradients, the maximal gradient is {\it parallel} to the magnetic field direction. Figure \ref{fig:fast-2} illustrates the $AM$ of fast modes VChGs. 

In contrast to the slow decreasing $AM$ for rotated Alfven and slow mode VChGs, the $AM$ of rotated fast mode VChGs decreases more significantly when the channel width passes through unity for both low and high $\beta$ case. This suggests a decrease of channel width suppresses the unwanted contribution of fast modes in velocity gradient calculations which makes thin channels preferable for tracing magnetic fields. The difference of the behavior of gradients from fast mode on one side and the Alfven and slow modes on the other side, makes it possible to separate the contributions from these modes. This issue will be discussed elsewhere (see also \S \ref{subsec:modesep})

Numerical simulations (e.g. \citealt{Cho2002CompressiblePlasmasb,CL03}) indicate that fast modes are subdominant at least for the cases of incompressible driving of turbulence. Our study indicates that in terms of VChGs one can expect a further suppression of the fast mode contribution in thin channels.

\section{Comparison with the correlation anisotropies within channel maps}
\label{sec:anisotropy}
It was demonstrated in \cite{2002ASPC..276..182L} that the correlation anisotropies in channel maps can trace magnetic fields. Further studies of this effect are provided in (Esquivel \& Lazarian 2005, Esquivel et al. 2015). We believe that correlation anisotropies can be very informative in terms of determining the relative contribution of compressible versus incompressible modes (see \citealt{2016MNRAS.461.1227K,2017MNRAS.464.3617K}). However, in terms of tracing of the detailed magnetic field structure they are inferior to the velocity gradients. Indeed, for correlation function to be informative, one requires a large number of sampling points to acquire enough statistics. As a result, one can expect that the correlation anisotropy requires a lot more data points for tracing magnetic field vector than the VChGs. Incidentally, a similar effect was demonstrated in LYLC for a test using the synchrotron intensity gradients and YL17b for a test using the VCGs. Here we test this statement for the VChGs.

\begin{figure}
\centering
\includegraphics[width=0.48\textwidth]{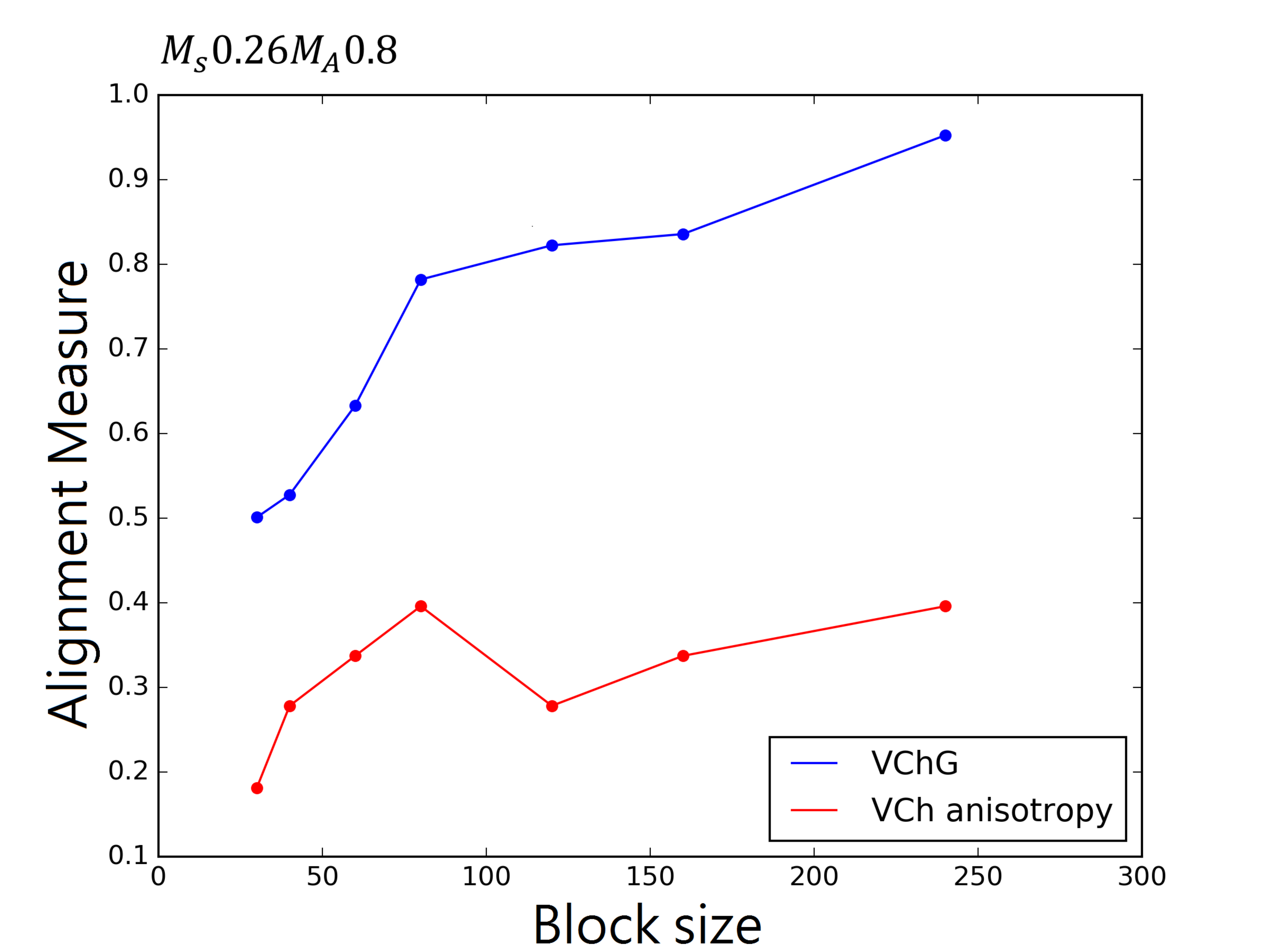}
\caption{\label{fig:cfa-1} A comparison of $AM$  obtained with the VChGs and the correlation function anisotropy for the case of Alfven modes. }
\end{figure}

In Figure \ref{fig:cfa-1} we show the VChGs alignment measure $AM$ versus the alignment measure of the directions of the maximal anisotropy (elongation) of the correlation functions of the channel map intensities. We clearly see that the channel map intensities are not good for detailed magnetic field tracing, which is consistent with our previous studies (see YL17, LYLC, YL17b).

The relative efficiency of velocity gradients is easy to understand. The velocity gradients correspond to the angular velocities of turbulent eddies. These individual eddies are aligned with the local magnetic field. On the contrary, the correlation functions are the statistical measures that are well defined only after the averaging. Thus the averaging over small patches does not produce good statistics necessary for the correlation functions to be well defined. The shortcomings of the technique of magnetic field tracing based on the correlation function anisotropies is also seen when the corresponding modifications of technique are compared with the VCGs and SIGs (see LYLC, YL17b). 

\section{Relative importance of velocity versus density fluctuations, and scale dependent gradients}
\label{sec:fluctuations}

Both velocities and densities contribute to the intensity fluctuations within channel maps, but their relative contribution is changing with the change of the channel map thickness (LP00).
To study the relative importance of density and velocity fluctuations, we use velocity and density data cubes obtained through our 3D MHD compressible simulations. The structures in these cubes are elongated in the same direction which we take to be x-direction. For our study we create the Position-Position-Velocity (PPV) cubes of synthetic spectral line by turning 90 decrees the density cube in respect to the velocity data cube, so the elongated structures in the density get perpendicular to the magnetic field. Mathematically, 
\begin{equation}
\label{eq:tc}
C_t(y,z) = \int_{|v_x|<\Delta v} dx \rho(x,y,z) v_x(x,z,y)
\end{equation}
 In other words, we are creating synthetic maps by transposing the velocity and density data by 90 degrees within the plane-of-sky plane, i.e. through the rotation along z-direction. 

In the PPV cube created this way the directions of anisotropy for the velocity and density are orthogonal and the resulting anisotropy  determines which contribution dominates.  When examining the anisotropy using the correlation function $R_2({\bf r}) = \langle C_t({\bf x}) C_t({\bf x}+{\bf r}) \rangle_{x}$ in the centroid map, the ratio between the axis perpendicular (the elongation direction of $C_t$) to magnetic field direction over that of parallel one (the elongation direction of $I$) will reflect the relative importance of velocity to density fluctuations to the centroids measured this way. 

\begin{figure}
\centering
\includegraphics[width=0.48\textwidth]{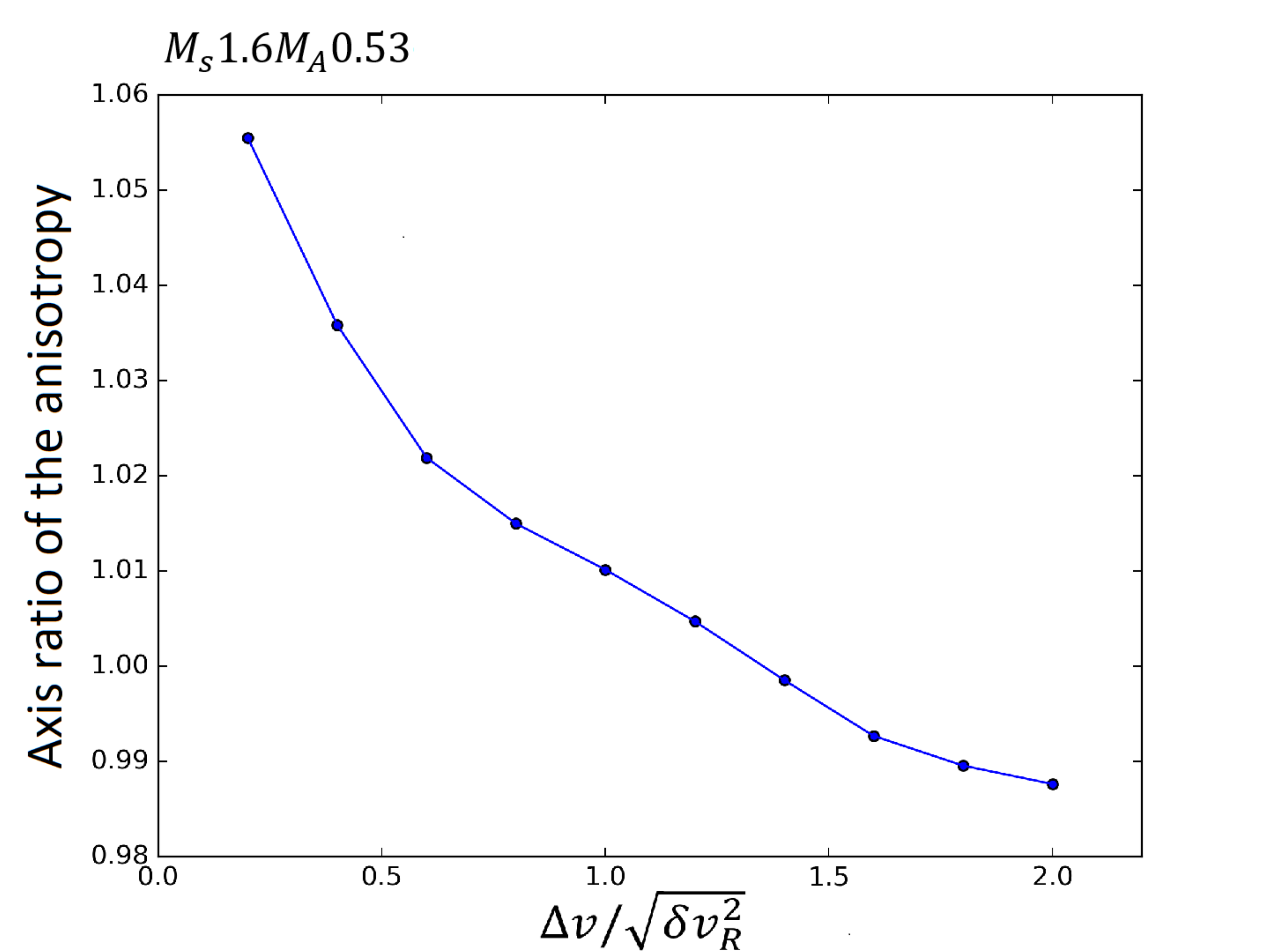}
\caption{\label{fig:fluc-1} The change of anisotropy (the axis ratio) as the velocity slice thickness changes for the case of centroid $C_t$ (see Eq. (\ref{eq:tc}) for the 90 degree rotated velocity and density data cubes. }
\end{figure}

Figure \ref{fig:fluc-1} shows the decreasing trend of the aforementioned axis ratio as velocity slice thickness increases. It is evident that even though the density fluctuations can contaminate the contribution of the velocity fluctuations within the channel maps, the overall anisotropy comes from the velocities provided that the channel map is sufficiently thin. Therefore using a thin slice  one can reduce the contribution from density data (see Figure \ref{fig:fluc-1}.)

\begin{figure}
\centering
\includegraphics[width=0.48\textwidth]{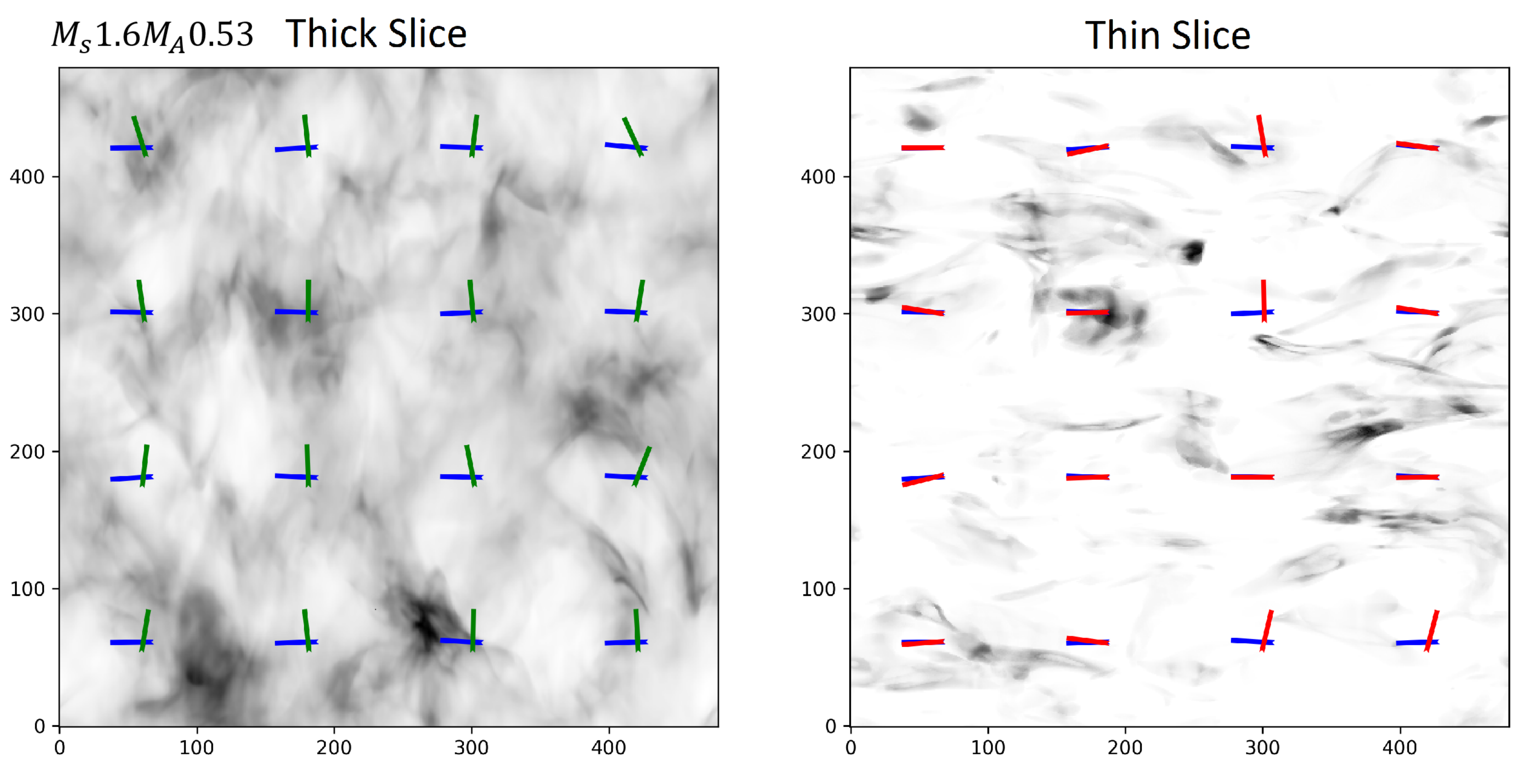}
\caption{\label{fig:transC} Illustration of gradients calculated for the thick (left panel) and thin (right panel) slices for the case when the density and velocity are turned 90 degrees in respect to each other. The rotated gradients are shown in green for the left panel and in red for the right panel. The magnetic field directions in both panels are shown in blue. }
\end{figure}

 We use the same set up employing density and velocity cubes turned 90 degrees to each other to explore the effects of channel thickness for the VChGs. We study {\it transposed channel map} by picking the slices satisfying Eq. \ref{criterion} in a PPV cube constructed with transposed density. Figure \ref{fig:transC} illustrates a thick and a thin slice result from such construction. The thin slice use contribution of PPV cubes with channel width $\sim$ 0.8 of $\delta v$, while the thick slice is just the integration of the transposed density on the full spectral line. It is very obvious from visual inspection that, thin slice structures are parallel to magnetic field in the velocity cube, while that of thick slice reflects the direction of magnetic field in the density cube. The latter is reflected in the AM being negative. In agreement with Figure \ref{fig:fluc-1} we see that the velocity information dominates over density one in when the velocity channel is thin.

\begin{figure}
\centering
\includegraphics[width=0.48\textwidth]{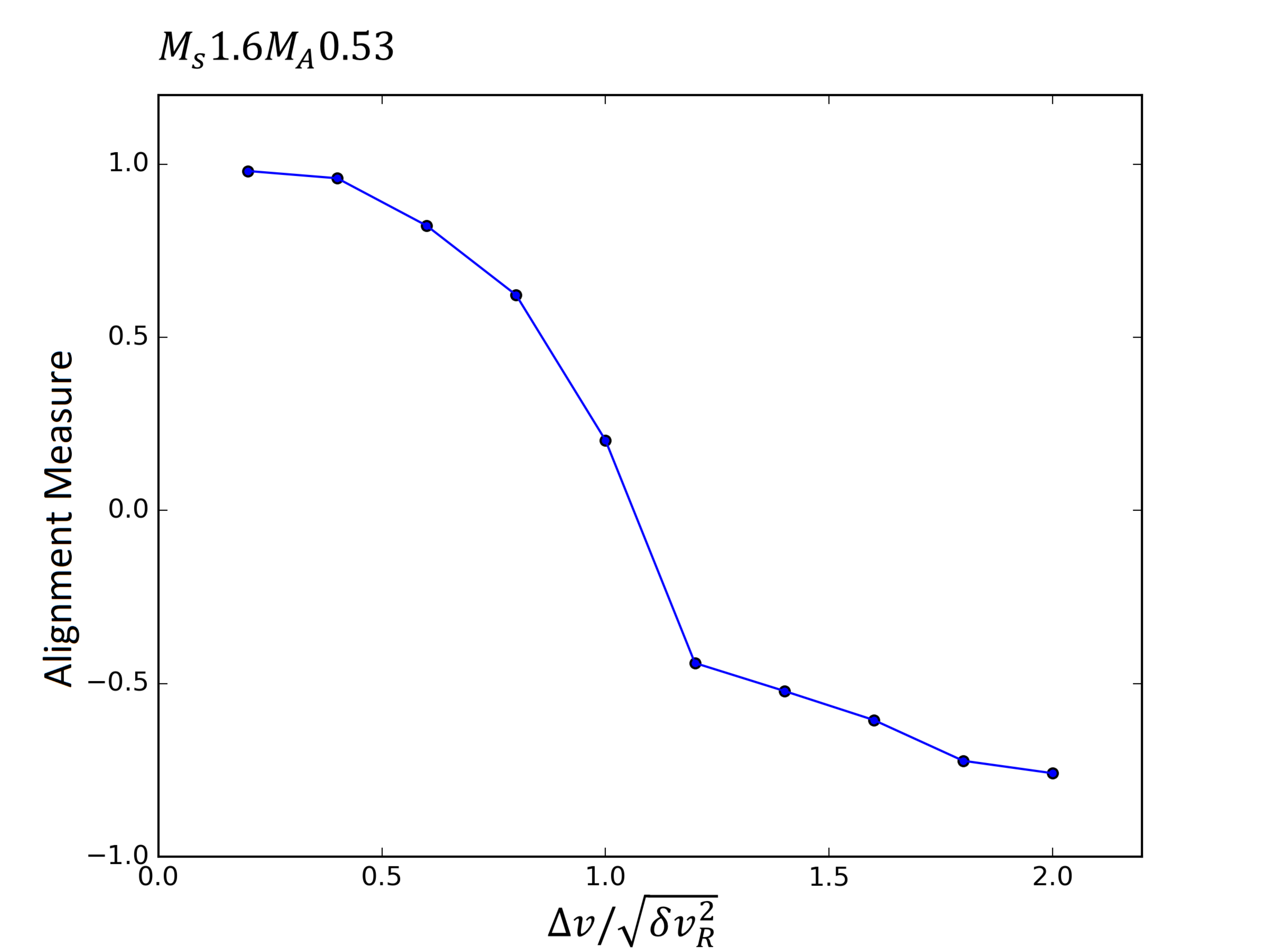}
\caption{\label{fig:transCAM} The alignment measure of the for VChGs  as the channel map width increases. }
\end{figure}

A more quantitative result can be seen in Figure \ref{fig:transCAM}, which is a  plot of the alignment measure. When the channel width $\Delta v/\sqrt{\delta v_R^2}$ is smaller than unity, velocity contribution is dominant and therefore the AM is positive. On the other hand, the AM for thick channel width maps are negative, indicating contribution from density is dominant. The point where $AM=0$ correspond to the channel width being unity. 

\section{Reduced velocity centroids obtained from the data}
\label{sec:reducedcentroid}

Our study above suggests that the decrease of channel width improves the alignment of the gradient direction and the magnetic field. However, as we discussed earlier, thermal line width provides a lower limit on the effective thickness of the velocity channels. To study velocities on the scales less than the thermal velocity broadening one may invoke velocity centroids, which properties have been studied e.g. in Esquivel \& Lazarian (2005), Kandel et al. (2017b).  The potential disadvantage of traditional centroids is that the entire spectral line is used while in some cases different parts of the line reflect magnetic fields in spatially different regions. This is the case, for instance, for the HI measurements where the galactic rotation curve provides rough information about the location of the emitting material. In addition, if the center of spectral line is saturated due to absorption effects, it is also good to use only the informative part of the line. Thus it is important to explore whether velocity centroids can be made a more flexible tool that can be used to study only a part of the spectral line. Below we experiment with such a measure. 

\begin{figure*}
\centering
\includegraphics[width=0.33\textwidth]{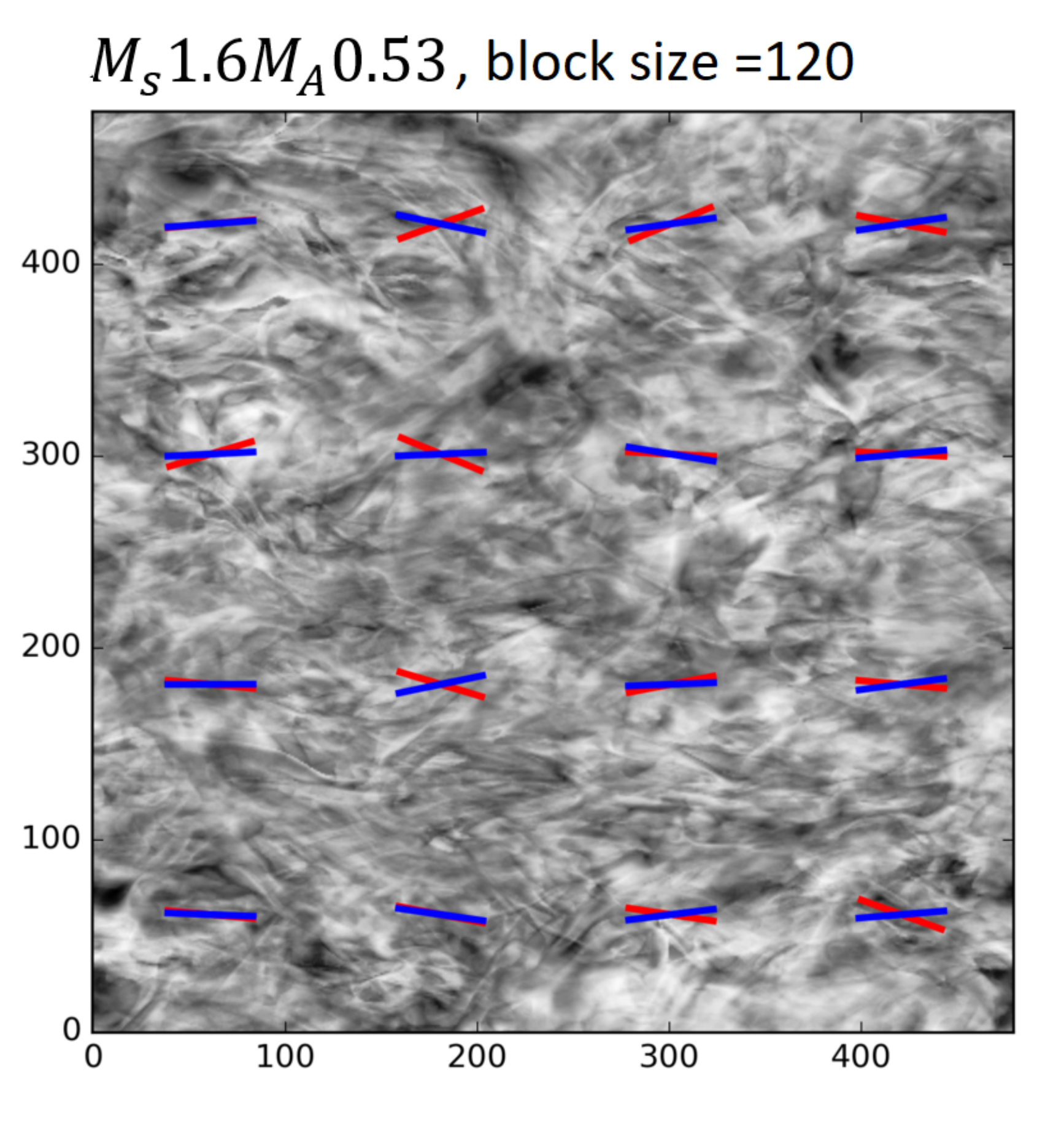}
\includegraphics[width=0.33\textwidth]{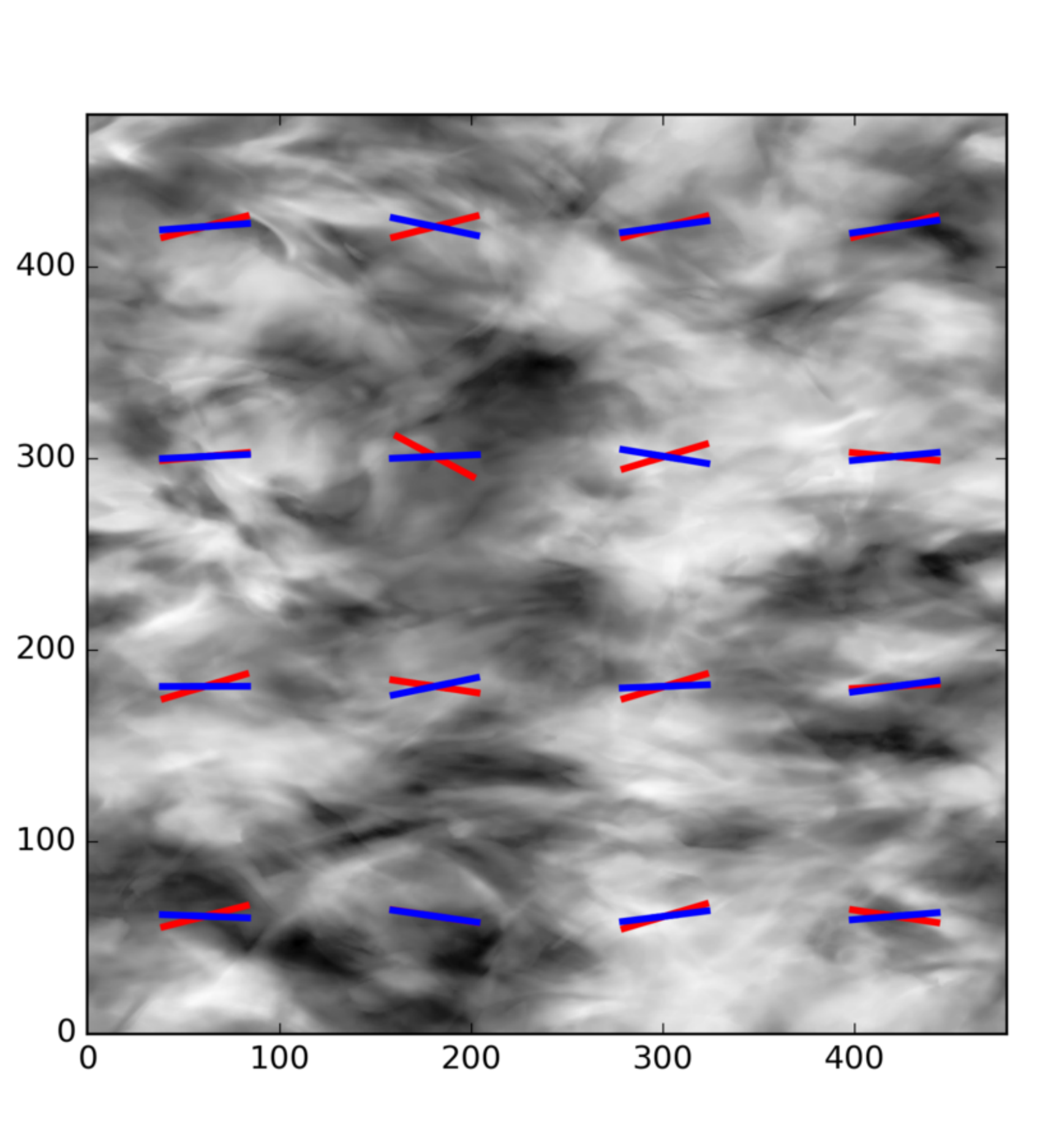}
\includegraphics[width=0.33\textwidth]{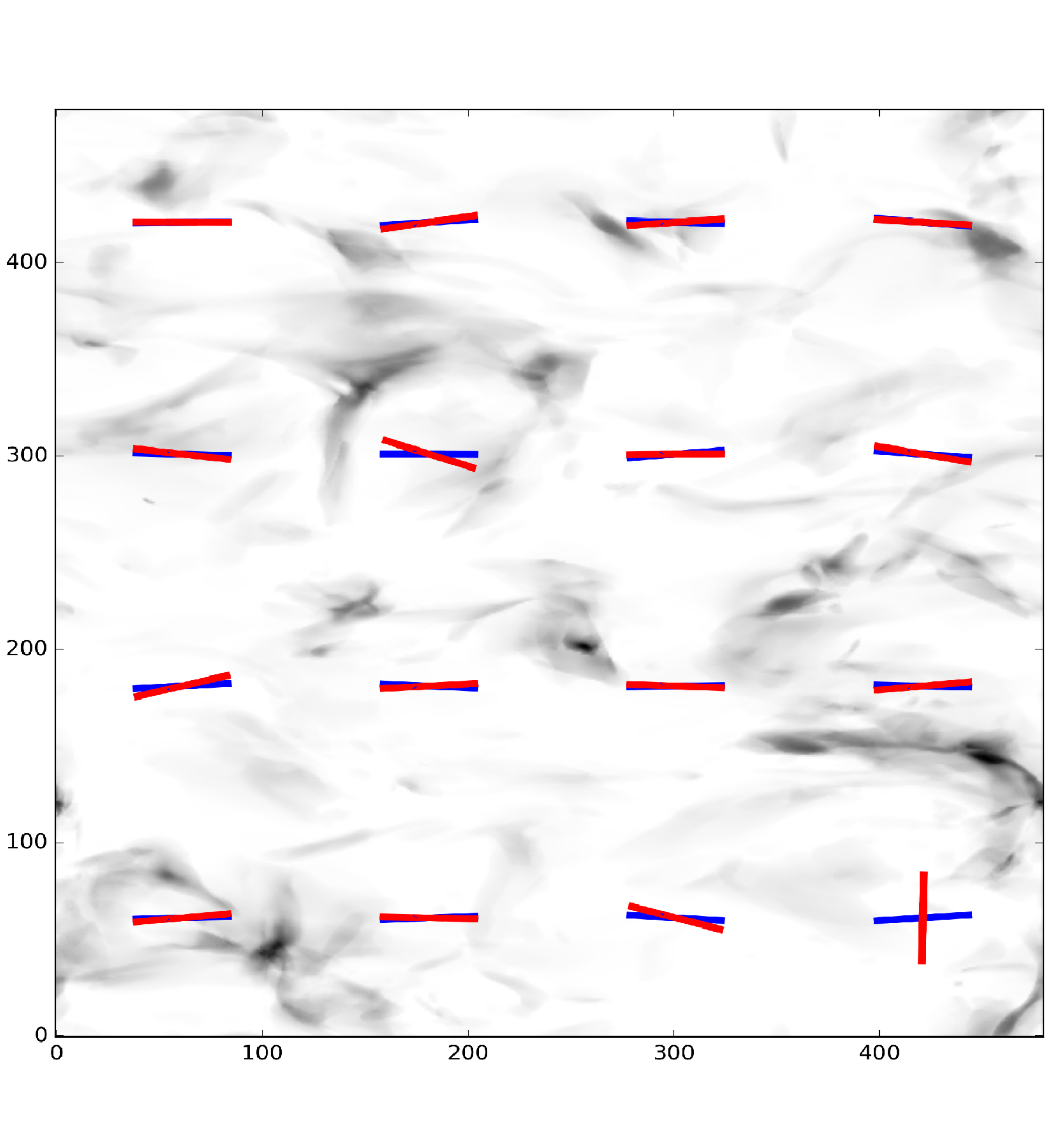}
\caption{\label{fig:rc-1} Maps of the reduced centroid for the central part of the line (left, $\delta v<0.2 \sqrt{\delta v_R^2}$), middle its middle part (middle,$0.2 \sqrt{\delta v_R^2}<\delta v< \sqrt{\delta v_R^2}$) and its wing (right,$\delta v>\sqrt{\delta v_R^2}$ ). Gradient directions are shown are in red, while the projected magnetic field directions are shown in blue.}
\end{figure*}

Given the channel map data we  introduce a new measure, i.e.  Reduced Velocity Centroids that contain only part of the line:
\begin{equation}
C_{\Delta v, n} ({\bf x})=\int_{\Delta v} dv v^{n} \rho_{ppv} ,
\end{equation}
where the index $n$ determines the order of the centroid. Increasing $n$ enhances the effects of velocity, but, in practice, it also increases the noise. Reduced Velocity Centroids are useful both in the case of studying gradients from extended galactic disk data and also for the data from the wings of the absorption lines. In the study we explore how Reduced Velocity Centroid Gradients (RVCGs) trace magnetic field. To study the new measure we separate the line into three regions, namely the central part, the middle part and the wing.  Figure \ref{fig:rc-1} illustrates the $n=1$ reduced centroid from a selected Alfven mode velocity data cube.  The central portion is very much the same as the thin slice result, giving a good alignment in respect to the magnetic field. The middle portion of the reduced centroid also shows a fair alignment, but the number of data points get limited due to the limitations of the numerical resolution and the shot noise increase. Nevertheless, the gradient map after sub-block average still provides a nice fit to magnetic field direction. The wing part has also a good alignment to magnetic field. That actually tells us that one can use the part of the line to study magnetic fields. This can be portion of the line broadened by the galactic shear, e.g. the 21 cm line from the galactic plane hydrogen as we discuss below.\footnote{This can also be a part of the line corresponding to clouds at velocities formally forbidden by galactic rotation curve (see Stil et al. 2006), but arising from the turbulent stochastic nature of the velocity distribution in the galactic disk. High velocity clouds also present a case where RVCGs can be useful.} This provides a way to study the 3D structure of magnetic field and other phenomena (e.g. self-gravity and shocks) that are being traced by gradients.\footnote{Naturally, the mapping from the velocity space to the real space is distorted by turbulent velocities and therefore is accurate up to the turbulent velocity dispersion.} Using only wings potentially provides a way to study gradients using lines with high optical depth e.g. 12CO line. All of this should be discussed in detail in other publications, while here we provide a preliminary study of the new measure. 

In fact, in MHD turbulence we expect that the overall line profile to be determined by the largest eddies that produce most of the dispersion. The gradients are not expected to depend on this. Therefore gradients in all three parts of the line reflect the alignment of the small scale eddies that are aligned in respect to the magnetic fields local to them.

\section{Comparison with the observational data}
\label{sec:obs}
\subsection{Application of VChGs and RVCGs}

For observations of galactic HI the use of different portions of the line is not any more just a test of our theoretical concepts. Due to the galactic rotation, different parts of the galactic HI have different velocities in respect to the observer. Therefore one can trace the magnetic field directions using the VChGs and the RVCGs. We illustrate our techniques using three regions consists of both diffuse HI regions and also 13CO data from a self-gravitating molecular cloud Vela C (Fissel et al. 2016). The HI region we compare with are from \cite{Clark15} for the studies of fliamentary structures in velocity channel spaces coming from the Galactic Arecibo L-band Feed Array HI (GALFA-HI) survey. GALFA-HI is an survey of the Galaxy in the $21 cm$ neutral hydrogen line. The data is obtained with the Arecibo Observatory 305 meter telescope. With the large aperture of the telescope, one get the angular resolution $\sim4'$, which is similar to Planck's best resolution ($\sim 5'$). For Vela C we compare our calculation with Balloon-borne Large Aperture Submillimeter Telescope for Polarimetr (BLASTPol, Galitzki 2014, Fissel et.al 2016), which is a 1.8-meter Cassegrain inferometric telescope detecting linearly submillimeter dust polarization. 

\begin{figure*}[h]
\centering
\includegraphics[width=0.49\textwidth]{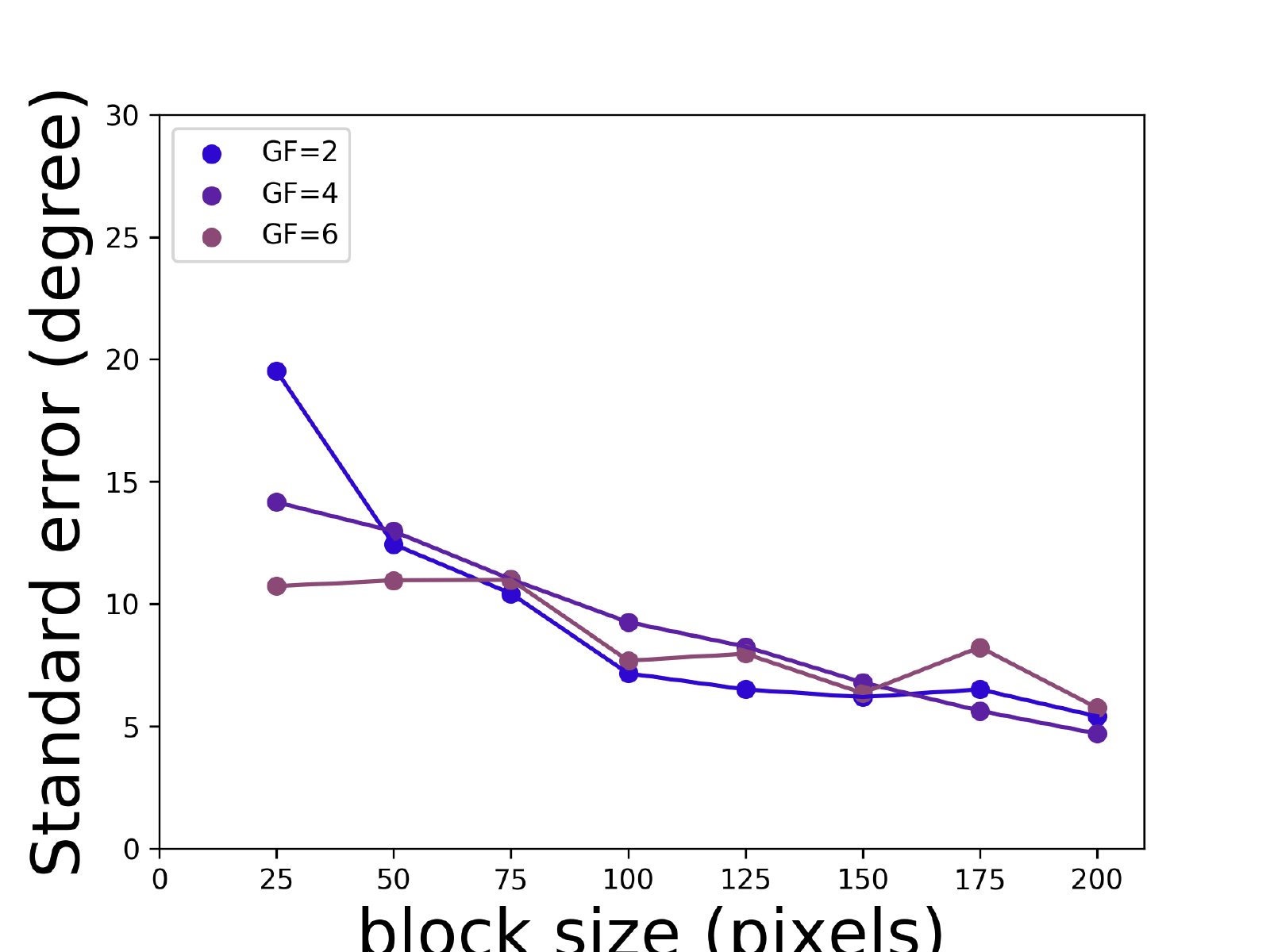}
\includegraphics[width=0.49\textwidth]{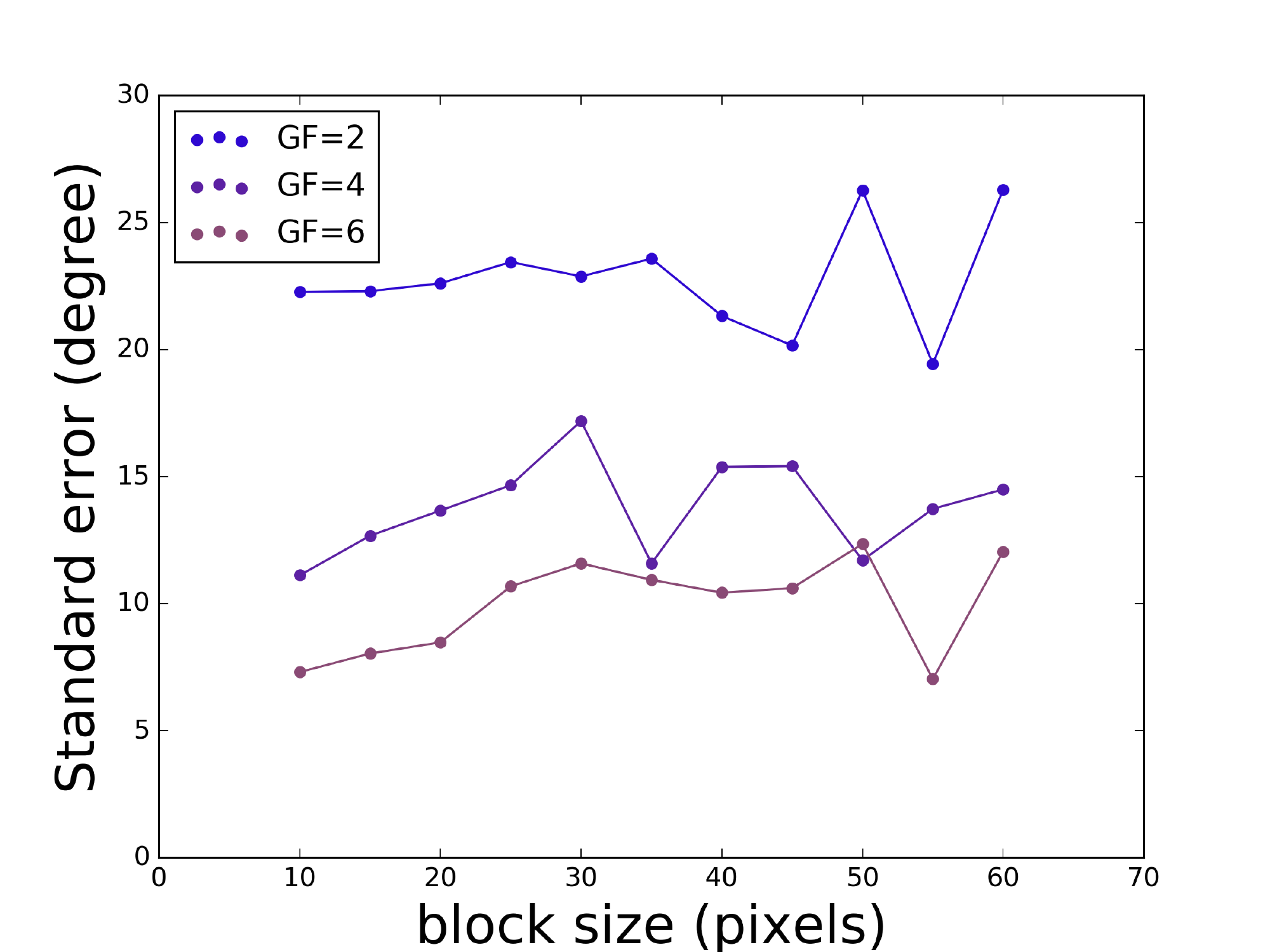}
\caption{\label{fig:AM_BS} Change of standard errors versus block size when applying different Gaussian Filters for GALFA-HI and Vela C. }
\end{figure*}

\begin{figure*}
\centering
\includegraphics[width=0.98\textwidth]{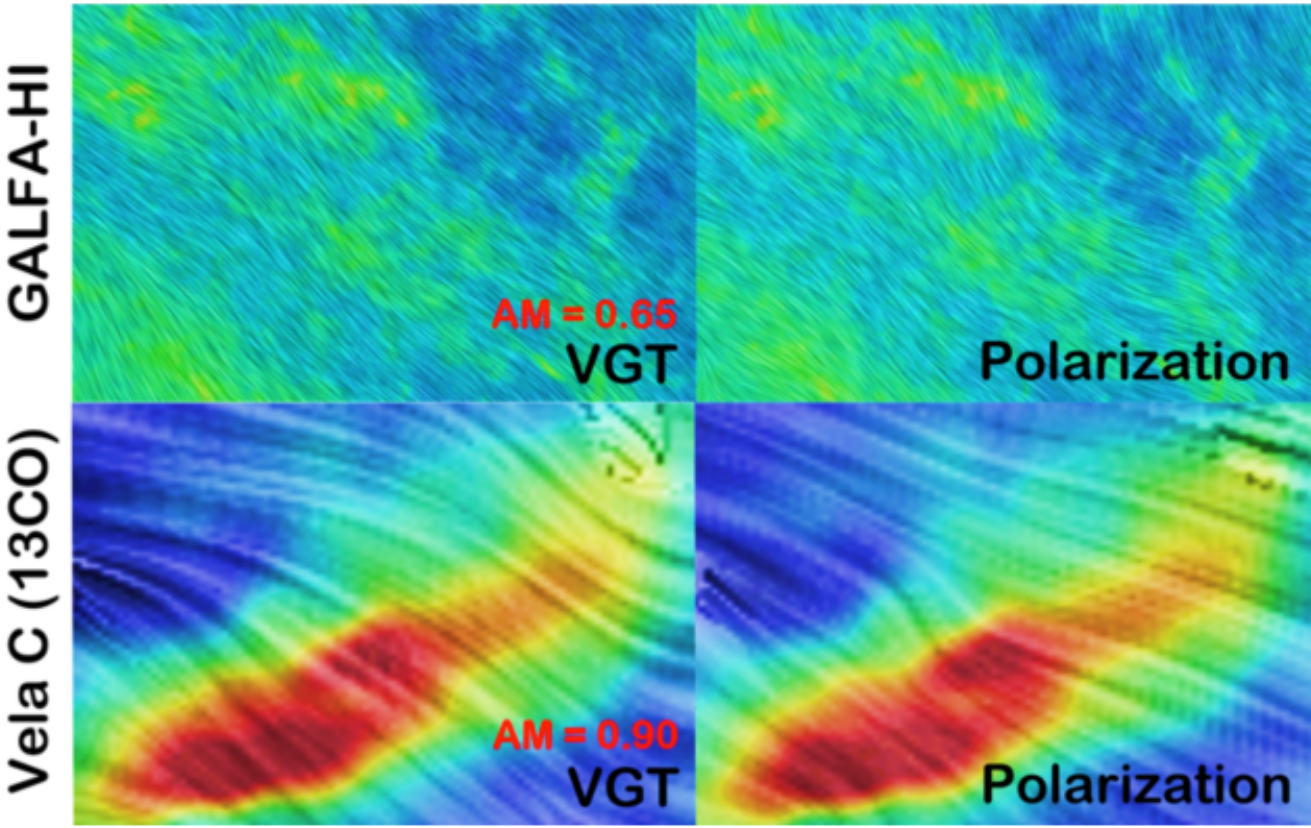}
\caption{\label{fig:GALFA} Magnetic field drapery patterns probed using VChGs (rotated $90\deg$ created by the line integral convolution method (LIC, Cabral \& Leedom 1993) for two observational examples GALFA-HI and Vela C (Fissel et al. 2016, reproduced from Yuen et al 2018). We here compare polarimetry data from different sources as the cloud density increases, and plot two of the magnetic field directions with LIC by rotating polarization for $90\deg$. The intensities are the {\it LIC-processed} texture that is proportional to the tracers intensity of the two maps.}
\end{figure*}

The regions we selected here are across $209.7^\circ - 196.6^\circ$, and declination (Dec.) across $22.5^\circ - 35.3^\circ$ for HI and $l\sim264.75-265.60$ and $b\sim 1.2-11.8$ for Vela C. .For the spectroscopic HI data, we stick with the channel with velocity ranges $0-2.944km/s$, which the velocity spectral line is peaked at. As a comparison, the dispersion of velocity across the whole map $10.3 km/s$, which means our selection of velocity channel is both thin (with a relative channel thickness of $\sim0.3$) and most representative compared to other channels. The analysis of Vela C is produce in Yuen et al. 2018. Here we reproduce one figure from that study to illustrate the ability of the VChGs to trace magnetic field in molecular clouds. Due to the thinner channels provided in the 13CO Vela C data (0.183 km/s), we use the relative channel thickness of $\sim 0.5$ for Vela C, which is about $1km/s$. We plot our VChGs for HI against Planck 353 Ghz polarization in the respective region by rotating both of them for $90\deg$. For Vela C, as it is a self-gravitating region, we keep the center part of VChGs unrotated while the boundary rotated for $90\deg$ (See YL17b for the criteria and the method of determining the gradient rotation density threshold) by estimating the volume density around the core region. We vary our block size and Gaussian kernel width (See Fig \ref{fig:AM_BS}) to acquire the optimal value ensuring the best combination of the alignment and the resolution. For the case of HI, the Gaussian filters have their respective gradient errors converge at block size of 100. For Vela C there is no such convergence plot, therefore we pick the lowest error point from the smallest Gaussian Filter to retain the data quality. The result is shown in Fig\ref{fig:GALFA} using the line integral convolution method (LIC, Cabral \& Leedom 1993). We find that the alignment measure is in general pretty well for the two regions.

Note that taking the thickness of the channel map of the order of the velocity dispersion one produces intensity-dominated channel maps, which provides a rough structure of the column densities with the grid size of the order of the large scale turbulent eddies. The gradient study of such a map of which is equivalent to the study using the IGs. Therefore the corresponding channel map gradients can be associated with the IGs that trace both the combination of the effects arising from magnetic field and shocks, as we discussed in YL17b. Combining this with the magnetic field information e.g. obtained with the VChGs is very synergetic for studying interstellar shocks propagating through galactic disk (See Sec 11.3 for a complete discussion)

\section{Additional effects and prospects of the research}
\label{subsec:selfgrav}

The technique of studying magnetic fields with velocity gradients shows many promising directions for further studies. We outline a few directions
without getting into much detail within this publication. 

\subsection{Effect of self-gravity}

\begin{figure}
\centering
\includegraphics[width=0.49\textwidth]{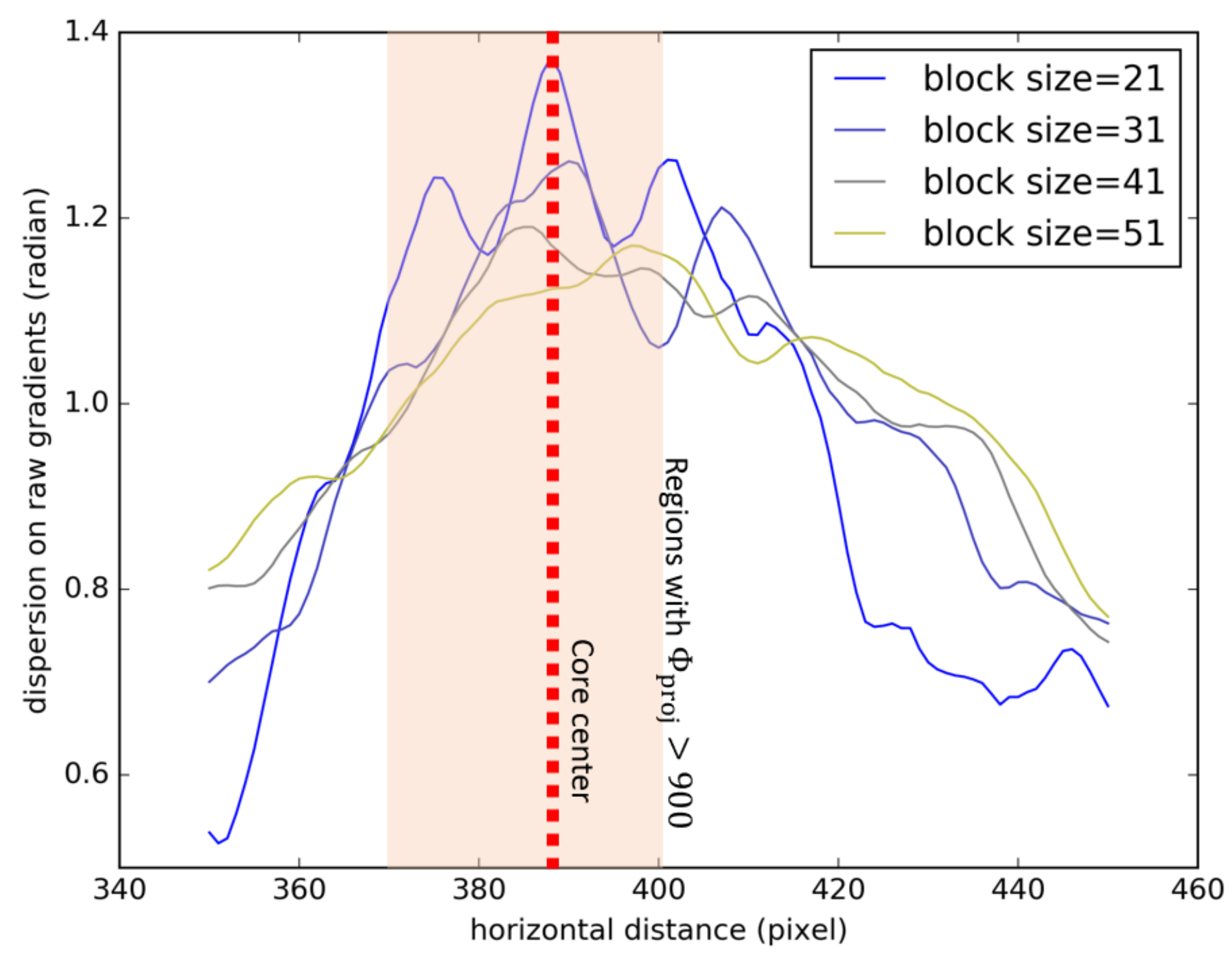}
\caption{\label{fig:grav_disp}  Dispersion of {\it raw} gradients with respect to block position (which we pick the center of the block to be the x-axis) for several choices of block size.}
\end{figure}

In the presence of self-gravity we expect that the gradients to change their alignment in relation to magnetic field. Naturally, if this happens over the block over which the calculation of the gradients is performed, this will result in the higher than average uncertainty in determining the direction of gradient. This is what we observe in our simulations, as illustrated in Fig \ref{fig:grav_disp} shows how the {\it raw} gradients (i.e. gradients before block-averaging) react when they come close to the self-gravitating core. Instead, we calculate the dispersion of raw gradients within a block of selected size, and move the block from the left of the core to the right. From Fig \ref{fig:grav_disp} we clearly see, as we come closer to the gravitational center (core center), the dispersion of gradients significantly increases. We also observe that the increase of dispersion is smoothed out and the peak of dispersion is deviated from the core center as we are using bigger block size. However, the average level of dispersion around the core region (about $\sim 1.2 radian$is still significantly higher than that of the diffuse region, which is $\sim 0.8 radian$. 

\begin{figure}
\centering
\includegraphics[width=0.49\textwidth]{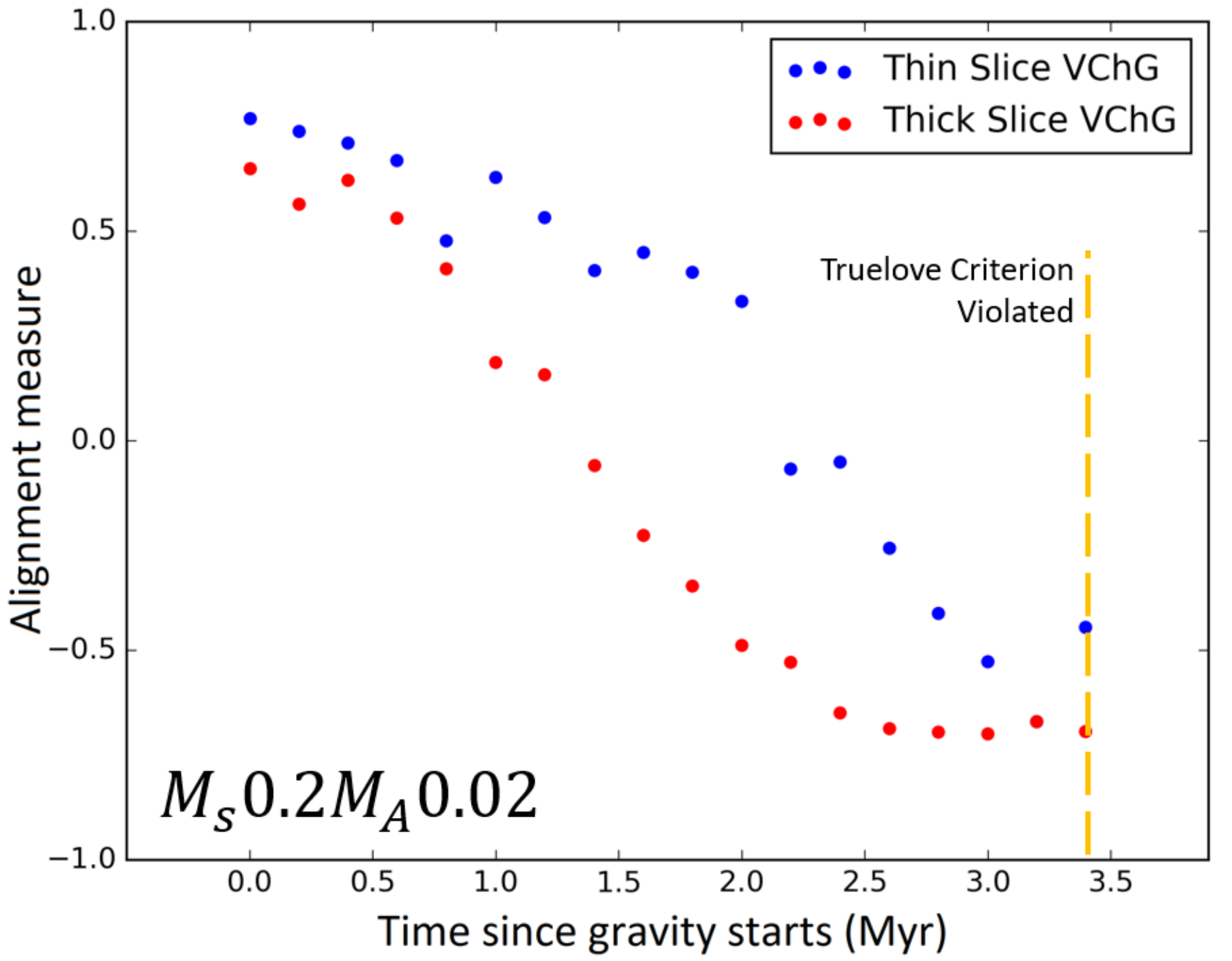}
\caption{\label{fig:grav} The variation of the AM of thick (red) and thin (blue) slices in time after the gravity is turned on.}
\end{figure}

In addition, our earlier work, in particular YL17b, identified the synergistic way of using simultaneously the VCGs and the IGs in order to reveal both regions dominated by self-gravity and shocks and to trace magnetic in the presence of these phenomena. In particularly,  in the presence of self-gravity both the VCGs and the IGs are changing their direction by 90 degrees, thus getting parallel to the magnetic field direction. However, the change direction of the IGs happens earlier than that of the VCGs and this provides the way of, first of all, identifying regions dominated by self-gravity, second, tracing magnetic fields within self-gravitating regions without any polarimetry data. Naturally, this effect is also present when we use the VChGs. The case thick velocity channels provides the information about the IGs, while the case of thin velocity channels provides the information about the velocities.  Figure \ref{fig:grav} illustrates the difference in the response with time of the gradients measured within thin and thick channels. The time is measured in simulations from the moment of self-gravity being turned on. Due to this effect in observations we expect the region of where the gradients in the thick velocity channels are turning 90 degrees in respect to magnetic field to be more extended than the region over which the gradients in thin velocity channels are turned 90 degrees in respect to the magnetic field. Therefore while at the regions far from the self-gravity center, the gradients in thin and thick velocity channels are going to be aligned, they will turn 90 degrees in respect to each other at some distance from the self-gravity center and closer to the self-gravity center they are expected to get back being aligned. The detection of such a change is a signature of a self-gravity effect which can be used to indicate when it is rotated 90 degrees gradients that trace magnetic fields.   

Turning of the gradients in thin velocity channels compared to thick velocity channels is also expected to happen within strong shocks. However, shocks can be distinguished from the self-gravity regions both through morphological differences and due to differences in the column densities. Thus in most cases, one can trace magnetic fields, shocks and identify regions of gravitational collapse through combining the gradients within thin and thick velocity channels.

\subsection{Use of interferometers and the effect of noise}

\begin{figure*}
\centering
\includegraphics[width=0.98\textwidth]{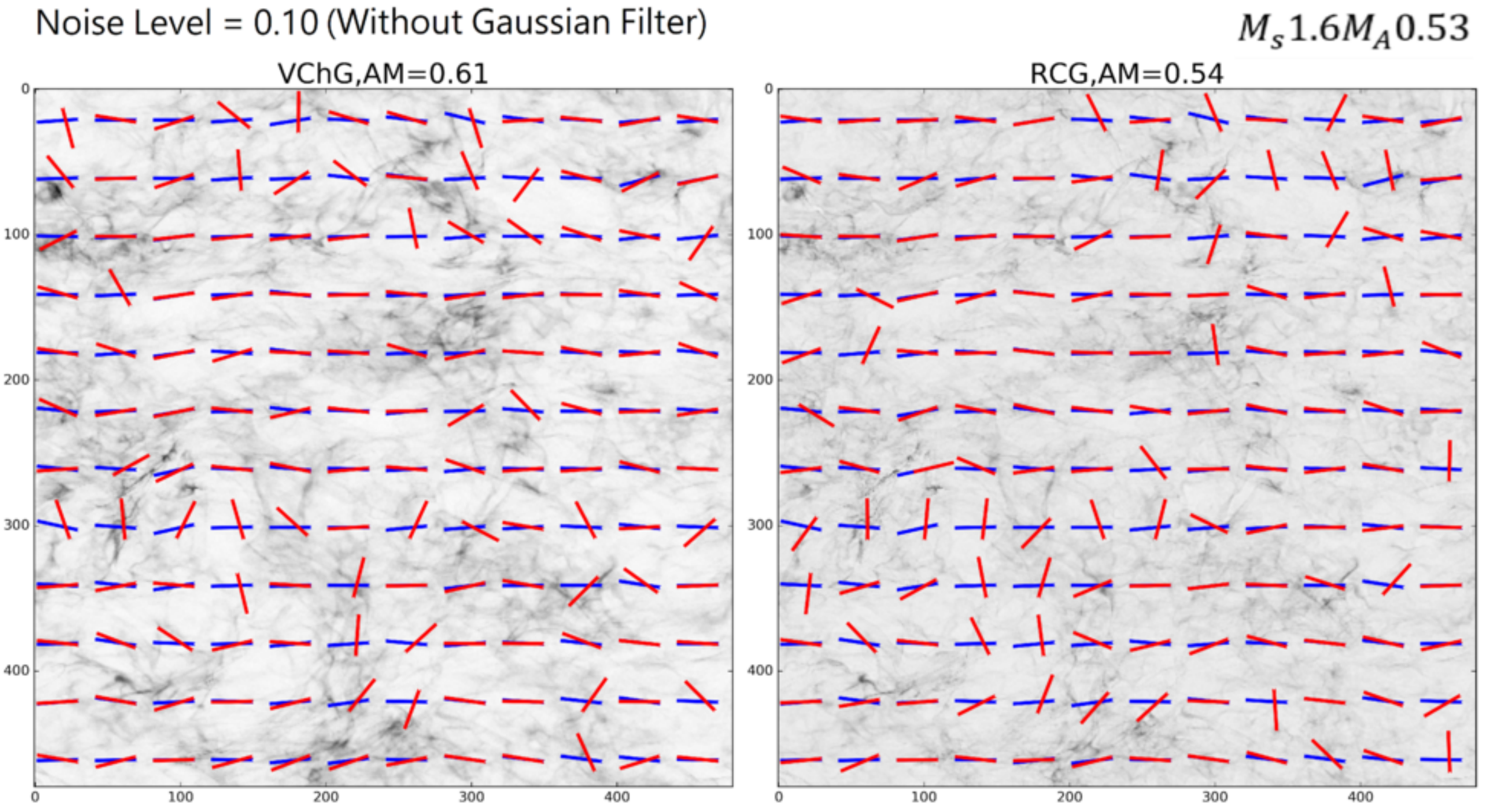}
\caption{\label{fig:noise} The alignment between gradients and projected magnetic field with white noises added to the original maps. }
\end{figure*}

\begin{figure*}
\centering
\includegraphics[width=0.98\textwidth]{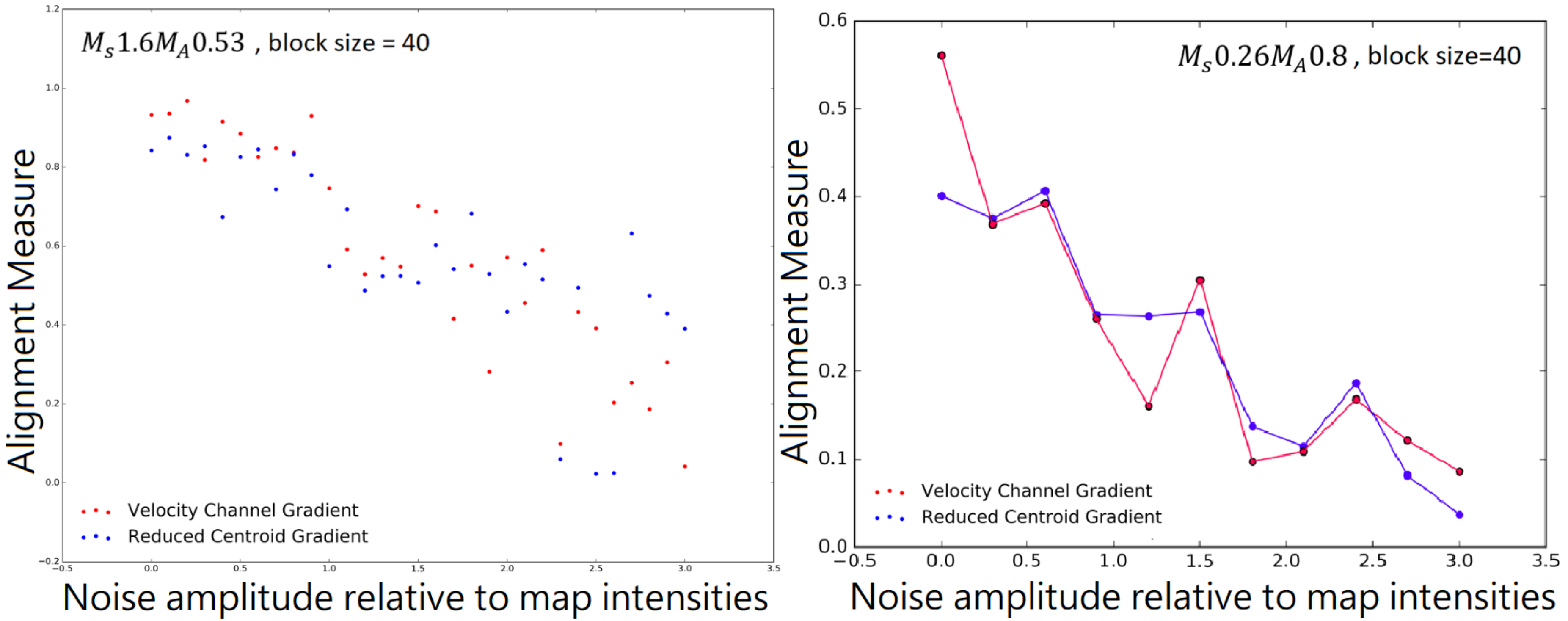}
\caption{\label{fig:noise-change} (Left) The alignment measure versus the amplitude of noise for VChGs and RVCGs respectively. A Gaussian Filter of width 2 pixels have been applied to all the map with noises. (Right) The same procedure as from the left but we keep density constant for low $M_s$ case.}
\end{figure*}

We would also like to examine the effect of noise for both RVCGs and VChGs.  After obtaining channel and reduced centroid maps from cube Ms1,6Ma0.53 at relative channel width of $0.2$, we add white noises as we did in \cite{LYLC17} on top of the maps, i.e. add white noises according to the root-mean-squared value of map intensities. Figure ~\ref{fig:noise} shows that with noise level being $0.1$ of the map intensities. One may already observe from the map that there are already some dirt altering the alignment of gradients to projected magnetic field. We therefore add a Gaussian filter with width of 2 pixels according to \cite{LYLC17} to negate the effect of noise. Figure ~\ref{fig:noise-change} shows the alignment measure versus the amplitude of noise for VChGs and RVCGs respectively after the Gaussian filter. The alignment remains fairly good until the ratio between noise amplitude relative to the map intensities is close to 1, and significantly drops to zero as the ratio is around 3. This simple test illustrates the strength of VChGs and RVCGs in predicting alignments in strong noise environments with the aid of small-width Gaussian filters.

Our results also show that RVCGs and VChGs are comparable in their performance both when the turbulent broadening is larger and smaller than the turbulent line width. Revealing the relative advantages of these two techniques requires a more detailed study and is beyond the scope of the present paper.

\begin{figure*}
\centering
\includegraphics[width=0.49\textwidth]{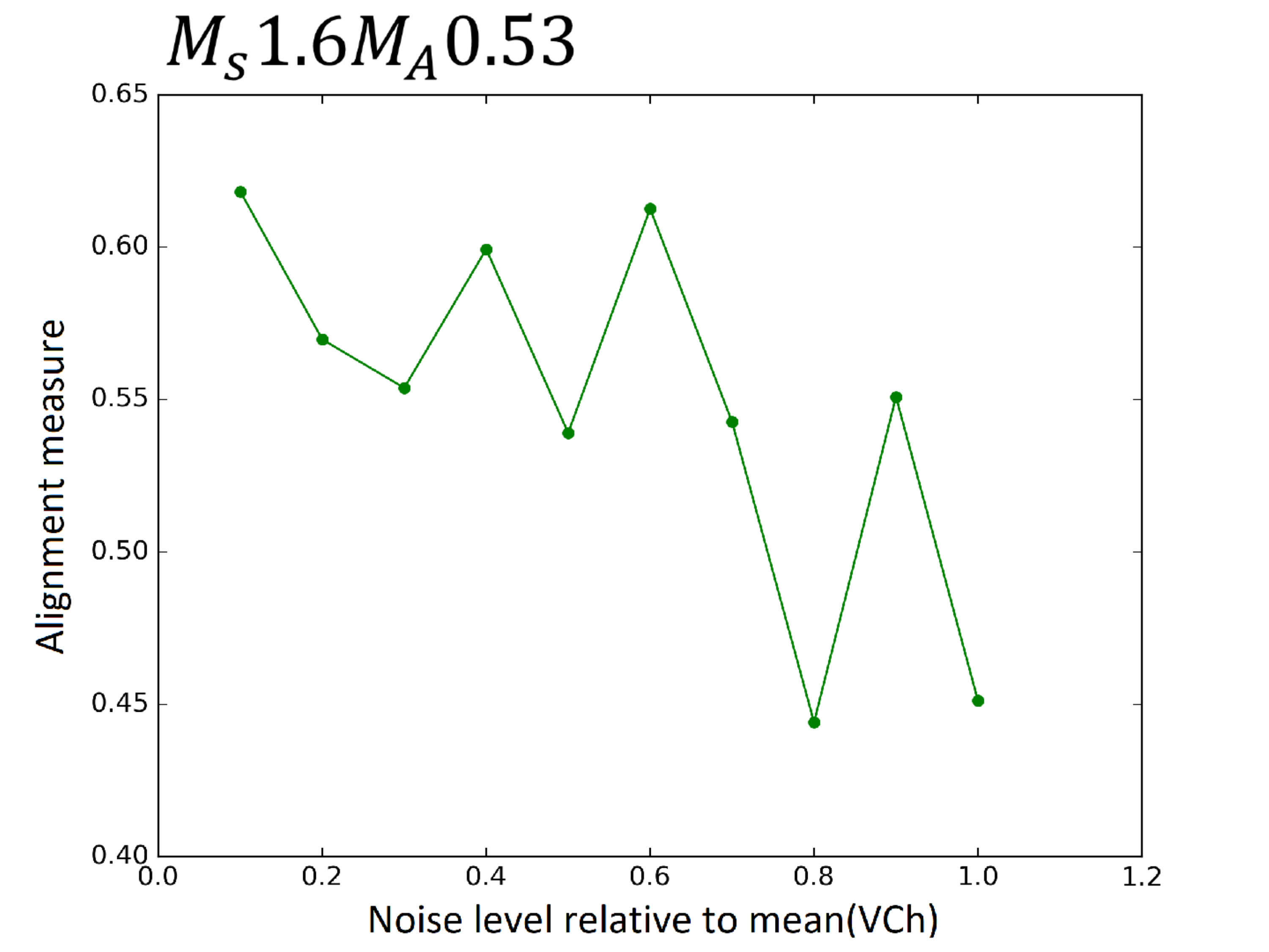}
\includegraphics[width=0.49\textwidth]{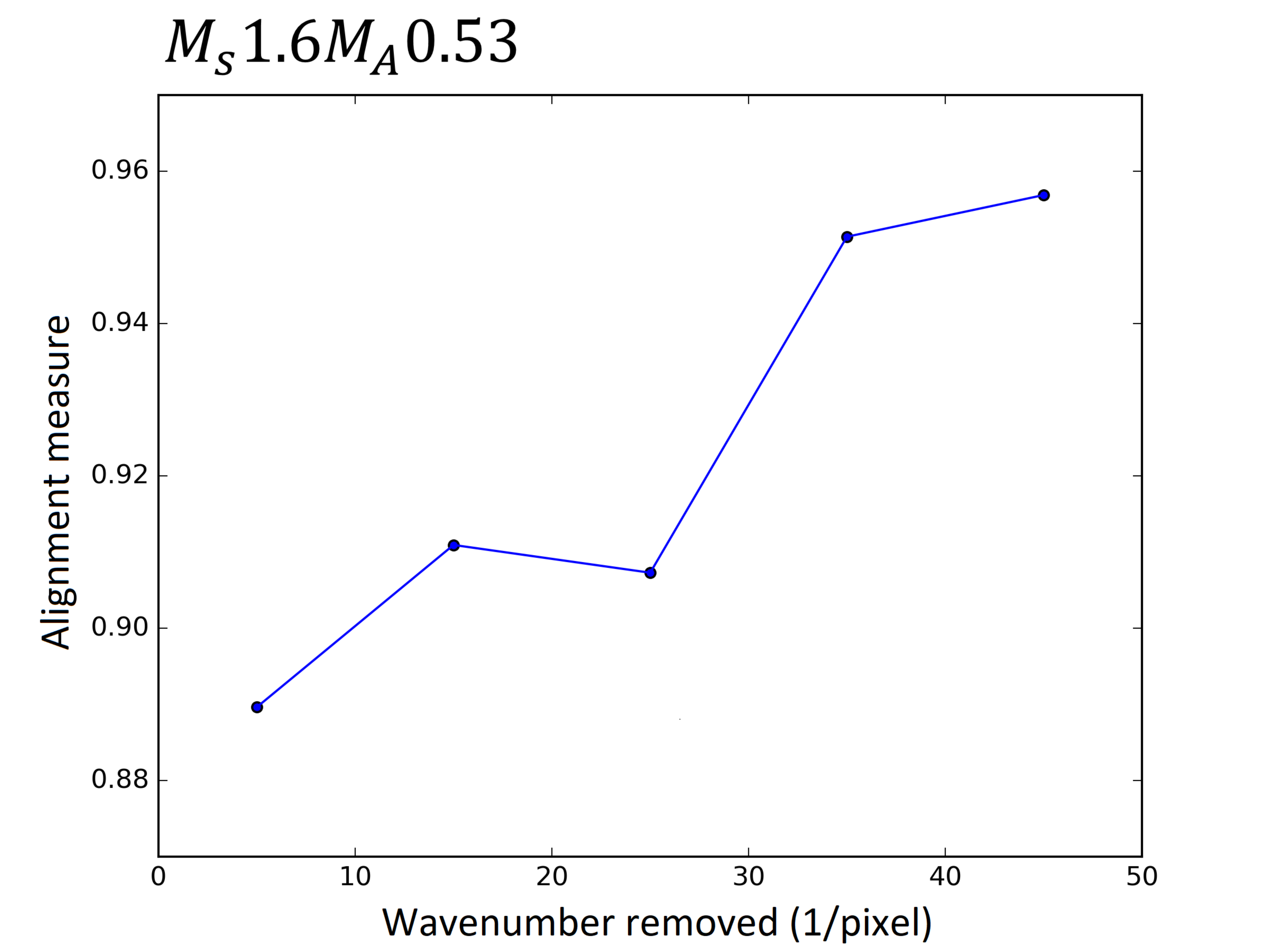}
\caption{\label{fig:practical} (Left) Illustration of the noise suppression technique works for VChGs. (Right) Effect of the low wavenumber removal on the VChGs.}
\end{figure*}

In Figure \ref{fig:practical} we show how the VChGs change in the presence of noise. We see that it is advantageous to remove spatial frequencies in order to decrease the noise aside from the Gaussian Filter method. Incidentally, similar results have been achieved in mimicking  LYLC and  YL17b with other types of gradients. We would like to note that the interferometric studies are frequently missing the low frequencies and therefore the filtering of the low frequencies is happening naturally in this case.

\subsection{Pure velocity caustics}

Both velocity and density gradients in subsonic turbulence are directed perpendicular to magnetic field. Therefore to isolate the effect of velocity we create PPV cubes using constant density. For such cubes all the intensity variations arise from the velocity crowding. To mimic subsonic turbulence we introduce the thermal line width which produces broadening larger than the turbulent velocities, in our case the thermal line width is 3.78 times of that of turbulent velocities. Our studies of velocity channel intensity correlations in \cite{2000ApJ...537..720L,2004ApJ...616..943L} suggests a significant suppression of the amplitude of the correlations in such a case and, as a result, high influence of noise. Thus in the  subsequent publications, e.g. \cite{2017MNRAS.464.3617K}, we claimed that velocity centroids provide a better way on studying velocity fluctuations when the thermal linewidth exceeds the turbulent line broadening. Nevertheless, for the gradients we do not see such a lose of information (See right of Figure \ref{fig:noise-change}), which we relate to the fact that for gradients the differences of the velocities are important and therefore the constant thermal width does not much affect them. This is an interesting effect that calls for further studies of the relative advantages of the channel map gradients and velocity centroid gradients. In particular, the studies of gradients of the velocity centroids may be advantageous in order to provide the complementary information.

\subsection{Physics of velocity gradients at different scales: ways to determine B-strength}

Real astrophysical flows include both regular and turbulent motions. However, an interesting property of turbulence that it increases the velocity gradients as the scale decreases. Therefore the velocity gradients induced by turbulence are expected to dominate. In addition, to increase the signal to noise, the contribution of the large scale motions can be removed using spatial filtering. 

Properties of MHD turbulence are different at different scales. For instance, if turbulence is injected at scale $L$ with velocities $v_L$ larger than the Alfven velocity $V_A$, up to the scale $l_{A}=L M_A^{-3}$, where $M_A=v_L/V_A$ is the Alfven Mach number, the turbulent motions are marginally affected by the magnetic field presence. Therefore if the velocity gradients are measured at scales larger than $l_A$ they cannot trace magnetic field. As a result, for super-Alfvenic motions the influence of gradients from scales larger than $l_A$ is pernicious in terms of magnetic field tracing. Motions at these large scales should be filtered out in order to increase the VChGs accuracy as a magnetic field tracer. 

At the same time the change of the properties gradients around the scale $l_A$ can be used to find the fluid magnetization. Indeed, by observing the change of the VChGs dispersion which changing the scale at which the gradients are calculated, one can determine $l_A$ and then calculate 
$M_A=(L/l_A)^{1/3}$. The calculations of $L$ can be done spectroscopically $L$ as demonstrated by Chepurnov et al. (2010) using the Velocity Coordinate Spectrum (VCS) technique suggested in Lazarian \& Pogosyan (2006). If $M_A$ is determined, the Alfven velocity can be obtained by associating the $v_L$ with the velocity dispersion that is measured spectroscopically. Note, that this technique is applicable to finding the magnetic field in super-Alfenic turbulence, i.e. turbulence with $M_A>1$ where the traditional Chandrasekhar-Fermi technique fails (see Falceta-Golzalez et al. 2008)

In practical terms, we expect to observe that the magnetic field to be organized over the patches of the scale $l_A$ with the mean magnetic field in different patches to change randomly  from one patch to another. As gradients reflect magnetic field direction we expect a similar coherent organization of velocity gradients over the patches of $l_A$.  

\begin{figure}[t]
\centering
\includegraphics[width=0.49\textwidth]{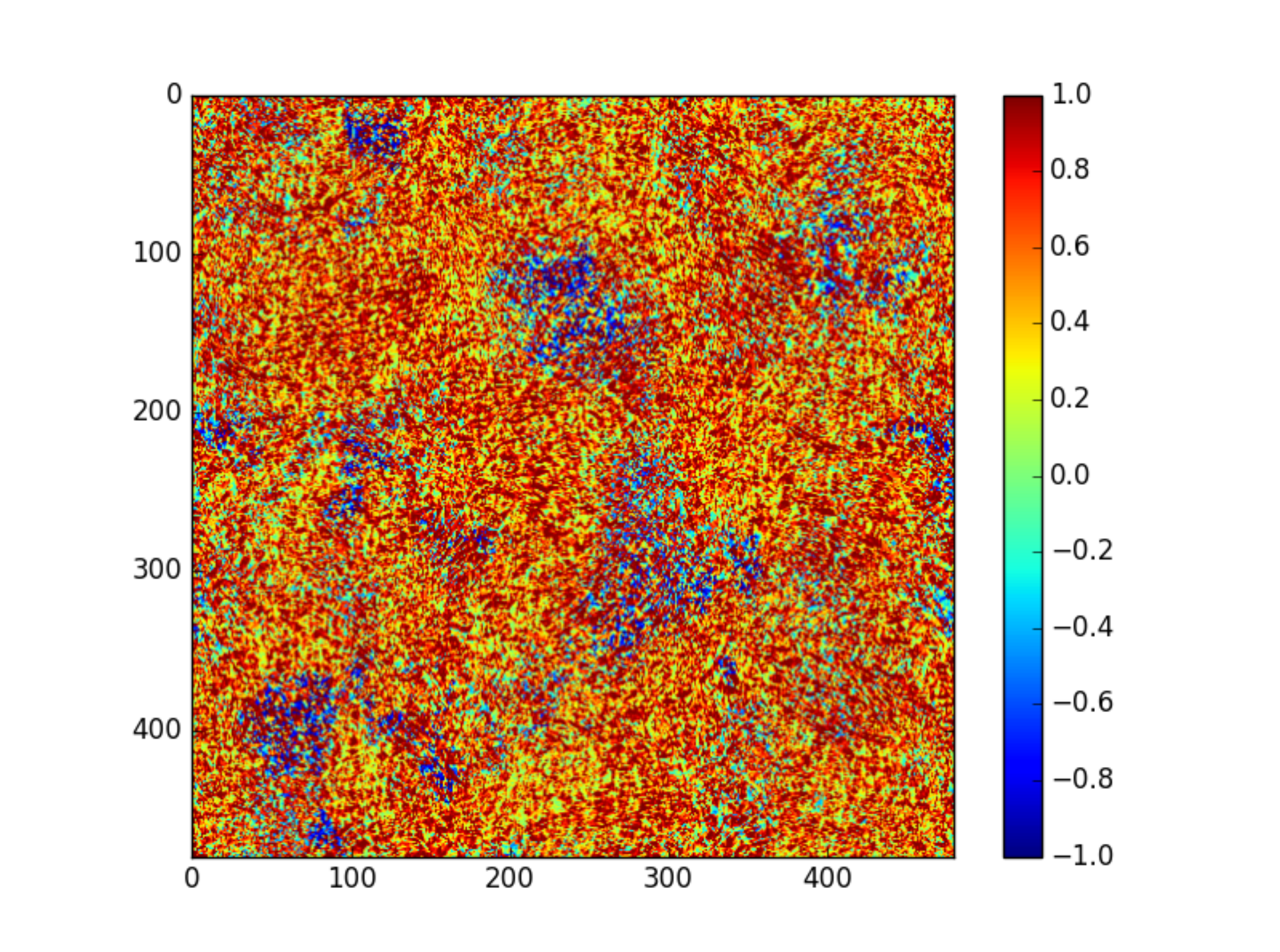}
\caption{\label{fig:islands} The distribution of alignment measure between {\it raw channel gradients} and projected magnetic field in our super-Alfvenic example Ms0.2Ma2.0. }
\end{figure}

The VChGs that we have discussed are tested with the simulations of MHD turbulence. The MHD approximation is applicable to partially ionized gas to the scales significantly larger than the scale of the neutral-ion decoupling (see Lithwick \& Goldreich 2001, Xu \& Lazarian 2017). At such scale the rate of turbulent variations is slow compared to the rate of neutral-ion collisions and therefore both species behave as one magnetized fluid. For Kolmogorov or GS95 scaling the rate of the eddy rotations changes as $v_l/l\sim l^{-2/3}$, i.e. increases with the decrease of the scale. Thus at a sufficiently small scale $l_{decoup}$ the ions do not collide with the ions frequently enough and get decoupled. For $l<l_{decoup}$ neutrals can start their own unmagnetized cascade and therefore their motion will stop reflecting magnetic field. 

We have discussed that the VChGs are tracing the magnetic field at the smallest scale that is resolved by the telescope. If this scale is larger than $l_{decoup}$ then the velocity gradients represent magnetic field as we discussed in this paper. The change in the orientation of the velocity gradients in the vicinity of $l_{decoup}$ can be used to establish this important scale. In particular, we expect to see the change of the relative orientation of the gradients obtained with neutrals with maps smoothed over scales $l>l_{decoup}$ and the high resolution maps resolving $l<l_{decoup}$. We also expect to see the differences of the velocity gradient orientation of neutrals and ions at scales $l<l_{decoup}$. Both effects can be used to establish $l_{decoup}$. The latter scale is related to the magnetic field strength as discussed in Xu et al. (2016) which presents a new way of magnetic field strength measurements. In other words, the VChGs present a tool that can be used both to trace magnetic field in molecular clouds and, given enough resolution, to test the neutral-ion decoupling scale. 

We believe that velocity gradients can trace magnetic field in most turbulent astrophysical environments. In particular, we expect the velocity gradients to trace magnetic fields in the accretion disks, where recent polarimetry showed that the grain alignment happens in respect to the radiation (see Rue et al. 2017, Lazarian \& Hoang 2007). 

Naturally, the changes of velocity gradient properties that we described for the case of the VChGs can also be studied with the VCGs and the RVCGs. 
We will present detailed testing of the ability of the velocity gradient technique to get the magnetic field strength elsewhere.

\subsection{Constraining VChGs directions using expectations from MHD turbulence theory}

It is important to make magnetic field tracing as precise as possible. In the sub-block averaging (YL17b), the blocks are created by evenly dividing the maps into localized regions and identifying the peaks of the Gaussian fits in these blocks as the most probable values within the block.  Theory of MHD turbulence provides ways to improve the procedures for magnetic field tracing. Below we experiment with two ways of improving the alignment and we view this as just initial steps to the MHD turbulence theory based automated procedure of velocity gradient tracing that employs Machine Learning (see Le Cun 1990). 

\begin{figure}[t]
\centering
\includegraphics[width=0.49\textwidth]{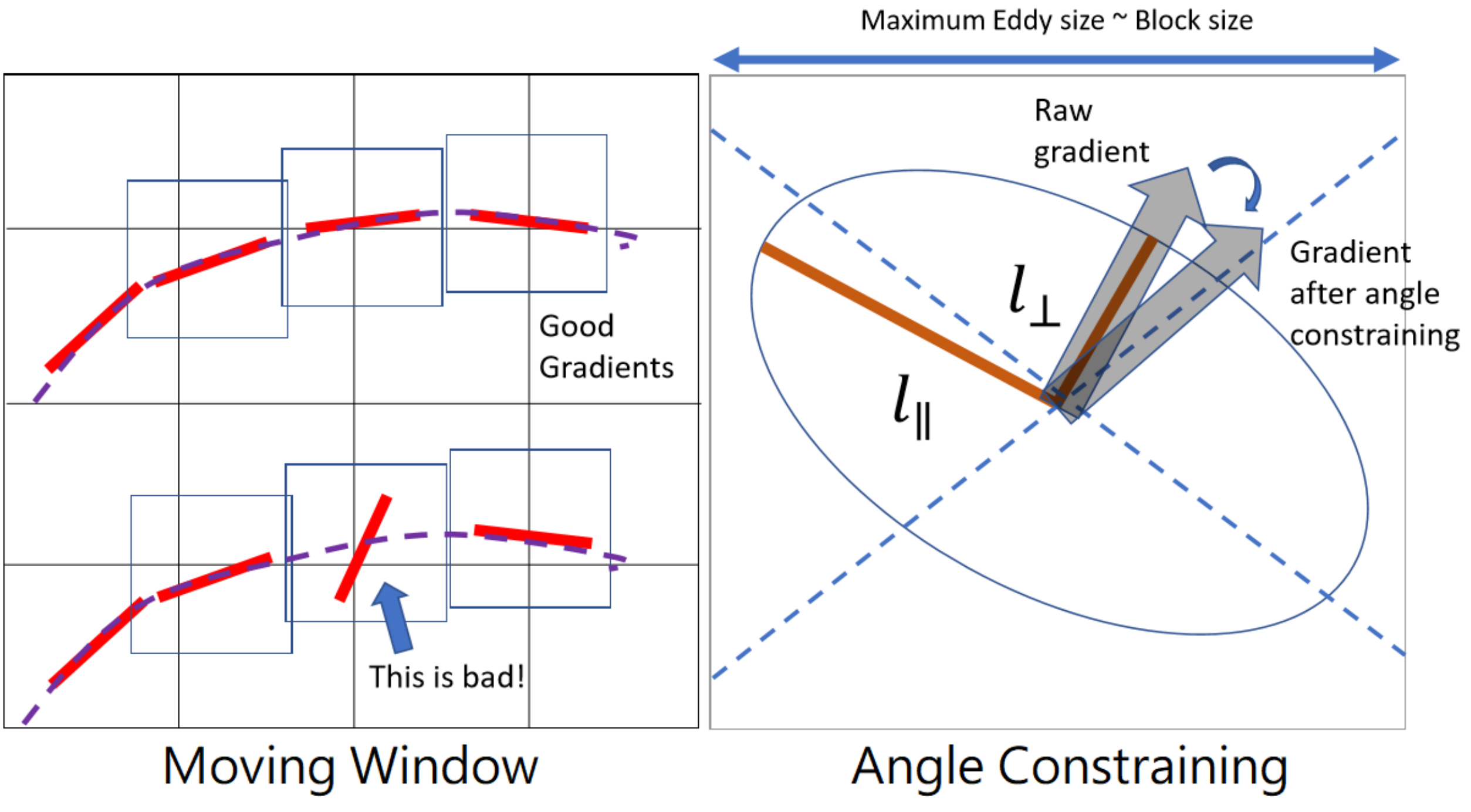}
\caption{\label{fig:newmethod} A pictorial illustration on how Moving Window (left) and Angle Constraining (right) should be pictorially in our gradient technique.  }
\end{figure}

\begin{figure}[t]
\centering
\includegraphics[width=0.49\textwidth]{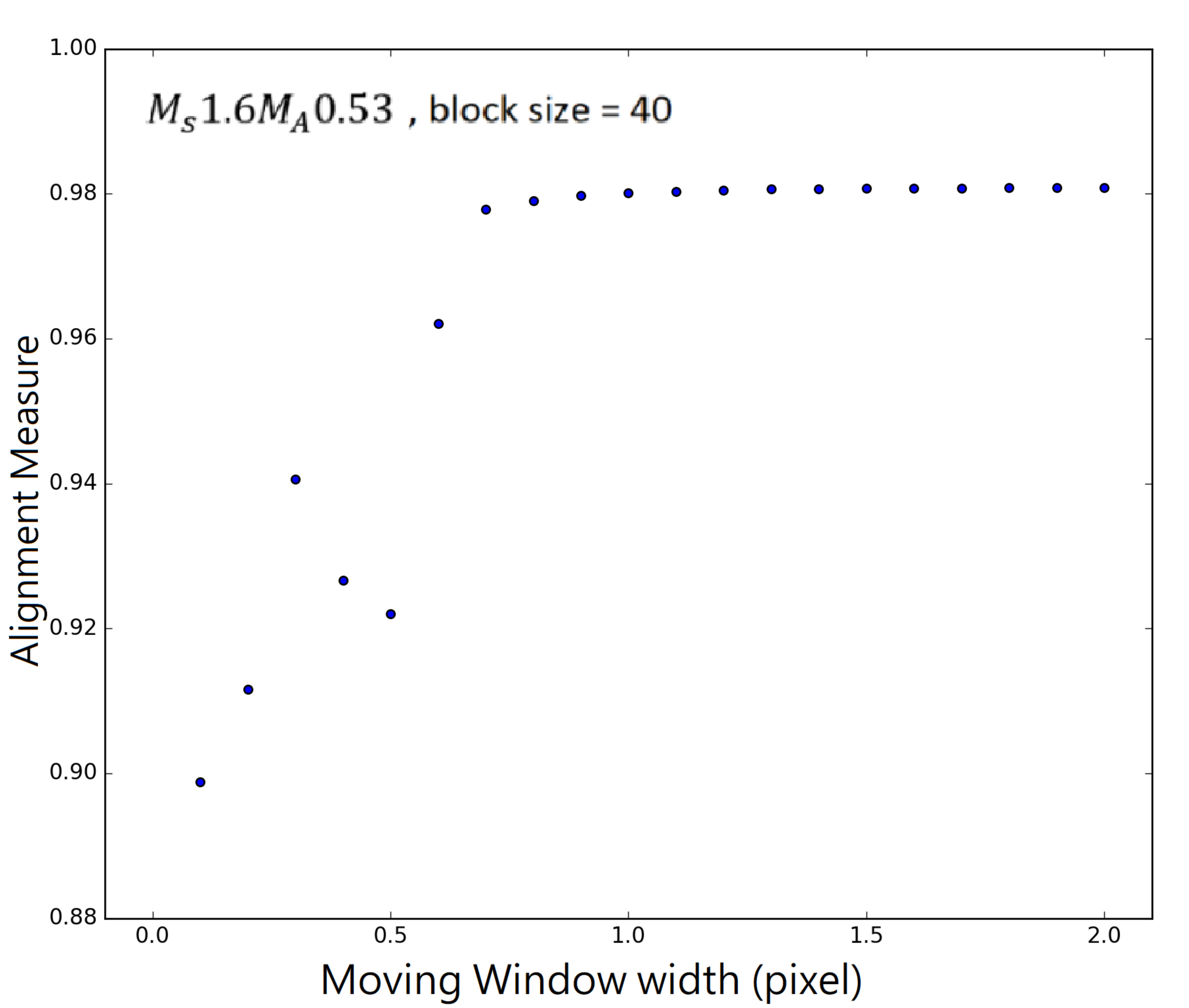}
\caption{\label{fig:MW} The effect of moving window to the alignment measure. }
\end{figure}

The Moving Window (MW) approach is an attempt to employ the sub-block averaging in a continues rather than a discrete manner. As magnetic field provides a continues representation it is advantageous to move the block and provide calculations as the block moves along the magnetic field lines that it traces. On the left of Fig \ref{fig:newmethod} shows pictorially how we improve the alignment: When there is an abnormal gradient vector compared to the neighboring vectors, we rotate the abnormal vector so that a smooth field line is formed. Mathematically the rotation can be handled by performing smoothing on both the cosines and sines of the {\it raw} gradient angle, which is a convolution of an averaging kernel with the raw cosine and sine data. We therefore pick the Gaussian kernel with different width to test how good the moving window is in our synthetic map. After the smoothing we apply the sub-block averaging to the processed gradient angle. Figure \ref{fig:MW} shows the strength of moving window with respect to the smoothing strength. One can see the alignment quickly rises until a window size of 0.75 pixel and being saturated after that.  Similar to the sub-block averaging,  one can estimate errors of the fitting in order to find the optimal moving window size. 

Another possible improvement is to use the expectations of the MHD theory in terms of applying the Angle Constraint (AC). Observations necessarily use the global system of reference related to the mean magnetic field. The probability distribution of magnetic field in terms of wavevectors parallel $k_{\|}$ and perpendicular $k_{\bot}$ to the mean magnetic field is provided in \cite{LP12}. This distribution predicts that the mean angle variations are the same at all scales. This variation can be determined with higher accuracy with the larger blocks and then can be used as a constraint when the direction of gradients is established within noisy small sub-blocks. Our application of the procedure of the AC provided a moderate improvement of the alignment measure of ~0.1 depending on the system's $M_A$. However, this procedure has minimal effect when $M_A$ is larger than unity. We believe that AC can be a useful part of the future algorithms of the VChGs calculation. 

\subsection{Dependence on the Alfven Mach numbers}

\begin{figure}[t]
\centering
\includegraphics[width=0.49\textwidth]{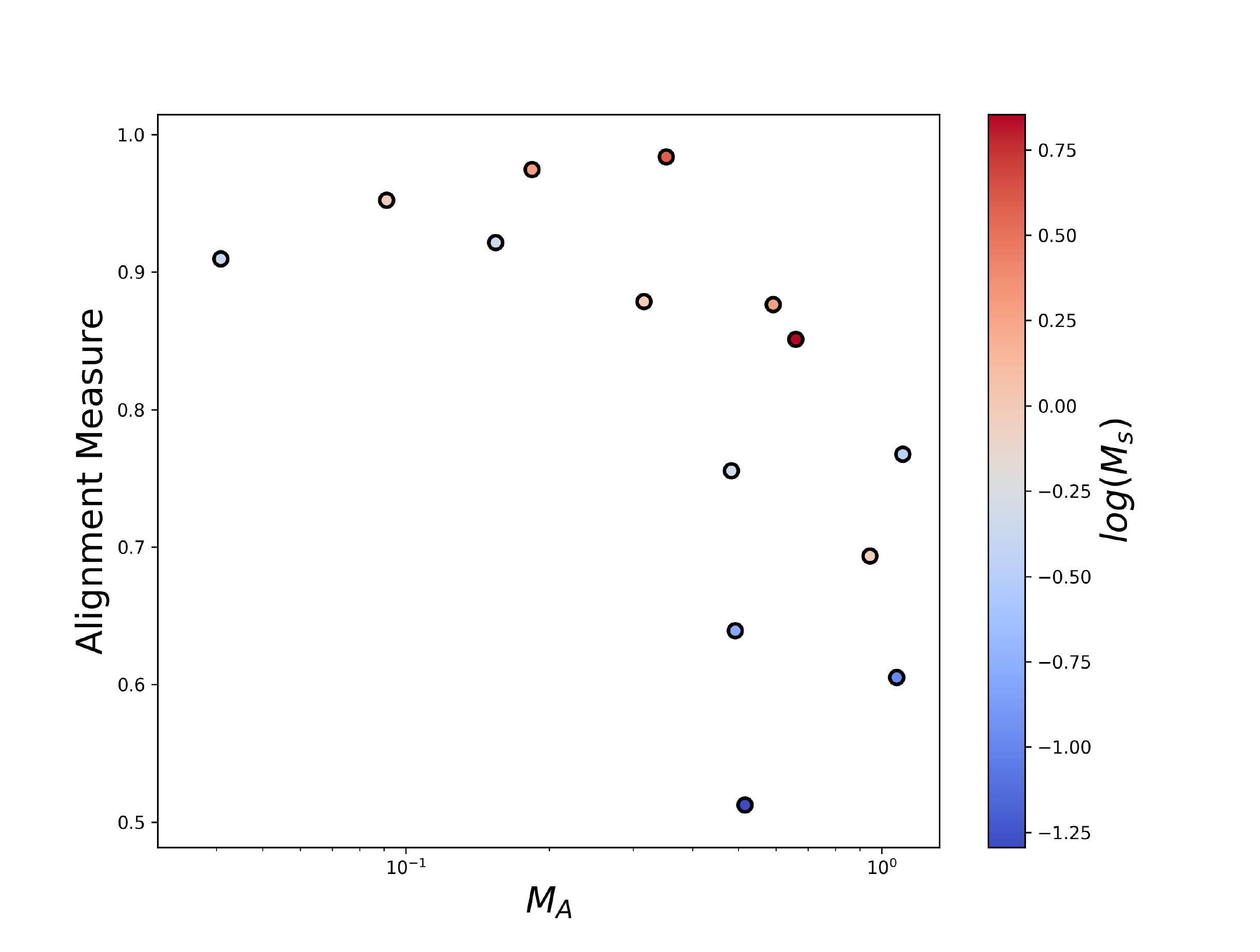}
\caption{\label{fig:maam} A plot showing the dependence of $M_A$ for alignment measure of channels (Selected relative channel width 0.2, block size 40, set B from YL17b.). The color of each point shows the logarithm of sonic mach number. }
\end{figure}

For observational tracing of magnetic field it is important to know what is to expect in terms of $AM$ dependence on the Alfven Mach number $M_A$. The Figure \ref{fig:maam} provides these dependence for our simulations. We see that the alignment decreases as $M_A$ approaches to unity. We attribute this to large angle variations of magnetic field along the line of sight. As $M_A$ increases, especially when the turbulence gets superAlfvenic, it is important to remove the low spatial frequencies. The corresponding procedure was shown to work in LYLC17, but here we present the direct application of the VChGs to the data without any spatial filtering. The procedures to improve the $AM$ for the VChGs are elaborated elsewhere. 

\section{Discussion}
\label{sec:discussion}

\subsection{Comparison with earlier studies}

The current study continues our series of research papers that are motivated by the modern understanding of the theory of MHD turbulence (see \citealt{2013SSRv..178..163B} for a review), in particular, the notion of the perpendicular cascade that is at the heart of this theory. Due to this cascade which can be visualized as hierarchy of turbulent eddies mixing magnetized plasmas perpendicular to the local direction of magnetic field (see \S \ref{sec:theory}), the gradients of velocity are expected to be maximal perpendicular to the local magnetic field, thus revealing its direction.  

The first paper that employed this theoretical arguments to trace magnetic fields using velocity centroids was GL17. This study attempted a quantitative approach, e.g. it introduced the alignment measure $AM$ that is used in the subsequent publications, including this one. However, the actual start of the quantitative tracing of magnetic fields with gradients relies on the block-averaging procedure introduced in YL17b. This procedure allowed determining both the pointwise direction of magnetic field and estimate the uncertainty of this direction. As a result, this procedure was used in all the papers that followed. In this paper, the block averaging approach used to identify the regions where the direction of gradients changes 90 degrees as a result of the effect of self-gravity. We noticed that the blocks over which this transition happens exhibit a much worse single direction fitting, which corresponds to higher level of fitting errors. We showed that my measuring the level of the errors it is possible to identify the regions of gravitational infall. 

A sister technique tracing magnetic fields using synchrotron intensity gradients (SIGs) was introduced in LYLC, where it was was shown that it is possible to establish the optimal size of the block for averaging providing the maximum resolution of the reliable magnetic field tracing. We employed this approach in the present paper. 

The correlation between density gradients and magnetic fields was empirically revealed by \cite{Soler2013}.  The technique did not attempt to trace the 
magnetic field but was intended for establishing the statistical correlation between the observed structures and the magnetic field direction. In terms of observational studies the technique in \cite{Soler2013} relied on the comparison of the observed intensity gradients and polarization and the study was answering the question whether the preferential orientation of density structures at a given column density is parallel or perpendicular to the magnetic field. \cite{Soler2013} introduced Histograms of Relative Orientation (HRO) that provided an empirical insight of the probability of the intensity gradients to be aligned with magnetic field depending on the density of the media. Note, that if plotted in the pictorial plane, the orientation of magnetic field and intensity gradient present a rather chaotic patten, which prevents one from tracing magnetic field with the gradients. It is in the statistical sense that the HRO establishes the change of the alignment with the ambient density. A detailed comparison of the HRO (see also Soler et al 2013,2017, Soler \& Hennebelle 2017) with our approach is provided YL17b.

The effect of partial alignment of density gradients with magnetic field in diffuse medium is easy to understand from the point of view of the gradient theory in this paper. The velocity mixing motions corresponding to the MHD cascade in diffuse media project their properties on the density structures. Thus low sonic Mach numbers such structures are preferentially elongated along magnetic field and velocity and density gradients behave similarly. However, the difference between density and velocity structures becomes significant as the Mach number increases (see Kowal et al. 2007). 
Thus the density gradients are inferior in tracing magnetic fields compared to velocity gradients. 

Note, that the procedures of calculation of gradients in HRO do not included block averaging and therefore the maps of density gradients obtained in that study exhibit chaotic variations of magnetic fields, which makes it impossible to trace magnetic fields even in low Mach number turbulence. By applying our block averaging procedure to densities we can introduce the Intensity Gradient (IG) technique that is different from the HRO. Within
this technique one can determine both the direction of the density gradients and the uncertainty over the image plane. This information is complementary to that from HRO. The sensitivity of density gradients to shocks within the IG technique can be used to identify shocks in the interstellar medium, as it discussed in YL17b. 

The preferential alignment of filaments within HI channel maps in the direction of magnetic field was established observationally in \cite{Clark15}. This work got significant resonance  as it suggested the possibility of using this alignment to improve the separation of CMB polarization from the galactic microwave foreground. In the work by \cite{Clark15} the thickness of channels was empirically chosen to be 3 km/s and the authors identify the structures with actual density filaments that exist in HI. Indeed, we were told (Clark, private communication) of the correspondence of the HI filaments in channel maps and the filaments observed in dust which suggests that the filaments are actual real physical objects rather than caustics. In this situation, it is necessary to study what is happening with the filaments as the channel map thickness increases. The theory in Lazarian \& Pogosyan (2000) suggests that \cite{Clark15} studies "thin channel maps" where most of the structures should be due to velocity caustics. Indeed, assuming that the typical turbulent velocity dispersion is $\sim 10 km/s$ and that the turbulence is Kolmogorov-type, i.e. $v_l\sim l^{1/3}$, one can estimate that the turbulent motions at scales larger than $L/(10/3)^3\approx 0.03L$, where $L$ is the injection scale, are in the "thin slice regime" (Lazarian \& Pogosyan 2000). Using the injection scale of $\sim 100$~pc obtained in Chepurnov et al. (2011), one concludes that on scales larger than 3 pc the intensity of fluctuations in the channel maps is produced to a significant degree by velocity caustics rather than the real physical entities, i.e. filaments. The calculations in \cite{Clark15} are produced on the maps smoothed up to $FWHM=30'$. Based on the galactic rotation curve, the structures from the channel maps with velocity ranges $-56km/s \sim -10km/s$ or $10km/s \sim 56km/s$ are mostly consists of intensity structures. However, the "velocity crowding effect" is expected to be more severe in channels maps at  $-10km/s \sim 10km/s$, in which the structures in the channel maps are coming from contributions from clouds in different physical locations along the line of sight.

What is peculiar about HI is still an issue of further investigations. Apparently, we do not observe any aligned features similar to filaments in our CO studies. At the same time velocity gradients trace successfully magnetic fields both in HI and denser regions like CO. We defer the issue of the nature of the filaments in \cite{Clark15} to future publications. 
 
We feel that the advantage of the velocity gradient technique is that it is rooted both in the MHD turbulence theory (see Brandenburg \& Lazarian 2013 for a review) and the theory of PPV statistics (see Lazarian \& Pogosyan 2002, Kandel et al. 2016). It is important that  the velocity gradients have a straightforward relation to the compressible MHD turbulence theory. For instance, using the VChGs, we explore the effect of three fundamental modes of MHD turbulence, i.e. Alfven, slow and fast, on the analysis. These modes affect the properties of the VChGs differently and our present opens a new avenue for the development of the gradient analysis. In particular, we feel that similar to the techniques of separating Alfven, slow and fast modes using correlation anisotropy as in the synchrotron study in \cite{LP12} and subsequent studies of velocity correlations in \cite{2016MNRAS.461.1227K,2017MNRAS.464.3617K}, the velocity gradients can be used to determine the regions with the prevalence of compressible or incompressible turbulence as well as determining the relative contribution of different modes. The corresponding procedures will be discussed elsewhere. 

\subsection{Towards a unified gradient technique}
\label{subsec:upic}
\begin{table*}[t]
 \centering
 \label{tab:gradient_table}
 \caption {A summary on the gradient alignment in different physical regime}
 \begin{tabular}{c c c c c c c}
  Papers & Gradients & Diffuse Alfven modes &Diffuse Slow modes &Diffuse Fast modes  & Self-grav. & Self-grav. rotation speed\\ \hline \hline
GL17,YL17 & VCGs & $\bot$ & $\bot$ & $\parallel$ &$\parallel$ & slow\\
YL17,YL17b & IGs  & $\bot$ & $\bot$ & $\parallel$ &$\parallel$ & fast\\
LYLC & SIGs & $\parallel$ & $\parallel$ & $\bot$ & $\parallel$ & not rotated\\
This work & thin VChGs &  $\bot$ & $\bot$ & $\parallel$ &$\parallel$ & slow\\
This work & thick VChGs & $\bot$ & $\bot$ & $\parallel$ &$\parallel$ & fast\\ \hline \hline
\end{tabular}
\end{table*}
The properties of magnetic field in magnetized turbulence are well established by a series of groundbreaking papers starting with GS95. In fact, the application of the turbulence theory to observations has been explored in a number of studies. In particular, using the analytical treatment of the PPV developed in LP00 and \cite{2004ApJ...616..943L,2006ApJ...652.1348L,2017MNRAS.464.3617K}, we provided the treatment of the anisotropies of correlation functions of the channel map intensities and of the velocity centroids in \cite{2017MNRAS.464.3617K}. These studies shed light on the properties of velocity gradients that we study in the present paper.  

Our series of papers starting with GL17 and YL17a provide a way on tracking magnetic field direction in realistically turbulent interstellar medium (see also LYLC,YL17b). We illustrated that the gradient technique is superior in terms of tracing magnetic field directions to the anisotropy tracing technique described in \S \ref{sec:anisotropy}. Nevertheless, these methods are complementary. Their synergy could provide both the detailed structure of magnetic field through using the velocity gradients as we discuss in the present paper and the information about the turbulence compressibility as it can be tested using the anisotropy of the correlation functions of the channel map intensities and velocity centroids (see Kandel et al. 2016, 2017). 

In our earlier papers we were focused on velocity centroids as the measures of velocity that are available from observations, the present paper introduces new measures, namely, gradients within channel maps or the VChGs and reduced velocity centroids gradients (RVCGs). These measures enhance the abilities of the gradient technique. In particular, both measures can analyze a part of a spectral line which has significant advantages, e.g. in the case of using galactic rotation curve to get the magnetic field direction variation along the line of sight. Other ways of constructing measures of turbulent velocity from observed spectral lines are also possible. In YL17b we showed advantages of using higher moments of velocity centroids when the observational data has low noise. Similarly, higher moments of reduced centroids were suggested for the RVCGs within this paper. The study of these different measures is another important avenue for the gradient research. These measures are affected by different degree by the fluctuations of density and this opens a way to study better disentangling density and velocity effects in the gradient studies. As we discussed in detail in YL17b the two fields have different properties and the difference is important for e.g. identifying the regions of collapse induced by self-gravity. 

The application the technique to different spectral lines corresponding allows one to study magnetic fields separately in different regions along the line of sight. This, in particular, is valuable for molecular clouds where by choosing the corresponding transitions one can study a cloud magnetic fields in depth layer by layer. 

These and other ways of gaining new information from gradients will be demonstrated elsewhere.In addition, velocity and intensity gradients (IGs) can be used together with the synchrotron intensity gradients (SIGs) and this was demonstrated in LYLC.

The procedures that we developing for the VChGs are also applicable to the IGs, the VCGs and the SIGs. In fact, the block averaging makes our IG technique very different from that in Soler et al. (2013). The latter work s
Compared to the VCGs and the IGs, the VChGs include the advantages of both diagnostic tools, while being influenced by the interfering effects. In fact, as the thickness of the velocity channels increases the VChGs gradually transfer to the IGs. However, compared the IGs, the VChGs contain much more information as the transfer from the thin to thick gradients is taking place. At the same time, it is not right to think that VChGs completely overshadows the IGs. For instance, the IGs can be applied to dust emission, where the is no lines to study using the VChGs. 

The VChGs and the RVCGs provide a useful tool for studying self-absorbing media (see \citealt{GLB17} for a discussion of the VCGs in self-absorbing media) and lines with complex structure, e.g. velocity lines have multi-peaks. Both the VChGs and the RVCGs allow using a part of a spectral line. For strongly self-adsorbing media the useful parts could be the wings. At the same time, for the lines with multi-peaks, the complexity of the line may present the complexity of the spatial structure of the emitting region. In the latter case studies of the separate peaks can provide the information about the magnetic field within a complex region. Detailed studies of the utility of the VChGs and the RVCGs as well as their complementary nature will be provided elsewhere. 

\subsection{Prospects of the technique}

The galactic rotation opens new prospects for studying the detailed structure of magnetic field in the Milky Way. The position of the Solar System within the galactic disk makes it impossible for us to use far infrared polarimetry to study magnetic fields of most of the molecular clouds. Indeed, for most of such studies the line of sight inevitably crosses more than one cloud. This confusion, combined with the failure of grain alignment at large optical depths (see Lazarian 2007) makes a polarimetric study most of the star formation hotbeds impossible. In comparison, the approach based on using gradients can employ the galactic rotation curve to separate the contribution from different clouds.  Moreover, the velocity gradients can probe the magnetic connection of the diffuse gas and molecular clouds. Such studies are impossible with far infrared polarimetry due to the signal from the diffuse media being too weak.

Velocity gradients add up differently compared to the Stokes parameters that are used in polarization. Therefore for exploring magnetic field in the galactic plane where we may expect significant direction variation along the line of sight, the VChGs and the RVCGs are advantageous compared to the VCGs. Using the former tools one can subdivide the line into velocity segments of the order of the turbulent velocity dispersion of the order of $\sim 7$ km/s and calculate the magnetic field direction for these segments. Within these segments the thin channel maps are used for calculating the VChGs and RVCGs. To compare with the polarization we then can use the mock Stock parameters similar as it is done in \cite{Clark15} for the filaments.  

In terms of resolving the spacial structure of magnetic field using the galactic rotation curve, for the atomic hydrogen studies the spatial resolution depends  both on the direction of study, the velocity range as well as on the turbulent velocity dispersion. The latter provides the grid size $\delta V$ that should be multiplied to the visual shear along the line of sight $(dV_{gal}/dz)^{-1}$ in the direction of the observations.  

If  the direction of magnetic field is changing along the line of sight the VChGs and the RVCGs can provide a coarse graded picture of the changes of the magnetic field direction perpendicular to the line of sight. In this way we can distinguish the variations of regular magnetic field over the cells that are larger than $V_L (dV_{gal}/dz)^{-1}$.  

There is a particular domain where velocity gradients present the only way for magnetic fields tracing. This is the case of high velocity clouds. They are tenuous compared to the interstellar medium along the line of sight and therefore any polarization associated with them is not possible to detect. However, these clouds are vivid in the velocity space, which gives a way to use either the VChGs or the RVCGs to map their magnetic fields. 

Another domain where dust polarimetry fails to trace magnetic field is related to circumstellar accretion disks. The dust there is aligned by radiative
torques, which according to the alignment theory in Lazarian \& Hoang (2007) align grains in the vicinity of stars in respect to the radiation direction
rather than the magnetic field direction (see also Tazaki et al. 2017). Therefore in spite of the ability of ALMA to resolve some of such disks, it
cannot really study their magnetic field structure. We expect that velocity gradient will be able to trace magnetic fields in circumstellar accretion disks. 

The change of the channel thickness changes the relative contributions of the velocity and density fluctuations into the channel map. Taking the thickness of the channel map of the order of the velocity injection one produces intensity-dominated channel maps and therefore the gradient study of such maps is equivalent to the study using the IGs. 

It is also important that the VChGs can be applied to interferometric data with in the situation when no single dish observations are available. This extends the application of the technique to distant and extragalactic objects in an important way.

We stated above that the observational information on turbulence is available only in the system of the mean field.  There are exception to this, however.  For instance, if only a thin slice of turbulence is seen due to the dust absorption effects (see \citealt{2017MNRAS.470.3103K}), the effects of the local system of reference may get important. Similarly, if the object under study and the turbulence scale are comparable, then the dispersion of magnetic field is going to decrease with the scale.

The synergetic use of the polarimetry and the velocity gradients can be very beneficial. For instance, velocity gradients can be employed to study magnetic fields in the disk of the galaxy, where the traditional far infrared polarimetry suffers from the effects of confusion as many clouds can be along the same line of sight. The galactic rotation curve can help isolate different clouds in the velocity space and allow studying their magnetic fields separately. 

The available large telescopes and interferometers can provide much better resolution that the far infrared telescopes on the balloons. The comparison of the polarimetry and velocity gradients can provide the information about the regions of the gravitational collapse as the velocity gradients will change their direction for such regions. As this low resolution identification of the gravity-dominated regimes is done, the velocity gradients may be used to study the details of the magnetic field structure that are not available with the existing far infrared polarimetry. 

As different molecules are produced at different depths inside clouds, the velocity gradients can study 3D magnetic field structure of the clouds. This provides a new dimension for the magnetic field studies. 

It is important to note, that the far infrared polarimetry and velocity gradients do not provide identical information. First of all, the magnetic fields that are traced by the polarimetry are weighted by the dust density, the latter is being proportional to the gas density. Within the VChG approach the contribution of the density fluctuations are reduced and potentially the directions of magnetic field measured by the VChGs can be closer to the actual projected magnetic field.\footnote{Velocity Centroid gradients of different orders (see YL17a) should have different weighting of the fluctuations of density along the line of sight. As a result, potentially, combining those with the VChGs one can get the actual projected magnetic field.} Moreover, the addition of gradients along the line of sight is happening differently from the addition of the Stocks parameters Q and U in the case of polarimetry. This opens a way of probing the 3D magnetic structure combining polarimetry and velocity gradients. We discuss this possibility in Yuen et al. (2018). 

While synergetic with polarimetry, velocity gradients present an independent way of magnetic field studies that does not require polarimetric information. Naturally, testing the velocity gradients with as much polarimetric data is advantageous to get more confidence in the new technique. Eventually, velocity gradients should be used on their own, however. 

We have identified (see YL17b) the 90 degree change in the relative orientation of velocity gradients and magnetic field directions in the case of self-gravitational collapse. Other situations when regular flows dominate turbulence can be present. Therefore it is important to provide more numerical studies of velocity gradients in expanding HI regions, supernovae explosions etc. in order to see whether one should expect the change of the direction of velocity gradients in other astrophysical settings. If this happens, it is important to test our approach based on calculating of the uncertainties of the fitting of velocity gradients as the way of identifying the change. 

Studies of the microwave foregrounds are extremely important for the attempts to detect and explore the polarization induced  enigmatic cosmological B-modes. Separating this polarization from the polarization arising from galactic foreground requires a breakthrough in understanding of galactic magnetic fields. Velocity gradients, e.g. the VChGs and RVCGs, as well as SIGs provide an independent way of mapping magnetic fields. The corresponding information can be used both independently to predict the foreground polarization, or preferably, as a prior for the polarization studies.

\section{Summary}
\label{sec:conclusion}
In this paper we have shown that a new measure, i.e. gradients calculated within velocity channel maps (VChGs) can trace magnetic field both in diffuse media and in molecular clouds. The VChGs of the thin channel maps that carry the information about turbulent velocities, while VChGs of the thick maps carry the information about the turbulent densities. The essence of emerging technique is to vary the channel thickness to get complementary information on both magnetic fields and shocks. As velocity gradients are a more direct tracers of magnetic field most of the paper is devoted to the VChGs calculated for the thin velocity channels. We compared the abilities of the VChGs and another new measure Reduced Velocity Centroid Gradients (RVCGs) that also traces magnetic fields with spectroscopic data.
In particular,
\begin{enumerate} 
\item We demonstrated  the alignments of VChGs obtained with basic MHD modes, i.e.
\begin{enumerate}
\item the VChGs from Alfven modes are perpendicular to magnetic field, and demonstrate the highest alignment;
\item the VChGs from Slow modes are also perpendicular to magnetic field but show somewhat reduced alignment compared to that from Alfven modes; their alignment drops faster as channel width increases;
\item the VChGs from Fast modes are parallel to magnetic field.
\end{enumerate}
\item We showed that the VChGs are more powerful in tracing magnetic field directions than the channel map correlation functions that we proposed earlier, namely, the VChGs provide more detailed information about magnetic field and can trace magnetic field for both supersonic and  subsonic turbulence.
\item We applied the VChGs to the observational HI data and compared the VChGs tracing of magnetic field with the {\it PLANCK} polarimetry data. 
\item We found that the RVCGs are comparable in their performance to the VChGs and both techniques can be used to trace magnetic field in diffuse interstellar gas and its interface with molecular clouds, molecular clouds, high velocity clouds etc.  
\item We believe that the VChGs are synergetic to other ways of magnetic field studies, in particular, to the far infrared polarimetry. Nevertheless, it is a independent way of studying magnetic field which can be used to trace magnetic fields on it own.
\item We claim that the VChGs and RVCGs can trace magnetic fields in situations when the traditional far-infrared polarimetry fails, e.g. due to the failure of dust to be aligned or due to the confusion effect typical for studying molecular clouds at low galactic latitudes. 
\item We demonstrated the advantages of the synergistic use of the different types of gradients (e.g. synchrotron intensity, spectral line intensity) paving a way for a new Gradient Technique of studying magnetic field ecosystem, shocks and self-gravitational collapse. The technique can provide the magnetic field structure that is valuable for disentangling galactic foregrounds and CMB polarization. 

\end{enumerate}

{\bf Acknowledgments.}  Elucidating discussions with Chris McKee and Susan Clarke are acknowledged. We thank Victor Lazarian for a number of suggestions in improving our presentation. We also thank Susan Clark and Laura Fissel in providing easy access to their data for our analysis. We thank the technical support from Ka Wai Ho for the 3D visualization of our data. AL acknowledges the support the NSF grant DMS 1622353 and AST 1715754.  The stay of KHY at UW-Madison is supported by the Fulbright-Lee Fellowship.  

\begin{thebibliography}{}
\providecommand\natexlab[1]{#1}
\providecommand\JournalTitle[1]{#1}
\bibitem[Andersson et al.(2015)]{2015ARA&A..53..501A} Andersson, B.-G., Lazarian, A., \& Vaillancourt, J.~E.\ 2015, \araa, 53, 501 
\bibitem[{Armstrong {et~al.}(1995)Armstrong, Rickett, \&
  Spangler}]{Armstrong1995ElectronMedium}
Armstrong, J.~W., Rickett, B.~J., \& Spangler, S.~R. 1995,
  \href{http://dx.doi.org/10.1086/175515}{\JournalTitle{The Astrophysical
  Journal}, 443, 209}

\bibitem[Beresnyak et al.(2005)]{2005ApJ...624L..93B} Beresnyak, A., Lazarian, A., \& Cho, J.\ 2005, \apjl, 624, L93 
\bibitem[Beck(2015)]{Beck15} Beck, R.\ 2015, Magnetic Fields in Diffuse Media, 407, 507 
\bibitem[Brandenburg \& Lazarian(2013)]{2013SSRv..178..163B} Brandenburg, A., \& Lazarian, A.\ 2013, \ssr, 178, 163 

\bibitem[{Burkhart {et~al.}(2012)Burkhart, Lazarian, \&
  Gaensler}]{Burkhart2012PropertiesMaps}
Burkhart, B., Lazarian, A., \& Gaensler, B.~M. 2012,
  \href{http://dx.doi.org/10.1088/0004-637X/749/2/145}{\JournalTitle{The
  Astrophysical Journal, Volume 749, Issue 2, article id. 145, 16 pp. (2012).},
  749}
\bibitem[Chepurnov \& Lazarian(2009)]{2009ApJ...693.1074C} Chepurnov, A., \& Lazarian, A.\ 2009, \apj, 693, 1074

\bibitem[{Chepurnov \& Lazarian(2010)}]{Chepurnov2010ExtendingData}
Chepurnov, A., \& Lazarian, A. 2010,
  \href{http://dx.doi.org/10.1088/0004-637X/710/1/853}{\JournalTitle{The
  Astrophysical Journal, Volume 710, Issue 1, pp. 853-858 (2010).}, 710, 853}

  
\bibitem[{Cho \& Lazarian(2002)}]{Cho2002CompressiblePlasmasb}
Cho, J., \& Lazarian, A. 2002,
  \href{http://dx.doi.org/10.1103/PhysRevLett.88.245001}{\JournalTitle{Physical
  Review Letters, vol. 88, Issue 24, id. 245001}, 88}

\bibitem[{Cho \& Lazarian(2003)}]{CL03}
---. 2003,
  \href{http://dx.doi.org/10.1046/j.1365-8711.2003.06941.x}{\JournalTitle{Monthly
  Notices of the Royal Astronomical Society, Volume 345, Issue 12, pp.
  325-339.}, 345, 325}

\bibitem[{Cho {et~al.}(2001)Cho, Lazarian, \&
  Vishniac}]{Cho2001SimulationsMedium}
Cho, J., Lazarian, A., \& Vishniac, E. 2001,
  \href{http://dx.doi.org/10.1086/324186}{\JournalTitle{The Astrophysical
  Journal, Volume 564, Issue 1, pp. 291-301.}, 564, 291}
  
\bibitem[Cho \& Vishniac(2000)]{2000ApJ...539..273C} Cho, J., \& Vishniac, E.~T.\ 2000, \apj, 539, 273 
\bibitem[Clark et al.(2015)]{Clark15} Clark, S.~E., Hill, J.~C., Peek, J.~E.~G., Putman, M.~E., \& Babler, B.~L.\ 2015, Physical Review Letters, 115, 241302 

\bibitem[{{Clarke} \& {Ensslin}(2006)}]{2006AJ....131.2900C}
{Clarke}, T.~E., \& {Ensslin}, T.~A. 2006,
  \href{http://dx.doi.org/10.1086/504076}{\JournalTitle{\aj}, 131, 2900}
  
  
\bibitem[Clemens et al.(2014)]{2014AAS...22422006C} Clemens, D.~P., Cashman, L., Hoq, S., Montgomery, J., \& Pavel, M.~D.\ 2014, American Astronomical Society Meeting Abstracts \#224, 224, 220.06 
\bibitem[Correia et al.(2016)]{2016ApJ...818..118C} Correia, C., Lazarian, A., Burkhart, B., Pogosyan, D., \& De Medeiros, J.~R.\ 2016, \apj, 818, 118 

\bibitem[{{Dolginov} \& {Mitrofanov}(1976)}]{1976Ap&SS..43..291D}
{Dolginov}, A.~Z., \& {Mitrofanov}, I.~G. 1976,
  \href{http://dx.doi.org/10.1007/BF00640010}{\JournalTitle{\apss}, 43, 291}

\bibitem[Draine(2009)]{D09} Draine, B.~T.\ 2009, \ssr, 143, 333 

\bibitem[{Draine(2011)}]{Draine2011PhysicsMedium}
Draine, B.~T. 2011, {Physics of the interstellar and intergalactic medium}
  (Princeton University Press), 540

\bibitem[{{Draine} \& {Weingartner}(1996)}]{1996ApJ...470..551D}
{Draine}, B.~T., \& {Weingartner}, J.~C. 1996,
  \href{http://dx.doi.org/10.1086/177887}{\JournalTitle{\apj}, 470, 551}

\bibitem[{{Esquivel} \& {Lazarian}(2005)}]{EL05}
{Esquivel}, A., \& {Lazarian}, A. 2005,
  \href{http://dx.doi.org/10.1086/432458}{\JournalTitle{\apj}, 631, 320}
 \bibitem[Esquivel \& Lazarian(2005)]{2005ApJ...631..320E} Esquivel, A., \& Lazarian, A.\ 2005, \apj, 631, 320 
 \bibitem[Esquivel et al.(2007)]{2007MNRAS.381.1733E} Esquivel, A., Lazarian, A., Horibe, S., et al.\ 2007, \mnras, 381, 1733 
 \bibitem[Eyink et al.(2011)]{2011ApJ...743...51E} Eyink, G.~L., Lazarian, A., \& Vishniac, E.~T.\ 2011, \apj, 743, 51 
\bibitem[Fernandez et al.(2014)]{2014MNRAS.440..298F} Fernandez, E.~R., Zaroubi, S., Iliev, I.~T., Mellema, G., \& Jeli{\'c}, V.\ 2014, \mnras, 440, 298
\bibitem[Esquivel \& Lazarian(2011)]{2011ApJ...740..117E} Esquivel, A., \& Lazarian, A.\ 2011, \apj, 740, 117 
\bibitem[{Gaensler {et~al.}(2011)Gaensler, Haverkorn, Burkhart, Newton-McGee,
  Ekers, Lazarian, McClure-Griffiths, Robishaw, Dickey, \&
  Green}]{Gaensler2011Low-Mach-numberGradients}
Gaensler, B.~M., Haverkorn, M., Burkhart, B., {et~al.} 2011,
  \href{http://dx.doi.org/10.1038/nature10446}{\JournalTitle{Nature, Volume
  478, Issue 7368, pp. 214-217 (2011).}, 478, 214}
\bibitem[Galtier et al.(2005)]{Gal2005} Galtier, S., Pouquet, A., \& Mangeney, A.\ 2005, Physics of Plasmas, 12, 092310

\bibitem[{{Ginzburg}(1981)}]{1981MoIzNRG}
{Ginzburg}, V.~L. 1981, \JournalTitle{Moscow Izdatel Nauka}

\bibitem[{Goldreich(1995)}]{GoldreichP.Sridhar1995GS95IITurbulence}
Goldreich, P.~;Sridhar, S. 1995,
  \href{http://dx.doi.org/10.1086/174600}{\JournalTitle{The Astronomical
  Journal}, 438, 763}
 \bibitem[Gonz{\'a}lez-Casanova \& Lazarian(2017)]{GL17} Gonz{\'a}lez-Casanova, D.~F., \& Lazarian, A.\ 2017, \apj, 835, 41 

\bibitem[Gonz{\'a}lez-Casanova et al.(2017)]{GLB17} Gonz{\'a}lez-Casanova, D.~F., Lazarian, A., \& Burkhart, B.\ 2017, arXiv:1703.03035 

\bibitem[{Gonz{\'{a}}lez-Casanova \&
  Lazarian(2016)}]{Gonzalez-Casanova2016VelocityFields}
Gonz{\'{a}}lez-Casanova, D.~F., \& Lazarian, A.
  \href{http://arxiv.org/abs/1608.06867}{2016}

\bibitem[{{Haverkorn} {et~al.}(2006){Haverkorn}, {Gaensler},
  {McClure-Griffiths}, {Dickey}, \& {Green}}]{2006ApJS..167..230H}
{Haverkorn}, M., {Gaensler}, B.~M., {McClure-Griffiths}, N.~M., {Dickey},
  J.~M., \& {Green}, A.~J. 2006,
  \href{http://dx.doi.org/10.1086/508467}{\JournalTitle{\apjs}, 167, 230}

\bibitem[{{Herron} {et~al.}(2016){Herron}, {Burkhart}, {Lazarian}, {Gaensler},
  \& {McClure-Griffiths}}]{2016ApJ...822...13H}
{Herron}, C.~A., {Burkhart}, B., {Lazarian}, A., {Gaensler}, B.~M., \&
  {McClure-Griffiths}, N.~M. 2016,
  \href{http://dx.doi.org/10.3847/0004-637X/822/1/13}{\JournalTitle{\apj}, 822,
  13}
\bibitem[Hayes et al.(2006)]{2006ApJS..165..188H} Hayes, J.~C., Norman, M.~L., Fiedler, R.~A., et al.\ 2006, \apjs, 165, 188 
\bibitem[Heyer et al.(2008)]{2008ApJ...680..420H} Heyer, M., Gong, H., Ostriker, E., \& Brunt, C.\ 2008, \apj, 680, 420-427 
\bibitem[{{Higdon}(1984)}]{1984ApJ...285..109H}
{Higdon}, J.~C. 1984,
  \href{http://dx.doi.org/10.1086/162481}{\JournalTitle{\apj}, 285, 109}

\bibitem[{{Hill} {et~al.}(2008){Hill}, {Benjamin}, {Kowal}, {Reynolds},
  {Haffner}, \& {Lazarian}}]{2008ApJ...686..363H}
{Hill}, A.~S., {Benjamin}, R.~A., {Kowal}, G., {et~al.} 2008,
  \href{http://dx.doi.org/10.1086/590543}{\JournalTitle{\apj}, 686, 363}

\bibitem[{{Iroshnikov}(1964)}]{I64}
{Iroshnikov}, P.~S. 1964, \JournalTitle{\sovast}, 7, 566
\bibitem[Jokipii(1966)]{J66} Jokipii, J.~R.\ 1966, \apj, 146, 480 

\bibitem[Kandel et al.(2017)]{2017MNRAS.470.3103K} Kandel, D., Lazarian, A., \& Pogosyan, D.\ 2017, \mnras, 470, 3103

\bibitem[Kowal et al.(2007)]{2007ApJ...658..423K} Kowal, G., Lazarian, A., \& Beresnyak, A.\ 2007, \apj, 658, 423 
\bibitem[{Kowal \& Lazarian(2010)}]{Kowal2010VelocityScalingsb}
Kowal, G., \& Lazarian, A. 2010,
  \href{http://dx.doi.org/10.1088/0004-637X/720/1/742}{\JournalTitle{The
  Astrophysical Journal, Volume 720, Issue 1, pp. 742-756 (2010).}, 720, 742}

\bibitem[{{Kraichnan}(1965)}]{K65}
{Kraichnan}, R.~H. 1965,
  \href{http://dx.doi.org/10.1063/1.1761412}{\JournalTitle{Physics of Fluids},
  8, 1385}
\bibitem[Kandel et al.(2016)]{2016MNRAS.461.1227K} Kandel, D., Lazarian, A., \& Pogosyan, D.\ 2016, \mnras, 461, 1227 
\bibitem[Kandel et al.(2017a)]{2017MNRAS.464.3617K} Kandel, D., Lazarian, A., \& Pogosyan, D.\ 2017, \mnras, 464, 3617 
\bibitem[Kandel et al.(2017b)]{2017MNRAS.470.3103K} Kandel, D., Lazarian, A., \& Pogosyan, D.\ 2017, \mnras, 470, 3103

\bibitem[{{Laing} {et~al.}(2008){Laing}, {Bridle}, {Parma}, \&
  {Murgia}}]{2008MNRAS.391..521L}
{Laing}, R.~A., {Bridle}, A.~H., {Parma}, P., \& {Murgia}, M. 2008,
  \href{http://dx.doi.org/10.1111/j.1365-2966.2008.13895.x}{\JournalTitle{\mnras},
  391, 521}

\bibitem[Lazarian et al.(2002)]{2002ASPC..276..182L} Lazarian, A., Pogosyan, D., \& Esquivel, A.\ 2002, Seeing Through the Dust: The Detection of HI and the Exploration of the ISM in Galaxies, 276, 182 
\bibitem[{Lazarian(2006)}]{Lazarian2006}
Lazarian, a. \href{http://dx.doi.org/10.1086/505796}{2006, 4}

\bibitem[{{Lazarian}(2007)}]{2007JQSRT.106..225L}
{Lazarian}, A. 2007,
  \href{http://dx.doi.org/10.1016/j.jqsrt.2007.01.038}{\JournalTitle{\jqsrt},
  106, 225}

\bibitem[{{Lazarian}(2009)}]{2009SSRv..143..357L}
---. 2009,
  \href{http://dx.doi.org/10.1007/s11214-008-9460-y}{\JournalTitle{\ssr}, 143,
  357}
\bibitem[Lazarian et al.(2015)]{2015ASSL..407..311L} Lazarian, A., Eyink, G.~L., Vishniac, E.~T., \& Kowal, G.\ 2015, Magnetic Fields in Diffuse Media, 407, 311 

\bibitem[{{Lazarian}(2016)}]{L16}
---. 2016,
  \href{http://dx.doi.org/10.3847/1538-4357/833/2/131}{\JournalTitle{\apj},
  833, 131}
 \bibitem[Lazarian \& Pogosyan(2000)]{2000ApJ...537..720L} Lazarian, A., \& Pogosyan, D.\ 2000, \apj, 537, 720 
\bibitem[Lazarian \& Pogosyan(2004)]{2004ApJ...616..943L} Lazarian, A., \& Pogosyan, D.\ 2004, \apj, 616, 943 
\bibitem[Lazarian \& Pogosyan(2006)]{2006ApJ...652.1348L} Lazarian, A., \& Pogosyan, D.\ 2006, \apj, 652, 1348 
\bibitem[{{Lazarian} \& {Pogosyan}(2012)}]{LP12}
{Lazarian}, A., \& {Pogosyan}, D. 2012,
  \href{http://dx.doi.org/10.1088/0004-637X/747/1/5}{\JournalTitle{\apj}, 747,
  5}

\bibitem[{{Lazarian} \& {Pogosyan}(2016)}]{LP16}
---. 2016,
  \href{http://dx.doi.org/10.3847/0004-637X/818/2/178}{\JournalTitle{\apj},
  818, 178}

\bibitem[{Lazarian \& Vishniac(1999)}]{Lazarian1999ReconnectionField}
Lazarian, A., \& Vishniac, E.~T. 1999,
  \href{http://dx.doi.org/10.1086/307233}{\JournalTitle{The Astrophysical
  Journal, Volume 517, Issue 2, pp. 700-718.}, 517, 700}
  
\bibitem[Lazarian et al.(2017)]{LYLC17} Lazarian, A., Yuen, K.~H., Lee, H., \& Cho, J.\ 2017, arXiv:1701.07883
\bibitem[Lee et al.(2016)]{2016ApJ...831...77L} Lee, H., Lazarian, A., \& Cho, J.\ 2016, \apj, 831, 77
\bibitem[{Lithwick \& Goldreich(2001)}]{Lithwick2001CompressiblePlasmas}
Lithwick, Y., \& Goldreich, P. 2001,
  \href{http://dx.doi.org/10.1086/323470}{\JournalTitle{The Astrophysical
  Journal, Volume 562, Issue 1, pp. 279-296.}, 562, 279}

\bibitem[{{Liu} {et~al.}(2009){Liu}, {Zakamska}, {Greene}, {Strauss}, {Krolik},
  \& {Heckman}}]{2009ApJLiu}
{Liu}, X., {Zakamska}, N.~L., {Greene}, J.~E., {et~al.} 2009,
  \href{http://dx.doi.org/10.1088/0004-637X/702/2/1098}{\JournalTitle{\apj},
  702, 1098}

\bibitem[{{Loeb} \& {Wyithe}(2008)}]{2008PhRvL.100p1301L}
{Loeb}, A., \& {Wyithe}, J.~S.~B. 2008,
  \href{http://dx.doi.org/10.1103/PhysRevLett.100.161301}{\JournalTitle{Physical
  Review Letters}, 100, 161301}

\bibitem[Mac Low \& Klessen(2004)]{MK04} Mac Low, M.-M., \& Klessen, R.~S.\ 2004, Reviews of Modern Physics, 76, 125 
\bibitem[Maron \& Goldreich(2001)]{2001ApJ...554.1175M} Maron, J., \& Goldreich, P.\ 2001, \apj, 554, 1175 

\bibitem[{{Matthaeus} {et~al.}(1983){Matthaeus}, {Montgomery}, \&
  {Goldstein}}]{1983PhRvL..51.1484M}{Matthaeus}, W.~H., {Montgomery}, D.~C., \& {Goldstein}, M.~L. 1983,\href{http://dx.doi.org/10.1103/PhysRevLett.51.1484}{\JournalTitle{Physical
  Review Letters}, 51, 1484}
\bibitem[McKee \& Ostriker(2007)]{MO07} McKee, C.~F., \& Ostriker, E.~C.\ 2007, \araa, 45, 565 

\bibitem[Nixon \& Aguado (2008) ]{NA08} {Nixon}, M.S. \& {Aguado} A.S., 2008,
\href{http://dx.doi.org/10.1016/B978-0-08-050625-8.50007-9}{Academic Press, 88.}

\bibitem[{{Montgomery} \& {Turner}(1981)}]{1981PhFl...24..825M}
{Montgomery}, D., \& {Turner}, L. 1981,
  \href{http://dx.doi.org/10.1063/1.863455}{\JournalTitle{Physics of Fluids},
  24, 825}

\bibitem[{{Pacholczyk}(1970)}]{1970ranp.book.....P}
{Pacholczyk}, A.~G. 1970, {Radio astrophysics. Nonthermal processes in galactic
  and extragalactic sources}

\bibitem[Planck Collaboration et al.(2016)]{Planck15X} Planck Collaboration, Adam, R., Ade, P.~A.~R., et al.\ 2016, \aap, 594, A10 
\bibitem[Planck Collaboration et al.(2016)]{Planck15.01} Planck Collaboration, Adam, R., Ade, P.~A.~R., et al.\ 2016, \aap, 594, A1

\bibitem[{{Schnitzeler} {et~al.}(2007){Schnitzeler}, {Katgert}, \& {de
  Bruyn}}]{2007A&A...471L..21S}
{Schnitzeler}, D.~H.~F.~M., {Katgert}, P., \& {de Bruyn}, A.~G. 2007,
  \href{http://dx.doi.org/10.1051/0004-6361:20077635}{\JournalTitle{\aap}, 471,
  L21}

\bibitem[{{Shebalin} {et~al.}(1983){Shebalin}, {Matthaeus}, \&
  {Montgomery}}]{1983JPlPh..29..525S}
{Shebalin}, J.~V., {Matthaeus}, W.~H., \& {Montgomery}, D. 1983,
  \href{http://dx.doi.org/10.1017/S0022377800000933}{\JournalTitle{Journal of
  Plasma Physics}, 29, 525}

\bibitem[Soler et al.(2013)]{Soler2013} Soler, J.~D., Hennebelle, P., Martin, P.~G., et al.\ 2013, \apj, 774, 128 
\bibitem[{{Takamoto} \& {Lazarian}(2016)}]{2016ApJ...831L..11T}
{Takamoto}, M., \& {Lazarian}, A. 2016,
  \href{http://dx.doi.org/10.3847/2041-8205/831/2/L11}{\JournalTitle{\apjl},
  831, L11}
\bibitem[Tazaki et al.(2017)]{2017ApJ...839...56T} Tazaki, R., Lazarian, A., \& Nomura, H.\ 2017, \apj, 839, 56

\bibitem[{{Westfold}(1959)}]{1959ApJ...130..241W}
{Westfold}, K.~C. 1959,
  \href{http://dx.doi.org/10.1086/146713}{\JournalTitle{\apj}, 130, 241}
\bibitem[Xu et al.(2015)]{Siyao15} Xu, S., Lazarian, A., \& Yan, H.\ 2015, \apj, 810, 44 
\bibitem[Yan \& Lazarian(2002)]{YL02} Yan, H., \& Lazarian, A.\ 2002, Physical Review Letters, 89, 281102 
\bibitem[Yan \& Lazarian(2008)]{YL08} Yan, H., \& Lazarian, A.\ 2008, \apj, 673, 942-953 
\bibitem[Yan \& Lazarian(2012)]{YL12} Yan, H., \& Lazarian, A.\ 2012, Numerical Modeling of Space Plasma Slows (ASTRONUM 2011), 459, 40 
\bibitem[Yuen \& Lazarian(2017a)]{YL17} Yuen, K.~H., \& Lazarian, A.\ 2017, arXiv:1701.07944 
\bibitem[Yuen \& Lazarian(2017b)]{YL17b} Yuen, K.~H., \& Lazarian, A.\ 2017, arXiv:1703.03026 
\bibitem[Zhang et al.(2016)]{2016ApJ...825..154Z} Zhang, J.-F., Lazarian, A., Lee, H., \& Cho, J.\ 2016, \apj, 825, 154 

\end{thebibliography}

\appendix
\section{Notations Related to Gradient Technique in This Paper}
 
\begin{table}[h]
 \centering
 \label{tab:notion_table}
  \begin{tabular}{c c}
 Acronyms & Meaning \\ \hline \hline
VCG & Velocity Centroid Gradient \\
VChG &Velocity Channel Gradient\\
RVCG & Reduced Velocity Centroid Gradient\\
AM & Alignment Measure\\
MW & Moving Window, an approach of improving accuracy of the gradients.\\
AC & Angle Constraint, an approach of using the MHD turbulence theory to improve the accuracy of gradient techinque. \\
SIG& Synchrotron Intensity gradient\\ \hline \hline
\end{tabular}
\caption{Acronyms used in this work}
\end{table}

\end{document}